\documentclass[aps,prd,twocolumn,showkeys,amsmath,amssymb,longbibliography,superscriptaddress]{revtex4-2}

\usepackage[utf8]{inputenc}
\usepackage[english]{babel}
\usepackage{graphicx}
\usepackage{xcolor}
\usepackage{amsmath,amsfonts,amssymb}
\usepackage[separate-uncertainty=true]{siunitx}
\DeclareSIUnit\bar{bar}
\usepackage[hypertexnames=false,colorlinks]{hyperref}
\usepackage{braket}
\usepackage{url}
\usepackage{xfrac}
\usepackage{multirow}
\usepackage[caption=false]{subfig}
\usepackage{minitoc}
\usepackage{xpatch}

\makeatletter
\patchcmd{\@ssect@ltx}
    {\addcontentsline{toc}{#1}{\protect\numberline{}#8}}
    {}
    {}
    {}
\makeatother
\pdfpageattr {/Group << /S /Transparency /I true /CS /DeviceRGB>>}

\definecolor{navygray}{RGB}{110,140,170}
\hypersetup{
citecolor={navygray}, 
linkcolor={navygray},
urlcolor={navygray},
}

\begin{document}

\title{Observation of Josephson Harmonics in Tunnel Junctions}

\author{Dennis~Willsch}
\thanks{The first two authors contributed equally.}
\affiliation{J\"ulich Supercomputing Centre, Forschungszentrum J\"ulich, 52425 J\"ulich, Germany}

\author{Dennis~Rieger}
\thanks{The first two authors contributed equally.}
\affiliation{IQMT, Karlsruhe Institute of Technology, 76344 Eggenstein-Leopoldshafen, Germany}
\affiliation{PHI, Karlsruhe Institute of Technology, 76131 Karlsruhe, Germany}

\author{Patrick~Winkel}
\affiliation{IQMT, Karlsruhe Institute of Technology, 76344 Eggenstein-Leopoldshafen, Germany}
\affiliation{PHI, Karlsruhe Institute of Technology, 76131 Karlsruhe, Germany}
\affiliation{Departments of Applied Physics and Physics, Yale University, New Haven, CT, USA}
\affiliation{Yale Quantum Institute, Yale University, New Haven, CT, USA}

\author{Madita~Willsch}
\affiliation{J\"ulich Supercomputing Centre, Forschungszentrum J\"ulich, 52425 J\"ulich, Germany}
\affiliation{AIDAS, 52425 J\"ulich, Germany}

\author{Christian~Dickel}
\affiliation{Physics Institute II, University of Cologne, 50937 K\"oln, Germany}

\author{Jonas~Krause}
\affiliation{Physics Institute II, University of Cologne, 50937 K\"oln, Germany}

\author{Yoichi~Ando}
\affiliation{Physics Institute II, University of Cologne, 50937 K\"oln, Germany}

\author{Rapha\"el~Lescanne}
\affiliation{LPENS, Mines Paris-PSL, ENS-PSL, Inria, Universit\'e PSL, CNRS, Paris, France}
\affiliation{Alice \& Bob, 53 Bd du G\'en\'eral Martial Valin, 75015 Paris, France}

\author{Zaki~Leghtas}
\affiliation{LPENS, Mines Paris-PSL, ENS-PSL, Inria, Universit\'e PSL, CNRS, Paris, France}

\author{Nicholas~T.~Bronn}
\affiliation{IBM Quantum, IBM~T.~J.~Watson Research Center, Yorktown Heights, NY 10598, USA}

\author{Pratiti~Deb}
\affiliation{IBM Quantum, IBM~T.~J.~Watson Research Center, Yorktown Heights, NY 10598, USA}

\author{Olivia~Lanes}
\affiliation{IBM Quantum, IBM~T.~J.~Watson Research Center, Yorktown Heights, NY 10598, USA}

\author{Zlatko~K.~Minev}
\affiliation{IBM Quantum, IBM~T.~J.~Watson Research Center, Yorktown Heights, NY 10598, USA}

\author{Benedikt~Dennig}
\affiliation{IQMT, Karlsruhe Institute of Technology, 76344 Eggenstein-Leopoldshafen, Germany}
\affiliation{PHI, Karlsruhe Institute of Technology, 76131 Karlsruhe, Germany}

\author{Simon~Geisert}
\affiliation{IQMT, Karlsruhe Institute of Technology, 76344 Eggenstein-Leopoldshafen, Germany}

\author{Simon~G\"unzler}
\affiliation{IQMT, Karlsruhe Institute of Technology, 76344 Eggenstein-Leopoldshafen, Germany}

\author{S\"oren~Ihssen}
\affiliation{IQMT, Karlsruhe Institute of Technology, 76344 Eggenstein-Leopoldshafen, Germany}

\author{Patrick~Paluch}
\affiliation{IQMT, Karlsruhe Institute of Technology, 76344 Eggenstein-Leopoldshafen, Germany}
\affiliation{PHI, Karlsruhe Institute of Technology, 76131 Karlsruhe, Germany}

\author{Thomas~Reisinger}
\affiliation{IQMT, Karlsruhe Institute of Technology, 76344 Eggenstein-Leopoldshafen, Germany}

\author{Roudy~Hanna}
\affiliation{PGI-9, Forschungszentrum J\"ulich and JARA J\"ulich-Aachen Research Alliance, J\"ulich, Germany}
\affiliation{RWTH Aachen University, 52062 Aachen, Germany}

\author{Jin~Hee~Bae}
\affiliation{PGI-9, Forschungszentrum J\"ulich and JARA J\"ulich-Aachen Research Alliance, J\"ulich, Germany}

\author{Peter~Sch\"uffelgen}
\affiliation{PGI-9, Forschungszentrum J\"ulich and JARA J\"ulich-Aachen Research Alliance, J\"ulich, Germany}

\author{Detlev~Gr\"utzmacher}
\affiliation{PGI-9, Forschungszentrum J\"ulich and JARA J\"ulich-Aachen Research Alliance, J\"ulich, Germany}
\affiliation{RWTH Aachen University, 52062 Aachen, Germany}

\author{Luiza~Buimaga-Iarinca}
\affiliation{CETATEA, INCDTIM, 400293 Cluj-Napoca, Romania}

\author{Cristian~Morari}
\affiliation{CETATEA, INCDTIM, 400293 Cluj-Napoca, Romania}

\author{Wolfgang~Wernsdorfer}
\affiliation{IQMT, Karlsruhe Institute of Technology, 76344 Eggenstein-Leopoldshafen, Germany}
\affiliation{PHI, Karlsruhe Institute of Technology, 76131 Karlsruhe, Germany}

\author{David~P.~DiVincenzo}
\affiliation{PGI-2, Forschungszentrum J\"ulich, 52425 J\"ulich, Germany}
\affiliation{RWTH Aachen University, 52062 Aachen, Germany}

\author{Kristel~Michielsen}
\affiliation{J\"ulich Supercomputing Centre, Forschungszentrum J\"ulich, 52425 J\"ulich, Germany}
\affiliation{AIDAS, 52425 J\"ulich, Germany}
\affiliation{RWTH Aachen University, 52062 Aachen, Germany}

\author{Gianluigi~Catelani}
\affiliation{PGI-11, Forschungszentrum J\"ulich, 52425 J\"ulich, Germany}
\affiliation{Quantum Research Center, Technology Innovation Institute, Abu Dhabi 9639, UAE}

\author{Ioan~M.~Pop}
\thanks{Corresponding author: Ioan Pop}
\email{ioan.pop@kit.edu}
\affiliation{IQMT, Karlsruhe Institute of Technology, 76344 Eggenstein-Leopoldshafen, Germany}
\affiliation{PHI, Karlsruhe Institute of Technology, 76131 Karlsruhe, Germany}

\date{\today}

\begin{abstract}
{\bfseries
Approaches to developing large-scale superconducting quantum processors must cope with the numerous microscopic degrees of freedom that are ubiquitous in solid-state devices. State-of-the-art superconducting qubits employ aluminum oxide (AlO$_x$) tunnel Josephson junctions as the sources of nonlinearity necessary to perform quantum operations. Analyses of these junctions typically assume an idealized, purely sinusoidal current-phase relation. However, this relation is only expected to hold in the limit of vanishingly low-transparency channels in the AlO$_x$ barrier. Here we show that the standard current-phase relation fails to accurately describe the energy spectra of transmon artificial atoms across various samples and laboratories. Instead, a mesoscopic model of tunneling through an inhomogeneous AlO$_x$ barrier predicts percent-level contributions from higher Josephson harmonics. By including these in the transmon Hamiltonian, we obtain orders of magnitude better agreement between the computed and measured energy spectra. The presence and impact of Josephson harmonics has important implications for developing AlO$_x$-based quantum technologies including quantum computers and parametric amplifiers. As an example, we show that engineered Josephson harmonics can reduce the charge dispersion and the associated errors in transmon qubits by an order of magnitude, while preserving their anharmonicity.}

\end{abstract}

\maketitle

\begin{figure}
  \centering
  \includegraphics[width=\columnwidth]{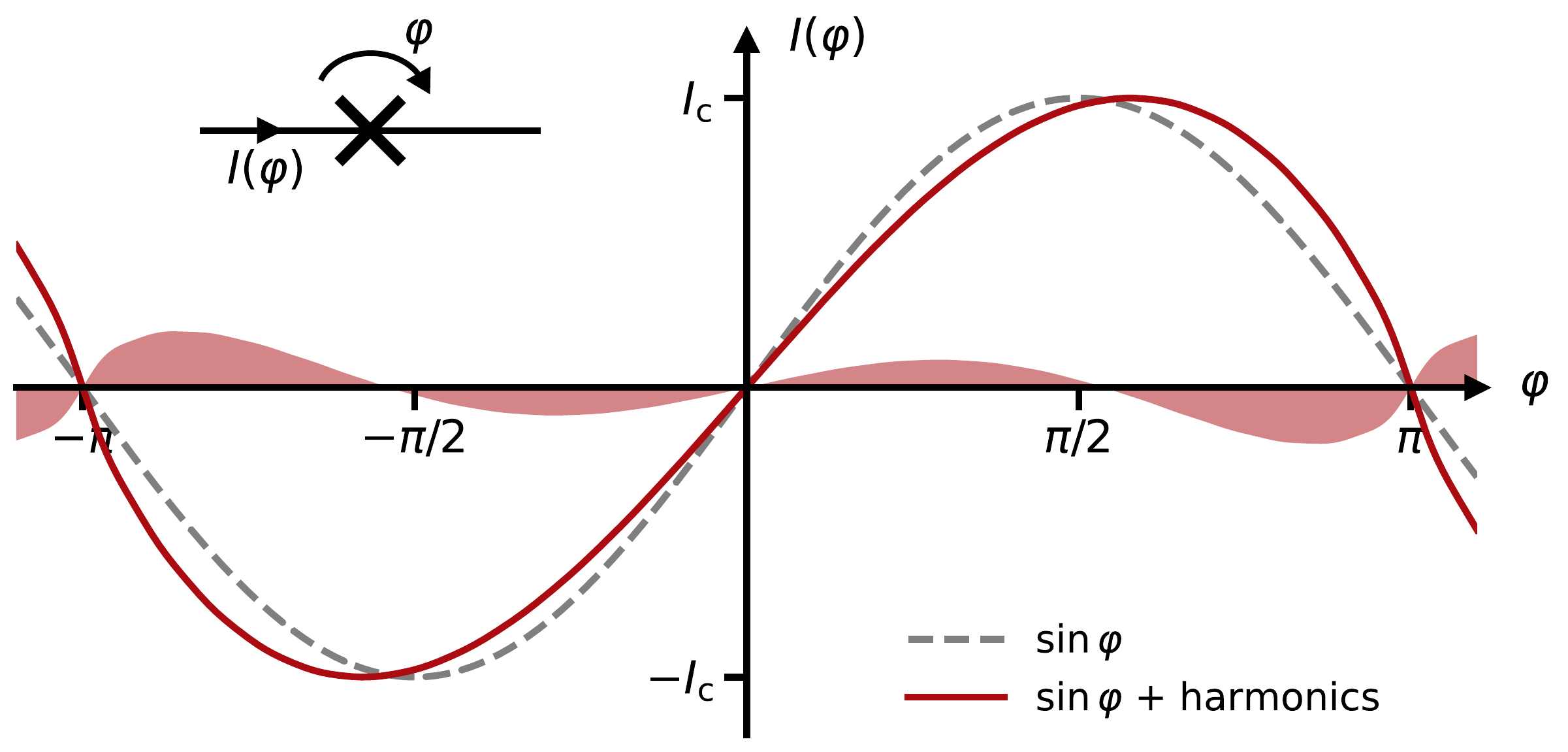}
  \caption{\textbf{Josephson harmonics are relevant for the current-phase relation (C$\varphi$R) of tunnel junctions.}
  The non-linear C$\varphi$R is the fingerprint of a Josephson junction (JJ), which relates the supercurrent $I(\varphi)$ to the phase $\varphi$ (inset).
  For tunnel JJs, the C$\varphi$R has been considered to be purely sinusoidal (dashed gray line, cf.~Eq.~\eqref{eq:SinusoidalCphiR}) with the maximum given by the critical current $I_\text{c}$.
  However, as we show in this work, even in tunnel JJs, the underlying microscopic complexity of the charge transport can manifest in the contribution of higher harmonics to the C$\varphi$R.
  As an example, the red line shows a C$\varphi$R consistent with measured data (cooldown 1 of the KIT sample; cf.~main text), which includes the harmonics expected from a mesoscopic model assuming an inhomogeneous AlO$_x$ barrier.
  The shaded red area shows the difference to the purely sinusoidal C$\varphi$R. 
  We provide C$\varphi$Rs for all other measured samples in Supplementary Fig.~\ref{suppfig:potentialsandcprs}.
  }
  \label{fig:1}
\end{figure}

The Josephson effect~\cite{Josephson1962,Josephson1974} is the keystone of quantum information processing with superconducting hardware: 
It constitutes a unique source of low-loss non-linearity, which is essential for the implementation of superconducting quantum bits, and it plays a similarly fundamental role as the non-linear current-voltage relation of diodes in semiconductor circuitry.
In particular, tunnel Josephson junctions (JJs), formed by two overlapping superconducting films separated by a thin insulating barrier, have enabled superconducting hardware to become one of the leading platforms for the realization of fault-tolerant quantum computers~\cite{GoogleQuAI2021, Krinner2022, Sivak2023QuantumErrorCorrectionBreakEven, Wendin2023QIPPerspective}. JJs are also at the heart of quantum limited amplification~\cite{Roy2016Aug}, metrological applications~\cite{K_Gramm_1976} such as the definition of the voltage~\cite{Shapiro1963} and a possible future current standard~\cite{Crescini2022}, and they enable quantum detectors such as the microwave photon counter~\cite{Albertinale2021}. With the advancement~\cite{Google2019QuantumSupremacy,King2023QuantumCriticalDynamics5000SpinGlass,Kim2023EvidenceForTheUtilityOfQuantumIBM} of superconducting artificial atom technology, the measurement and understanding of subtle features in the Josephson effect, similarly to the fine structure discovered in natural atoms, is increasingly relevant in setting the accuracy of both circuit control and circuit models.

Although the mesoscopic dimensions of JJs imply the existence of many conduction channels, for tunnel junctions this complexity is usually condensed into a single effective parameter, the critical current $I_\text{c}$, in the well-known Josephson C$\varphi$R (see gray line in Fig.~\ref{fig:1}),
\begin{equation}
  I(\varphi) = I_\text{c} \sin\varphi\,,\label{eq:SinusoidalCphiR}
\end{equation}
where $\varphi$ is the superconducting phase difference across the junction. This simplification is remarkable given the fact that other types of junctions, such as weak links, point contacts and ferromagnetic JJs, generally exhibit non-sinusoidal C$\varphi$Rs containing higher Josephson harmonics $\sin(2\varphi)$, $\sin(3\varphi)$, etc.~\cite{Likharev1986JosephsonJunctions,Golubov2004,Goldobin2007JJSecondHarmonicSFS,DeLange2015CurrentPhaseRelationFromSpectrum,Kringhoj2018SemiconductorNanowireJJ,Stoutimore2018SecondHarmonicSFSJJ,Bargerbos2020PointContactChargeDispersion}.
Here we show that Josephson harmonics are also relevant for tunnel JJs (see Fig.~\ref{fig:1}).

\begin{figure*}[t]
  \centering
  \includegraphics[width=\textwidth]{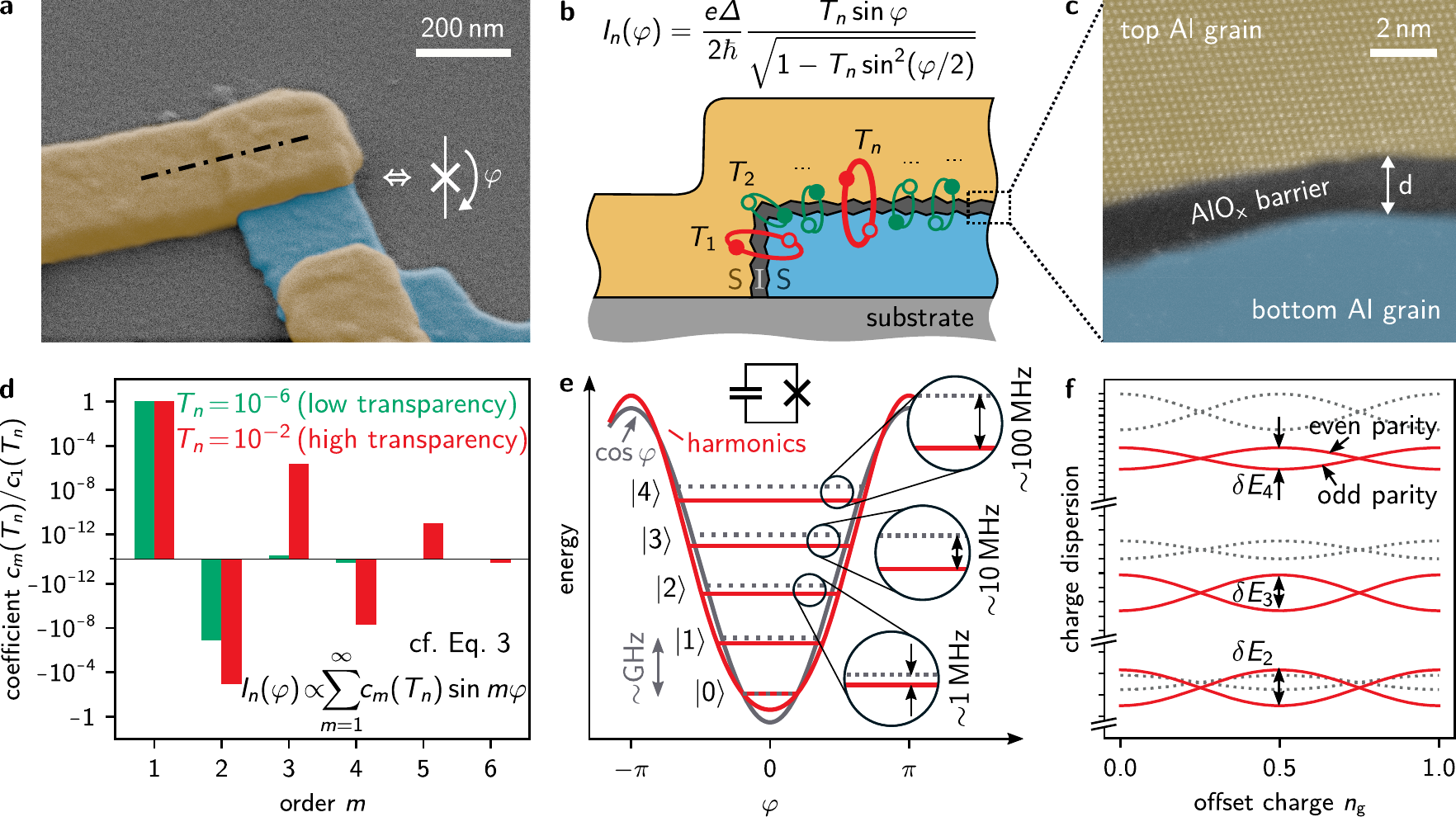}
  \caption{\textbf{Josephson harmonics result from junction barrier inhomogeneity.} \textbf{a}~False-colored scanning electron microscope (SEM) image of a typical Al-AlO$_x$-Al Josephson junction (JJ) fabricated at KIT. The bottom and top electrodes are colored in blue and yellow, respectively. The inset shows the circuit symbol for a JJ with phase difference $\varphi$ across the barrier.
  \textbf{b}~Cross-section schematic of the superconductor-insulator-superconductor (SIS) JJ at the location indicated by the dash-dotted line in panel~a. The supercurrent $I_n(\varphi)$ of each conduction channel $n=1,\ldots,N$ depends on its transparency~$T_n$ (cf.~Eq.~\eqref{eq:ChannelCurrent}). We sketch a distribution of multiple low and a few high transparencies $T_1, \dots, T_\text{N}$ in green and red, respectively.
  \textbf{c}~False-colored high-angle annular dark field scanning transmission electron microscope (HAADF-STEM) image centered on the AlO$_x$ tunnel barrier of a typical JJ fabricated at KIT, with average thickness $d\approx\SI{2}{\nano\meter}$ as indicated by the white arrow. Individual columns of atoms of the Al grain in the top electrode are visible due to zone axis alignment, which is not the case for the bottom Al electrode (additional STEM images with thickness variations and structural defects such as grain boundaries are shown in Supplementary Fig.~\ref{suppfig:tem}). 
  \textbf{d}~Normalized Fourier coefficients $c_m(T_n)$ of the JJ current-phase relation (cf.~Eq.~\eqref{eq:ChannelCurrent}) for a low ($10^{-6}$, green) and high ($10^{-2}$, red) transparency channel. Note the alternating sign for even and odd order $m$ and the fact that high transparency channel coefficients (in red) remain relevant to higher order.
  \textbf{e}~Sketch of how the higher-order terms in the JJ Hamiltonian modulate the potential and shift the energy levels (red) of superconducting artificial atoms compared to a purely $\cos\varphi$ potential (gray). In our manuscript we focus on transmon devices, which consist of a large capacitor in parallel to the JJ (refer to the circuit schematic inset). The discrepancy between the models generally increases at higher levels.
  \textbf{f}~The higher-order Josephson harmonics also influence the charge dispersion of the transmon levels vs.~offset charge $n_\text{g}$. The two branches per energy level correspond to a change between even and odd charge parity (i.e.~quasiparticle tunneling~\cite{Catelani2014ChargeParity,Serniak2019OffsetChargeSensitiveTransmon}; cf.~Supplementary Fig.~\ref{suppfig:cologne_sample_figure} in Section~\ref{sec:samplesKoeln}).
  }
  \label{fig:2}
\end{figure*}

\begin{figure*}[t]
  \centering
  \includegraphics[width=\textwidth]{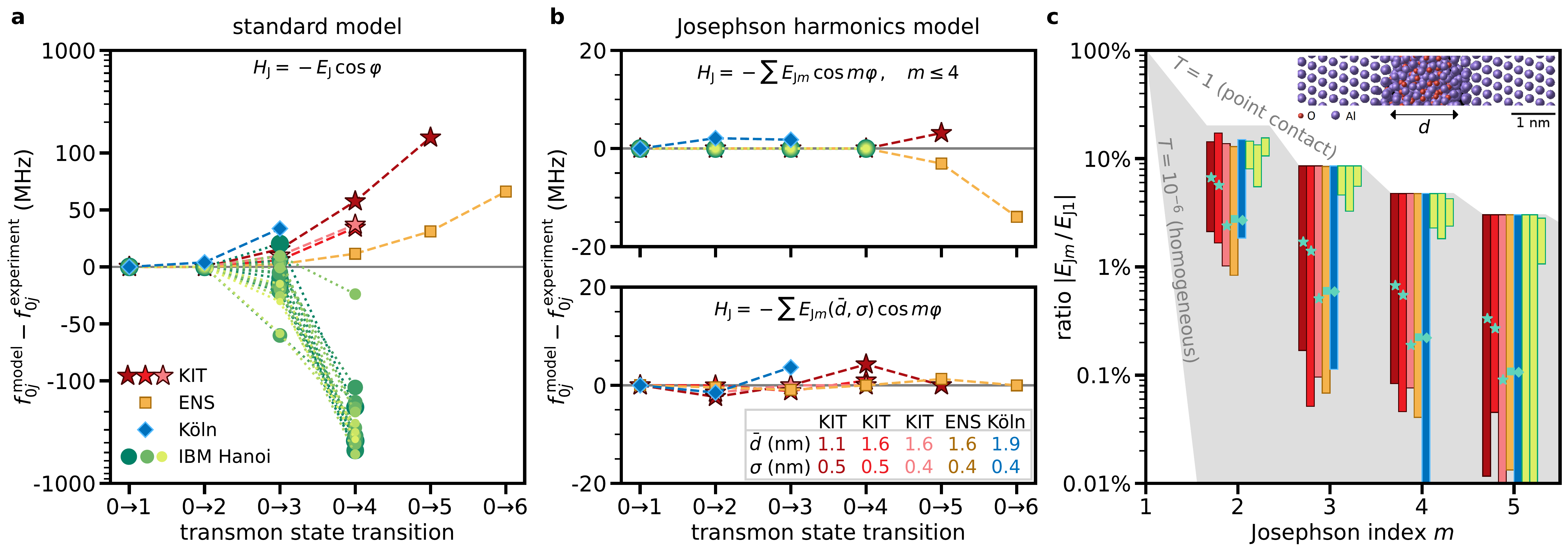}
  \caption{\textbf{Standard transmon model fails to describe the measured frequency spectra.}
  \textbf{a}~Differences between the frequencies $f_{0j}^{\mathrm{model}}$ predicted by the standard transmon model in Eq.~\eqref{eq:standardmodel} and the measured transitions $f_{0j}^{\mathrm{experiment}}$.
  The markers indicate the different experiments at KIT (red stars), ENS (yellow squares), K\"oln (blue diamonds), and IBM (green circles).
  For the KIT experiment, we show results for three successive cooldowns of the same sample (CD 1--3, dark red to bright red, respectively).
  For the K\"oln experiment we chose a set of measured transitions at a fixed magnetic field (cf.~blue arrow in Fig.~\ref{fig:4}a).
  For the IBM experiment we show results for 20 qubits in the IBM Hanoi device, using different marker sizes and shades of green.
  Measurement imprecisions are on the order of $\SI{1}{MHz}$ and not visible in the figure.
  Note that the scale on the vertical axis is linear between $\pm\SI{100}{\mega\hertz}$ and logarithmic onward.
  Dashed and dotted lines are guides to the eye.
  \textbf{b}~Same as panel a, with $f_{0j}^{\mathrm{model}}$ given by the Josephson harmonics Hamiltonian in Eq.~\eqref{eq:harmonicsmodel}. The top panel shows a model truncated at $E_{\mathrm{J}4}$. The bottom panel shows the mesoscopic model of tunneling through an inhomogeneous AlO$_x$ barrier, where $E_{\mathrm{J}m}(\bar d,\sigma)$ is parameterized in terms of the average barrier thickness $\bar d$ and the standard deviation $\sigma$ (see Eq.~\eqref{eq:dsigmamodel}; the fit values are listed in the table inset). 
  \textbf{c}~
  Ranges of the Josephson harmonics ratios $\vert E_{\text{J}m}/E_{\text{J}1}\vert$ that are
  consistent with the measured spectra.
  The ranges are represented by colored vertical bars using the same coloring as in panel~a.
  For the IBM Hanoi device, we show the ranges for qubits 0--2 from left to right (ranges for the other qubits are shown in Supplementary Section~\ref{sec:scan}).
  The shaded gray area highlights the region between two limiting cases: the fully open quantum point contact with unit transparency and a homogeneous barrier with $T_n=10^{-6}$ for all $n$.
  Turquoise markers on the vertical bars indicate the harmonics ratios calculated from the mesoscopic model, where the average thickness $\bar d$ and the standard deviation $\sigma$ are given in panel b.
  The inset shows an Al-AlO$_x$-Al junction obtained from molecular dynamics simulations (cf.~Supplementary Fig.~\ref{suppfig:md_model_stages}) with average barrier thickness $\bar d=\SI{1.5}{\nano\metre}$ (cf.~Fig.~\ref{fig:2}c).
  }
  \label{fig:3}
\end{figure*}

\begin{figure*}[t]
  \centering
  \includegraphics[width=\textwidth]{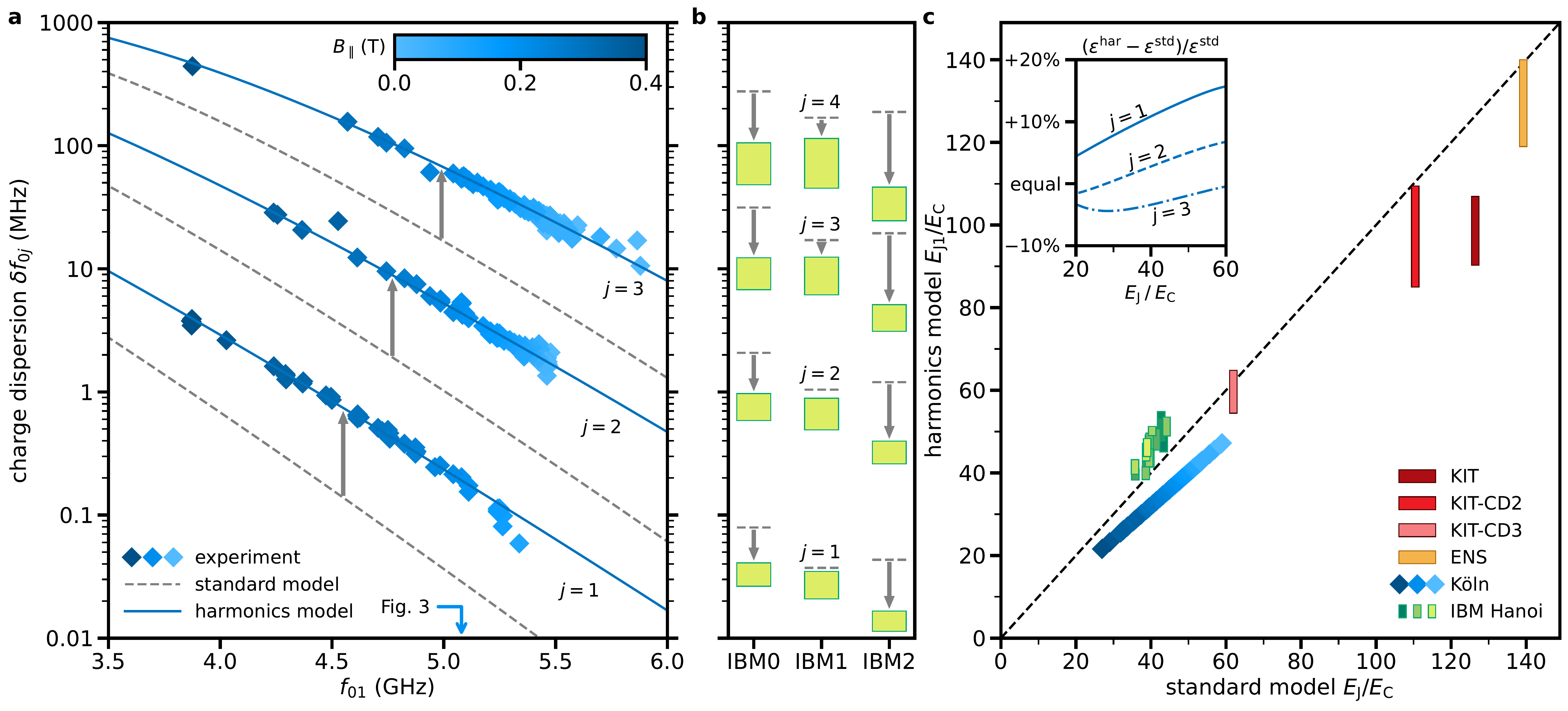}
  \caption{
  \textbf{Influence of Josephson harmonics on the charge dispersion.}  \textbf{a}~Measured charge dispersion $\delta f_{0j}$ (blue diamonds) of states $j=1,2,3$ for the experiment in K\"oln, plotted as a function of the $f_{01}$ frequency.
  All transition frequencies are tuned as the Josephson energy is suppressed by up to $\SI{35}{\percent}$ by means of an in-plane magnetic field $B_\parallel$ swept to \SI{0.4}{\tesla}.
  The standard model Eq.~\eqref{eq:standardmodel} shown in dashed gray underestimates the charge dispersion by a factor of $2$ to $7$ (gray arrows), while the Josephson harmonics model Eq.~\eqref{eq:harmonicsmodel} plotted in solid blue overlaps with the measured data.
  Note that both are computed with the same parameters used for Fig.~\ref{fig:3}; the Josephson energy is reduced with increasing magnetic field and the other parameters such as the $E_{\mathrm{J}m}/E_{\mathrm{J}1}$ ratios are kept constant.
  The blue arrow indicates $f_{01}=5.079\,\mathrm{GHz}$, corresponding to the dataset shown in Fig.~\ref{fig:3}.
  \textbf{b}~Evidence that Josephson harmonics can reduce the charge dispersion by an order of magnitude (gray arrows). 
  The dashed gray lines represent the standard model predictions.
  In contrast, the green bars show results from all Josephson harmonics models.
  The data corresponds to IBM qubits 0--2 (cf.~green bars in Fig.~\ref{fig:3}c) for the levels $j=1,2,3,4$; results for all other samples are shown in Supplementary Fig.~\ref{suppfig:chargedispersion}. 
  \textbf{c}~The values of $E_{\mathrm{J}1}/E_\mathrm{C}$ change compared to the standard model $E_\text{J}/E_\text{C}$, which constitutes the main correction to the predicted charge dispersions in panels~a and b.
  The bars represent the range of suitable ratios $E_{\mathrm{J}1}/E_\mathrm{C}$ (cf.~Fig.~\ref{fig:3}c) for the successive cooldowns of the KIT sample (red bars), the ENS sample (yellow bar), the K\"oln sample (blue diamonds, using the same color coding as in panel~a), and the IBM Hanoi device (green bars).
  The dashed diagonal indicates the case in which the ratios $E_{\mathrm{J}1}/E_\mathrm{C}$ of the harmonics model and $E_{\mathrm{J}}/E_\mathrm{C}$ of the standard model are equal.
  The inset shows the correction $(\varepsilon^{\mathrm{har}} - \varepsilon^{\mathrm{std}})/\varepsilon^{\mathrm{std}}$ to the relative charge dispersion $\varepsilon=\delta f_{0j}/f_{01}$ for fixed $E_\mathrm{J}^{\mathrm{std}}/E_\mathrm{C}^{\mathrm{std}}=E_{\mathrm{J}1}^{\mathrm{har}}/E_\mathrm{C}^{\mathrm{har}}$ for the K\"oln sample, where $\varepsilon^{\mathrm{std}}$ is given by the standard charge dispersion~\cite{koch2007transmon} and $\varepsilon^{\mathrm{har}}$ is computed using the Josephson harmonics model.  
  }
  \label{fig:4}
\end{figure*}

To understand the limits of the approximation Eq.~\eqref{eq:SinusoidalCphiR} for tunnel junctions, we have to take a closer look at commonly used Al-AlO$_x$-Al JJs, fabricated by shadow evaporation~\cite{Dolan1977} and schematized in Fig.~\ref{fig:2}a-c, which reveals a complex microscopic reality. The C$\varphi$R of the junction is obtained by summing the supercurrents of $N$ conduction channels, $I(\varphi)=\sum_{n=1}^N I_n(\varphi)$. Each channel (cf.~Fig.~\ref{fig:2}b) has a transparency-dependent C$\varphi$R~\cite{Beenakker1992,Golubov2004} which can be expressed as a Fourier series:
\begin{equation}
  I_n(\varphi) \propto \frac{T_n\sin\varphi}{\sqrt{1-T_n\sin^2(\varphi/2)}}=\!\:\sum_{m=1}^\infty\!\: c_m(T_n)\sin (m\varphi)\,.\label{eq:ChannelCurrent}
\end{equation}
The conduction channel transparency $T_n$ is defined as the tunnel probability for an electron impinging on the insulating barrier of channel $n$ and $c_m(T_n)$ are the order $m$ Fourier coefficients for $I_n(\varphi)$. These coefficients alternate in sign and decay in magnitude with increasing order $m$ (cf.~Fig.~\ref{fig:2}d). The ratio $\vert c_{m+1}/c_m\vert$ of successive coefficients increases with $T_n$ (cf.~Supplementary Section~\ref{sec:theorybackground}): the more transparent a channel, the more relevant is the contribution of higher harmonics. To put it simply, in higher-transparency channels it is more likely for Cooper pairs to tunnel together in groups of $m$, which corresponds to the $\sin(m \varphi)$ terms in the C$\varphi$R.

In the limit $T_n \to 0$, only the $\sin\varphi$ term of Eq.~\eqref{eq:ChannelCurrent} survives. If all channels in a JJ are in this limit, we recover the purely sinusoidal C$\varphi$R of Eq.~\eqref{eq:SinusoidalCphiR}, with the critical current of the junction $I_\text{c}$ proportional to the sum of transparencies. Assuming a perfectly homogeneous barrier, for a typical junction with $\sim\si{\square\micro\meter}$ area and resistance comparable to the resistance quantum, one expects $N\sim 10^6$ and $T_n\sim 10^{-6}$ \cite{Glazman2021, Kittel2004}, leading to negligible (below $10^{-6}$) corrections to the purely sinusoidal C$\varphi$R. 

But is this the reality? Here we argue that in the presence of contaminants, atomic scale defects~\cite{Fritz2019} and random crystalline orientations of the grains in contact, evidenced by STEM images and molecular dynamics simulations (see Fig.~\ref{fig:2}c and Supplementary Section~\ref{sec:moleculardynamics}), we have reasons to doubt it. 
In fact, about two decades ago, AlO$_x$ barrier inhomogeneity motivated the transition in magnetic junctions to more uniform oxides such as MgO~\cite{DaCosta2000MagneticJunctionsAlOxInhomogeneityProblem,Parkin2004GMRHomogeneousMgOJunctions,Yuasa2004GMRHomogeneousMgOJunctions}. Consequently, we expect a distribution of transparencies in AlO$_x$~\cite{Cyster2020AtomicStructureAlOxJunctions,Cyster2021SimulatingFabricationAlOxJunctions} with possibly a few relatively high-transparency channels~\cite{Aref2014NonuniformAlOxBarriers,Zeng2015ObservationThicknessAlOxJosephsonJunctions} introducing measurable corrections to the C$\varphi$R (see Fig.~\ref{fig:1}). The microscopic structure of each barrier is therefore imprinted on the C$\varphi$R of the JJ, and the challenge is how to experimentally access this information.

For our study of tunnel JJs, we use transmon devices~\cite{koch2007transmon}, in which a JJ is only shunted by a large capacitor to form a non-linear oscillator with the potential energy defined by the C$\varphi$R of the junction (cf.~Fig.~\ref{fig:2}e). The resulting individually addressable transition frequencies in the microwave regime can be measured using circuit quantum electrodynamics techniques~\cite{Blais2021}. We compare the spectra of multiple samples to the prediction of the standard transmon Hamiltonian based on a sinusoidal C$\varphi$R (Eq.~\eqref{eq:SinusoidalCphiR}) and find increasing deviations for the higher energy levels of all samples, as sketched in Fig.~\ref{fig:2}e,f. Only by accounting for higher harmonics in the C$\varphi$R are we able to accurately describe the entire energy spectrum. 
A similar methodology was used in \cite{DeLange2015CurrentPhaseRelationFromSpectrum} to reconstruct the C$\varphi$R of a semiconductor nanowire Josephson element.
While our study focuses on transmon qubits, the conclusions we draw regarding the C$\varphi$R of tunnel junctions should trigger a reevaluation of the current models for tunnel-JJ-based devices used in quantum technology and metrology~\cite{Nigg2012BlackBoxCircuitQuantization,Ansari2019BlackBoxCircuitQuantization,Riwar2022CircuitQuantizationTimeDependentMagneticField,Blais2021,Miano2023HamiltonianFluxBiasedJJ}.

Since transmons are widely available in the community, we are able to measure and model the spectra of multiple samples from laboratories around the globe: fixed frequency transmons fabricated and measured at the Karlsruhe Institute of Technology (KIT, cf.~Supplementary Fig.~\ref{suppfig:HFSS}) in three cooldowns (CDs, cf.~Supplementary Fig.~\ref{suppfig:KIT_CD1}) and Ecole Normale Sup\'erieure (ENS) Paris (same device as in Ref.~\cite{Lescanne2019EscapeUnconfinedStatesFloquet}), a tunable transmon subject to an in-plane magnetic field at the University of Cologne (K\"oln, identical setup and similar device as in Ref.~\cite{Krause2022KoelnQubit}, cf.~Supplementary Fig.~\ref{suppfig:cologne_sample_figure}), and 20 qubits from the IBM Hanoi processor (IBM). All transmons are based on standard Al-AlO$_x$-Al tunnel junctions (cf.~Fig.~\ref{fig:2}) and are measured either in a 3D architecture or a 2D coplanar waveguide geometry (for detailed descriptions of each sample see Supplementary Section~\ref{suppsec:samples}). The spectroscopy data consists of (i) transition frequencies $f_{0j}$ into transmon states $j=1,2,\ldots$ up to $j=6$, each measured as $j$-photon transitions at frequencies $f_{0j}/j$, and (ii) the resonator frequencies $f_{\mathrm{res},j}$ depending on the transmon state $j=0,1$~(see~\hyperref[sec:methods]{Methods}).

In Fig.~\ref{fig:3}, we compare the measured transition frequencies to predictions $f_{0j}^{\mathrm{model}}$, obtained by exact diagonalization of two different model Hamiltonians. The first model is the \emph{standard transmon model}, which has served the community for over 15 years~\cite{koch2007transmon},
\begin{equation}
    H_\mathrm{std} = 4E_\mathrm{C} (n-n_{\mathrm{g}})^2 - E_\mathrm{J} \cos\varphi + H_\mathrm{res}\,,
    \label{eq:standardmodel}
\end{equation}
where $E_{\mathrm{C}}$ is the charging energy, $E_{\mathrm{J}}$ is the Josephson energy,  $n_{\mathrm{g}}$ is the offset charge, and the operators $n$ and $\varphi$ represent the charge normalized by twice the electron charge and the phase difference across the junction, respectively.
All models include the readout resonator Hamiltonian given by $H_\mathrm{res} = \Omega a^\dagger a + G n(a + a^\dagger)$, where $\Omega$ is the bare resonator frequency, $G$ is the electrostatic coupling strength, and $a^\dagger$ ($a$) is the bosonic creation (annihilation) operator. Including $H_\mathrm{res}$ ensures that dressing of the states due to transmon-resonator hybridization is taken into account~\cite{blais2004circuitqed,blais2007circuitqed,koch2007transmon,Blais2021}. 

We obtain the parameter set $(E_\mathrm{C},E_\mathrm{J},\Omega,G)$ of the standard transmon model Eq.~\eqref{eq:standardmodel} by solving the Inverse Eigenvalue Problem (IEP)~\cite{Friedland1977InverseEigenvalueProblems,Friedland1987InverseEigenvalueProblemNumericalMethods,Chu1998InverseEigenvalueProblems,ChuGolub2005InverseEigenvalueProblemsBook} for the measured spectroscopy data (see~\hyperref[sec:methods]{Methods}). For the K\"oln sample this data includes the offset charge dispersion (additional data for different magnetic fields is given in Supplementary Section~\ref{sec:iepkoeln}). 
We note that the IEP is the very same science problem that was historically solved to model the energy spectra natural atoms and molecules (see e.g.~\cite{DowningHouseholder1955InverseEigenvalueProblemMolecules,Toman1966MolecularSpectroscopyInverseEigenvalueProblem,Brussaard1977ShellModelApplicationsNuclearSpectroscopyIEP}), which led to the discovery of the fine structure.

In Fig.~\ref{fig:3}a we show that the standard transmon model Eq.~\eqref{eq:standardmodel} fails to describe the measured frequency spectra for all samples. The observed deviations are much larger than the measurement imprecision, for which we can set a conservative upper bound on the order of \SI{1}{\mega\hertz}. 
While the standard transmon model with two parameters can trivially match the $f_{01}$ and $f_{02}$ transitions, the measured $f_{03}$ can already deviate by more than \SI{10}{\mega\hertz}. The deviations are positive for the KIT, ENS and K\"oln samples, while the IBM transmons mostly show negative deviations (cf.~Supplementary Section~\ref{sec:Hamiltonian}).
It is important to remark that other corrections, such as the stray inductance in the JJ leads, hidden modes coupled to the qubit, the coupling between qubits as present on the IBM multi-qubit device, or an asymmetry in the superconducting energy gaps, while being relevant, cannot fully account for the measured discrepancy (see Supplementary Section~\ref{sec:alternativecorrections}). Notably, similar deviations can be found in previously published transmon spectra \cite{Peterer2015HigherTransmonStates,Xie2018Compact3DQuantumMemory,Xie2019PhDThesis,Krause2022KoelnQubit}, as we detail in Supplementary Fig.~\ref{suppfig:additionalresults} and Sections~\ref{sec:additionalevidence} and \ref{sec:chargedispersion}. 

In Fig.~\ref{fig:3}b, we demonstrate that orders of magnitude better agreement with our measured spectra can be achieved by using the \emph{Josephson harmonics model}:
\begin{equation}
    H_\mathrm{har} = 4E_\mathrm{C} (n-n_{\mathrm{g}})^2 - \sum_{m\ge1} E_{\mathrm{J}m}\cos (m\varphi) + H_\mathrm{res} \,.
    \label{eq:harmonicsmodel}
\end{equation}
In general the values $E_{\mathrm{J}m}$ are a fingerprint of each junction's channel-transparency distribution $\rho(T)$ with many degrees of freedom. Here we consider two simplified models (further models are discussed in Supplementary Section~\ref{sec:josephsonharmonics}): (i)~a phenomenological model truncated at $E_{\mathrm{J}4}$ (top panel) 
and (ii)~a mesoscopic model of tunneling through a non-uniform oxide barrier (bottom panel).
We note the phenomenological $E_{\mathrm{J}4}$ model guarantees agreement for the lowest 4 transitions (see~\hyperref[sec:methods]{Methods}), and while many samples have physically reasonable $E_{\mathrm{J}m}$ coefficients when truncating at $E_{\mathrm{J}4}$, a few JJs require terms up to $E_{\mathrm{J}6}$ (see Supplementary Section~\ref{sec:scan}).

The mesoscopic model allows us to derive $\rho(T;\bar d,\sigma)$ based on a Gaussian thickness distribution with average thickness $\bar d$ and standard deviation $\sigma$ (see Supplementary Section~\ref{sec:mesoscopicmodel}).
As a consequence, all Josephson harmonics for $m\ge2$ are parameterized in terms of the two parameters $\bar d$ and $\sigma$ according to
\begin{equation}
    E_{\mathrm{J}m}(\bar d,\sigma) \propto \int_0^1 \, c_m(T)\, \rho(T;\bar d,\sigma)\, \mathrm{d}T\,,
    \label{eq:dsigmamodel}
\end{equation}
where the Fourier coefficients $c_m(T)$  (cf.~Eq.~\eqref{eq:ChannelCurrent} and Fig.~\ref{fig:2}d) are weighted by the channel-transparency distribution $\rho(T;\bar d,\sigma)$.
In this model, relatively large ratios $\vert E_{\mathrm{J}m}/E_{\mathrm{J}1}\vert$ originate from higher transparency contributions from the narrower regions of the barrier (cf.~the STEM images in Supplementary Fig.~\ref{suppfig:tem}).
The model can describe the samples at KIT, ENS, and K\"oln (see Fig.~\ref{fig:3}b) but not the IBM device (see~Supplementary Section~\ref{sec:mesoscopicmodel}). 
The model parameters $\bar d$ and $\sigma$ (cf.~Fig.~\ref{fig:3}b) are comparable to results from molecular dynamics simulation and STEM pictures of the oxide barrier (see Supplementary Section~\ref{sec:moleculardynamics}).

In Fig.~\ref{fig:3}c, we indicate the ranges of $E_{\mathrm{J}m}$ coefficients consistent with the measured spectra. The bars represent the lower and upper limits of Josephson harmonics ratios $\vert E_{\mathrm{J}m}/E_{\mathrm{J}1}\vert$. The corresponding $\sin(m\varphi)$ contribution to the C$\varphi$R is given by $m\vert E_{\mathrm{J}m}/E_{\mathrm{J}1}\vert$ (cf.~Fig.~\ref{fig:1} for the KIT sample). 
The ratios lie between two limiting cases spanning the physical regime (shaded gray area):
(i)~the upper limit, $\vert E_{\mathrm{J}m}/ E_{\mathrm{J}1}\vert=3/(4m^2-1)$, corresponds to an open quantum point contact, i.e.~one channel with $T=1$, and (ii)~the lower limit, $\vert E_{\mathrm{J}m}/ E_{\mathrm{J}1}\vert\sim(T/4)^{m-1}/m^{3/2}$, corresponds to a perfectly homogeneous low-transparency barrier ($T_n=T=10^{-6}$ for all $n$). For the scanning routine, we include harmonics up to $E_{\mathrm{J}10}$ to obtain results within the physical regime and to see when truncation is possible (see~\hyperref[sec:methods]{Methods}). Remarkably, for all samples the $E_{\mathrm{J}2}$ contribution is in the few \% range even after considering additional corrections such as series inductance or gap asymmetry in the superconducting electrodes (see Supplementary Section~\ref{sec:alternativecorrections}).

The Josephson harmonics ratios computed from the mesoscopic model Eq.~\eqref{eq:dsigmamodel} are shown with turquoise markers.
Notice that the barrier evolved between cooldowns of the KIT sample due to aging (CD1 to CD2) and thermal annealing (CD2 to CD3) (cf.~Supplementary Section~\ref{suppsec:samplesKIT}).
Even for the most homogeneous barrier (CD3), the second-harmonic contribution is $E_{\mathrm{J}2}/E_{\mathrm{J}1}\approx\SI{-2.4}{\percent}$, implying that there would be at least one conduction channel with a transparency $T\ge0.29$ (see Supplementary Section~\ref{sec:theorybackground}). 
The methodology presented in Fig.~\ref{fig:3} can serve as a tool to characterize Josephson harmonics and tunnel barrier homogeneity, independent of circuit design.

Since the charge dispersion increases for higher transmon levels (even for the standard transmon Hamiltonian~\cite{koch2007transmon}, cf.~Fig.~\ref{fig:2}f) and is exponentially sensitive to the shape of the JJ potential (cf.~Fig.~\ref{fig:2}e), a natural question arises:
What are the consequences of the Josephson harmonics on the transmon's susceptibility to offset charges?
In Fig.~\ref{fig:4}a we show the measured charge dispersion $\delta f_{0j}$ of the K\"oln device for states $j=1,2,3$ vs.~the first transition frequency $f_{01}$, which is tuned by in-plane magnetic field $B_{\parallel}$ of up to $\SI{0.4}{\tesla}$ (see Supplementary Section~\ref{sec:samplesKoeln} for details). The charge dispersion predicted by the standard model (dashed gray) falls short of the measurements by a factor of 2 to 7 for the three measured transitions.
In contrast, when using the Josephson harmonics model, the computed charge dispersion matches the data (blue lines). We emphasize that for both models we use the same parameters as in the Fig.~\ref{fig:3} analysis (i.e., the standard model and the $E_{\mathrm{J}4}$ model), and vary the first Josephson energy to match the qubit frequency $f_{01}$ for different magnetic fields while keeping the $E_{\mathrm{J}m}/E_{\mathrm{J}1}$ ratios constant.

Interestingly, the presence of large Josephson harmonics, as in the case of the IBM qubits (see Fig.~\ref{fig:3}c), can also reduce the charge dispersion, which directly decreases charge noise decoherence. We show evidence for this in Fig.~\ref{fig:4}b, on the first three IBM qubits, for which the charge dispersion of the qubit transition can be a factor of 4 lower than expected from the standard transmon model.  
This observation indicates a possible optimization route in which Josephson harmonics are engineered (e.g.~by shaping the channel transparencies or by adding inductive elements in series) and the spectrum is steered towards regions of reduced charge dispersion and increased anharmonicity (see Supplementary Fig.~\ref{suppfig:engineering}). A recent work~\cite{bozkurt2023doublefourier} proposes a similar approach to engineer arbitrary shaped C$\varphi$Rs using networks of effective high transparency JJs, each of which is a series of tunnel JJs.

The main reason for the failure of the standard transmon model in describing the charge dispersion (when fitted to $f_{01}$ and $f_{02}$) is that it misjudges the value of $E_\mathrm{J}/E_\mathrm{C}$. 
To quantify this effect, in Fig.~\ref{fig:4}c we plot the values of $E_{\mathrm{J}1}/E_\mathrm{C}$ from the Josephson harmonics model against the value of $E_\mathrm{J}/E_\mathrm{C}$ from the standard model. Indeed, the $E_{\mathrm{J}1}/E_\mathrm{C}$ ranges for many of our measurements are not compatible with the standard model $E_\mathrm{J}/E_\mathrm{C}$ ratio (dashed diagonal).
We note that when evaluated for the same $E_\mathrm{J}/E_\mathrm{C}$, the Josephson harmonics correction to the charge dispersion is relatively small (see inset of Fig.~\ref{fig:4}c). 

In summary, we have shown that for ubiquitous AlO$_x$ tunnel junctions, the microscopic structure, currently underappreciated in its complexity, causes level shifts and modifies the charge dispersion in superconducting artificial atoms. In order to fully describe the measured transmon energy spectra, we amend the standard $\sin\varphi$ Josephson current-phase relation for tunnel junctions to include higher order $\sin(m\varphi)$ harmonics, with the relative amplitude of the $m=2$ term in the few \% range. We confirm this finding in various sample geometries from four different laboratories, and we argue that the source of the Josephson harmonics is the presence of relatively higher transparency channels with $T\gg 10^{-6}$ in the AlO$_\text{x}$ tunnel barrier. The methodology shown here can reveal percent-level deviations from a sinusoidal C$\varphi$R, which are hard to detect in more standard measurements based on asymmetric direct current superconducting quantum interference devices~\cite{DellaRocca2007MeasurementCurrentPhaseRelation}.

The observation of Josephson harmonics in tunnel junctions highlights the need to revisit established models for superconducting circuits. Our work directly impacts the design and measurement of transmon qubits and processors:
As an illustration, we show that by engineering Josephson harmonics, the dephasing due to charge noise can be reduced by an order of magnitude without sacrificing anharmonicity. These results ask for future research studying the implications of Josephson harmonics and associated Andreev bound states in other tunnel-JJ-based circuits, e.g.~fluxonium or generalized flux qubits~\cite{Nguyen2022Aug}.

In general, we expect the inclusion of the harmonics will refine the understanding of superconducting artificial atoms and will directly benefit, among others, quantum gate and computation schemes relying on higher levels~\cite{DiCarlo2009twoqubitcoupling,rigetti2010CR,chow2013microwaveconditionalphasegate,sank2016beyondRWA,Negirneac2021FluxTunableTransmonCZGateAtTheSpeedLimit,Rol2019QuantumGateHigherLevels,Roy2023SuperconductingQutrit}, quantum-non-demolition readout fidelities~\cite{Gusenkova2021QuantumNondemolitionFluxonium,Shillito2022DynamicsTransmonIonization,Cohen2022ClassicalChaosDrivenTransmons}, and frequency crowding mitigation in quantum processors~\cite{Hertzberg2021LaserAnnealingTransmonsIBM}. Josephson harmonics will probably also have to be accounted for in topological JJ circuits~\cite{Gyenis2021, Rymarz2021, Smith2022}, parametric pumping schemes employed in microwave amplifiers and bosonic codes~\cite{CampagneIbarcq2020, Lescanne2020}, amplification and mixing~\cite{Frattini2017SNAIL,Grimm2020SNAIL}, JJ metrological devices~\cite{K_Gramm_1976,Shapiro1963,Crescini2022,Roy2016Aug}, Floquet qubits~\cite{Huang2021FloquetQubitsDynamicalSweetSpots,Gandon2022FloquetQubit}, protected Josephson qubits~\cite{Gladchenko2008JosephsonRhombusChain,Gyenis2021,Smith2022}, etc., and they can be harnessed to realize Josephson diodes \cite{Fominov2022HigherHarmonicSquidDiode}. As devices become increasingly sophisticated with progressively smaller error margins, higher-order Josephson harmonics will need to be either
suppressed via the development of highly uniform and low transparency barriers,
or engineered and included as an integral part of the device physics.

\section*{Acknowledgements}
D.W., M.W. and K.M. thank Hans De Raedt and Hannes Lagemann for stimulating discussions.
C.D., J.K. and Y.A. thank Lucas Marten Janssen for assistance with the measurements and Philipp Janke for contributions to the data analysis.
I.M.P. thanks Uri Vool for providing comments on an early version of the manuscript.
C.D. and I.M.P. thank Leo DiCarlo for discussions that helped improve the manuscript.
L.B.-I., C.M., and I.M.P. thank Jared Cole for insightful discussions about simulating the AlO$_x$ structure.
D.W., M.W. and K.M. gratefully acknowledge the Gauss Centre for Supercomputing e.V. (www.gauss-centre.eu) for funding this project by providing computing time on the GCS Supercomputer JUWELS~\cite{JuwelsClusterBooster} at J\"ulich Supercomputing Centre (JSC).
D.W. and M.W. acknowledge support from the project J\"ulich UNified Infrastructure for Quantum computing (JUNIQ) that has received funding from the German Federal Ministry of Education and Research (BMBF) and the Ministry of Culture and Science of the State of North Rhine-Westphalia.
P.W., B.D., T.R. and G.C. acknowledge support from the German Ministry of Education and Research (BMBF) within the project GEQCOS (FKZ: 13N15683 and 13N15685).
D.R., S.Ge., S.G\"u., S.I., W.W., G.C. and I.M.P. acknowledge support from the German Ministry of Education and Research (BMBF) within the project QSolid (FKZ: 13N16151 and 13N16149).
C.D., J.K. and Y.A. acknowledge support from the European Research Council (ERC) under the European Union's Horizon 2020 research and innovation program (grant agreement No.~741121) and from Germany's Excellence Strategy - Cluster of Excellence Matter and Light for Quantum Computing (ML4Q) EXC 2004/1 - 390534769.
P.P. acknowledges support from the German Ministry of Education and Research (BMBF) within the QUANTERA
project SiUCs (FKZ: 13N15209).
P.D. was supported by the IBM Quantum Community Advocate internship program.
R.H., J.H.B., P.S. and D.G. acknowledge the support of Hitachi High-Technologies. 
This work has been supported financially by the German Federal Ministry of Education and Research (BMBF) via the TLE4HSQ project (Grant No.~13N15983). P.S. acknowledges financial support by the German Federal Ministry of Education and Research (BMBF) via the Quantum Futur project MajoranaChips (Grant No.~13N15264) within the funding program Photonic Research Germany.
L.B.-I. and C.M. acknowledge support from UEFISCDI Romania through the contract ERANET-QUANTERA QuCos 120/16.09.2019, and from ANCS through Core Program 27N/2023, project No.~PN 23 24 01 04.

\section*{Author contributions}

D.W., D.R., P.W., M.W., and I.M.P. conceived of the presented study.
D.W., D.R., P.W., M.W., and I.M.P. wrote the original draft.
D.R., P.W., B.D., S.G\"u., P.P., and T.R. performed the experiments on the KIT sample.
C.D. and  J.K. performed the experiments on the K\"oln sample.
R.L. and Z.L. performed the experiments on the ENS sample.
N.T.B. and P.D. performed the experiments on the IBM sample.
R.H., J.H.B., P.S., S.Ge. and S.I. acquired the STEM images and prepared the corresponding samples.
D.W., M.W., and G.C. performed the theoretical modeling and the numerical simulations.
L.B.-I. and C.M. performed the molecular dynamics simulations.
D.P.D., K.M., G.C., and I.M.P. supervised the work.
All authors analyzed the data and contributed to reviewing and editing the manuscript and the Supplementary Information.

\section*{Competing interests}
The authors declare no competing interests.	

\section*{Methods}
\label{sec:methods}

\textbf{Diagonalizing the Hamiltonians to obtain model predictions:}
We construct the matrices of $H_{\mathrm{std}}$ in Eq.~\eqref{eq:standardmodel} and $H_{\mathrm{har}}$ in Eq.~\eqref{eq:harmonicsmodel} by first diagonalizing the bare transmon matrix (excluding $H_{\mathrm{res}}$) in the charge basis $\{\ket n\}$, where $4E_\mathrm{C}(n-n_\mathrm{g})^2=\sum_n 4E_\mathrm{C}(n-n_\mathrm{g})^2 \ket{n}\!\bra{n}$ is diagonal and $-E_{\mathrm{J}m}\cos (m\varphi)=-\sum_nE_{\mathrm{J}m}/2\,(\ket{n}\!\bra{n+m}+\ket{n+m}\!\bra{n})$ has constant entries $-E_{\mathrm{J}m}/2$ on the $m^{\mathrm{th}}$ subdiagonal (we ensure enough terms by generally verifying that the predictions do not change if more terms are included).
This yields the transmon eigenenergies $E_j$ and eigenstates $\ket j$.
Then we diagonalize the joint transmon-resonator Hamiltonian $H_{\mathrm{std}/\mathrm{har}} = \sum_j E_j\ket{j}\!\bra{j} + \Omega a^\dagger a + \sum_{j,j'} G \ket{j}\!\bra{j}\!n\!\ket{j'}\!\bra{j'}(a+a^\dagger)$, where $a=\sum_k\sqrt{k+1}\ket{k}\!\bra{k+1}$.
To each resulting eigenenergy $E_{\overline{l}}$ and eigenstate $\ket{\overline{l}}$, we assign a photon label $k$ and a transmon label $j$ based on the largest overlap $\max_{k,j}\vert\braket{kj|\overline{l}}\vert$ (this only works for small $k$, cf.~Supplementary Section~\ref{sec:dressedstates}), which yields the dressed energies $\smash{E_{\overline{kj}}}$ and states $\smash{\ket{\overline{kj}}}$.
This procedure is done for both $n_\mathrm{g}=0$ and $n_\mathrm{g}=1/2$. From the resulting dressed energies $E_{\overline{kj}}(n_\mathrm{g})$, we compute the transmon transition frequencies $f_{0j}^{\mathrm{model}}(n_\mathrm{g}) = (E_{\overline{0j}}(n_\mathrm{g})-E_{\overline{00}}(n_\mathrm{g}))/2\pi$ and the resonator frequencies $f_{\mathrm{res},j}^{\mathrm{model}}(n_\mathrm{g})=(E_{\overline{1j}}(n_\mathrm{g})-E_{\overline{0j}}(n_\mathrm{g}))/2\pi$ (setting $\hbar=1$).
The predicted frequencies are then given by $f_{0j}^{\mathrm{model}}=(f_{0j}^{\mathrm{model}}(0) + f_{0j}^{\mathrm{model}}(1/2))/2$, $f_{\mathrm{res},j}^{\mathrm{model}}=(f_{\mathrm{res},j}^{\mathrm{model}}(0)+f_{\mathrm{res},j}^{\mathrm{model}}(1/2))/2$, and the charge dispersion is $\delta f_{0j}^{\mathrm{model}} = \vert f_{0j}^{\mathrm{model}}(0) - f_{0j}^{\mathrm{model}}(1/2)\vert$.
We consistently use $n=-N,\ldots,N$ with $N=14$ and thus $2N+1=29$ charge states, $j=0,\ldots,M-1$ with $M=12$ transmon states, and $k=0,\ldots,K-1$ with $K=9$ resonator states, where the $N$, $M$, and $K$ have been chosen by verifying that the model predictions do not change significantly by adding more states.

\textbf{Solving the IEP to obtain model parameters:}
The inverse problem~\cite{ChuGolub2005InverseEigenvalueProblemsBook,jaynes2003probability} to obtain the parameters
$\mathbf{x}^{\mathrm{std}}$ of the standard model Hamiltonian in Eq.~\eqref{eq:standardmodel}
and $\mathbf{x}^{\mathrm{har}}$ of the harmonics model Hamiltonian in Eq.~\eqref{eq:harmonicsmodel}, such that the linear combinations of eigenvalues $\mathbf{f}=(f_{01}^{\mathrm{model}},f_{02}^{\mathrm{model}},\ldots,f_{0N_f}^{\mathrm{model}},f_{\mathrm{res},0}^{\mathrm{model}},f_{\mathrm{res},1}^{\mathrm{model}})$ agree with the measured data, is an instance of the Hamiltonian parameterized IEP (HamPIEP, see Supplementary Section~\ref{sec:HamPIEP}). We solve the HamPIEP using the globally convergent Newton method~\cite{Dennis1996NumericalMethodsUnconstrainedOptimization} with cubic line search and backtracking~\cite{numericalrecipes} and the Broyden-Fletcher-Goldfarb-Shanno (BFGS) algorithm~\cite{NumericalOptimization} as implemented in TensorFlow Probability~\cite{TensorFlow282}.
The Jacobian $\partial \mathbf{f}/\partial\mathbf{x}$ is obtained by performing automatic differentiation through the diagonalization with TensorFlow.
For the $E_{\mathrm{J}4}$ model shown in Fig.~\ref{fig:3}b, the IEP is solved unambiguously for $\mathbf x=(E_\mathrm{J1},E_\mathrm{J2},E_\mathrm{J3},E_\mathrm{J4},\Omega,G)$ using the lowest 4 transmon transition frequencies, and we fix the values $E^{\mathrm{KIT}}_\mathrm{C}/h=\SI{242}{\mega\hertz}$, $E^{\mathrm{ENS}}_\mathrm{C}/h=\SI{180}{\mega\hertz}$, and $E^{\mathrm{IBM}}_\mathrm{C}/h=\SI{300}{\mega\hertz}$, respectively, to make the models consistent with further available information such as accurate finite-element simulations (cf.~Supplementary Section~\ref{suppsec:samplesKIT}) or knowledge of the transmon capacitance.
For the mesoscopic model (see Supplementary Section~\ref{sec:mesoscopicmodel}), the parameters $\mathbf x=(\bar d,\sigma,E_\mathrm{C},E_\mathrm{J},\Omega,G)$ are found by minimizing the function $\sum_{j=1}^{N_f} \vert f_{0j}^{\mathrm{model}}/j - f_{0j}^{\mathrm{experiment}}/j \vert + \sum_{j=0}^{1} \vert f_{\mathrm{res},j}^{\mathrm{model}} - f_{\mathrm{res},j}^{\mathrm{experiment}} \rvert$ using the BFGS algorithm. The initial values for the minimization are given by $\bar d=\SI{1.64}{\nano\metre}$ (taken from the molecular dynamics result in Supplementary Section~\ref{sec:moleculardynamics}),  $\sigma=\bar d/4$ (see also Supplementary Table~\ref{tab:literaturebarriers}), and $(E_\mathrm{C},E_\mathrm{J},\Omega,G)$ from the standard transmon model.
For the K\"oln data, where 288 data points have to be described by the same model parameters $\mathbf{x}$ (cf.~Fig.~\ref{fig:4}a) and only the Josephson energy is varied, we use cubic interpolation as a function of $f_{01}^{\mathrm{model}}$ and include only a few central points for the available frequencies in the solution of the IEP (the residuals are given in Supplementary Fig.~\ref{suppfig:koelnresiduals}).
All model parameters are provided in the repository~\cite{repository} accompanying this manuscript.

\textbf{Scanning the Josephson energies:}
To obtain the range of suitable Josephson energies $\{E_{\mathrm{J}m}\}$ (shown in Fig.~\ref{fig:3}c) that are consistent with a measured spectrum, we use an exhaustive scanning procedure.
A spectroscopy dataset of $N_f$ measured transition frequencies $f_{0j}$, $j=1,\ldots,N_f$, and two resonator frequencies $f_{\mathrm{res},0}$ and $f_{\mathrm{res},1}$ uniquely determines---via the HamPIEP---the values $\mathbf{x}=(E_\mathrm{J1},\ldots,E_{\mathrm{J}N_f},\Omega,G)$. 
We then scan the values of four additional ratios $\mathbf{y}=(E_{\mathrm{J}N_f+1}/E_{\mathrm{J}1},\dots,E_{\mathrm{J}N_f+4}/E_{\mathrm{J}1})$,
namely for each of these four $E_{\mathrm{J}m}/E_{\mathrm{J}1}$ over $16$ geometrically spaced values between the point contact limit $3(-1)^{m+1}/(4m^2-1)$ and $(-1)^{m+1}\min\{10^{-7},\vert E_{\mathrm{J}m+1}/E_{\mathrm{J}1}\vert\}$ (always skipping the first to ensure $\vert E_{\mathrm{J}m}/E_{\mathrm{J}1}\vert>\vert E_{\mathrm{J}m+1}/E_{\mathrm{J}1}\vert$). 
Additionally we include $\mathbf{y}=(0,0,0,0)$ to see if truncation at $E_{\mathrm{J}N_f}$ is allowed.
For each combination $\mathbf{y}$, we solve the HamPIEP for the spectroscopy data to obtain the unique solution $\mathbf{x}$. We call the ratios $\mathbf{e}=(1,E_{\mathrm{J}2}/E_{\mathrm{J}1},\ldots,E_{\mathrm{J}N_f+4}/E_{\mathrm{J}1})$ a trajectory that can reproduce the spectrum. However, the trajectory $\mathbf{e}$ may not be physical, since (i) some of the leading ratios $E_{\mathrm{J}m}/E_{\mathrm{J}}$ for $m\le N_f$ might be beyond the quantum point-contact limit, (ii) the Josephson energies might not be strictly decreasing in absolute value for increasing order $m$, or (iii) the signs might not be alternating. Note that this can also happen when the Josephson harmonics model Eq.~\eqref{eq:harmonicsmodel} is truncated at too low orders (see~Supplementary Section~\ref{sec:scan}). For all $E_{\mathrm{J}m}$, the maximum and minimum possible ratios $\vert E_{\mathrm{J}m}/E_{\mathrm{J}}\vert$ define the vertical bars in Fig.~\ref{fig:3}c. 

\section*{Data availability}

The spectroscopy data and the model parameters that support the findings of this study are available in the J\"ulich DATA repository at \url{https://doi.org/10.26165/JUELICH-DATA/LGRHUH}.

\section*{Additional information}

\textbf{Supplementary information} is available for this paper.

\textbf{Correspondence and requests for materials} should be addressed to I.M.P.

\let\oldaddcontentsline\addcontentsline
\renewcommand{\addcontentsline}[3]{}

\bibliography{bib}
\let\addcontentsline\oldaddcontentsline

\pagebreak
\clearpage
\onecolumngrid

\setcounter{section}{0}
\setcounter{table}{0}
\renewcommand{\thetable}{S\arabic{table}}
\setcounter{figure}{0}
\renewcommand{\thefigure}{S\arabic{figure}}
\setcounter{equation}{0}
\renewcommand{\theequation}{S\arabic{equation}}

\begin{center}
	\textbf{\large{Supplementary Information for \\\smallskip
		``Observation of Josephson Harmonics in Tunnel Junctions''}}\\
\end{center}

\tableofcontents

\clearpage 

\section{Theory}
\label{sec:theory}

We review the theory of superconducting tunnel junctions, starting from the general current-phase relation as a sum over conduction channels with certain transparencies.
Then we derive a closed-form expression for the higher-order Josephson energies and study several transparency distributions. We also introduce a mesoscopic model of tunneling through an inhomogeneous (i.e.~non-uniform) barrier, which can describe the non-negligible contributions from higher harmonics to the Josephson effect in regular SIS junctions.
Based on these results, we give a motivation for some phenomenological models for the relative size of $E_{\mathrm{J}m}/E_\mathrm{J}$ and show results for these alternative models as well as for other transmon data extracted from the literature. Additionally, we perform an exhaustive scan over the higher-order Josephson contributions, yielding the full range of suitable $E_{\mathrm{J}m}/E_\mathrm{J}$ for each sample, independent of the particular models discussed. 
Finally, we study potential alternative corrections, both on a circuit level---such as a stray inductance, additional hidden modes, or the coupling to other qubits---and on the junction level---such as an asymmetry in the superconducting electrodes. We explain why we have come to the conclusion that these corrections cannot account for the mismatch between the datasets and the standard transmon model.

\subsection{Theoretical background}
\label{sec:theorybackground}

The general formula for the current-phase relation in Josephson tunnel junctions can be written as a sum over conduction channels~\cite{Beenakker1992},
\begin{equation}\label{eq:currentphaserelation}
    I_S (\varphi)  = \frac{e\Delta}{2\hbar} \sum_{n=1}^N \frac{T_n \sin\varphi}{\sqrt{1-T_n \sin^2(\varphi/2)}}\,,
\end{equation}
where $\Delta$ is the superconducting energy gap, $N$ is the number of conduction channels, $T_n\in[0,1]$ are their transparencies (the transmission amplitudes $t_n$ through the channels determine both their transparencies, $T_n = |t_n|^2$, as well as the reflection coefficients for Andreev processes, see e.g.~Section~4.4 in \cite{Asano2021AndreevReflectionJJ} and Fig.~4.7 there), 
$\varphi$ is the phase difference across the Josephson junction, $e$ is the electron charge, and $\hbar$ is the reduced Planck quantum. Equation~\eqref{eq:currentphaserelation} is Beenakker's multichannel generalization of a formula found by Haberkorn et al.~\cite{Haberkorn1978TheoreticalStudyCurrentPhaseJosephson} (see also Ref.~\cite{Boettcher1997MultichannelDCJosephsonEffect} and \cite[Eq.~(17)]{Golubov2004}). We neglect here temperature-dependent corrections as they are exponentially small in $T/T_c \ll 1$.

For tunnel junctions, it is generally assumed that $T_n \ll 1$ for all $n$~\cite{Glazman2021,Kittel2004}, so keeping only the lowest order contribution would give
\begin{equation}
    I_S (\varphi) \simeq I_\text{c} \sin\varphi\,, \quad I_\text{c} = \frac{e\Delta}{2\hbar}\sum_n T_n = \frac{\pi \Delta}{2e} G\,,
\end{equation}
where $G_N=R_N^{-1}=e^2/(\pi\hbar)\sum_n T_n$ is the normal-state conductance of the junction.
The corresponding contribution $H_\mathrm{J}$ to the effective Hamiltonian is found using the relation
\begin{equation}\label{eq:IH}
    I(\varphi) = \frac{2e}{\hbar} \frac{d H_\mathrm{J}}{d\varphi}\,,
\end{equation}
and we recover the Josephson term of the standard transmon model,
\begin{equation}\label{eq:standard_Hj_and_Delta}
    H_\mathrm{J} \simeq - E_\mathrm{J} \cos\varphi\,,\quad E_\mathrm{J} = \displaystyle\frac{\Delta}{8} g\,,
\end{equation}
with $g=G_N/(e^2/h)$ the dimensionless conductance. 

\begin{figure*}
  \centering
  \includegraphics[width=\textwidth]{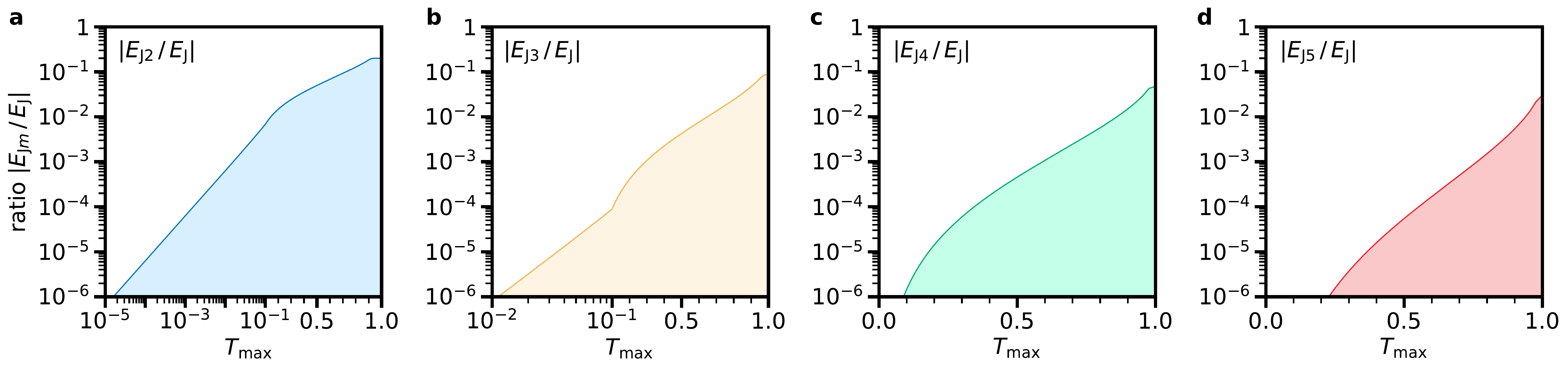}
  \caption{\textbf{Bound for the largest transparency $T_{\mathrm{max}}$ contributing to the conduction channels.} 
  For a given ratio $\vert E_{\mathrm{J}m}/E_{\mathrm{J}}\vert$ for 
  \textbf{a} $m=2$, \textbf{b} $m=3$, \textbf{c} $m=4$, \textbf{d} $m=5$, the line indicates a lower bound for the largest transparency, of which at least one conduction channel must contribute.
  The area indicates all values $\vert E_{\mathrm{J}m}/E_{\mathrm{J}}\vert$ that are possible given $T_{\mathrm{max}}$.
  Note the change between linear and logarithmic scale at $T_{\mathrm{max}}=0.1$ for the left two panels.
  An upper limit on the bound is given by the quantum point contact limit  (see Section~\ref{sec:single-channel-distribution}) which results in the flat part of the curve for $T_{\mathrm{max}}$ close to 1.}
  \label{suppfig:transparency}
\end{figure*}

However, in general, Eq.~\eqref{eq:currentphaserelation} contains higher harmonics, and the condition $T_n \ll 1$ is not always satisfied for all transmission channels $n$.
To obtain an expression for higher harmonic corrections, we integrate Eq.~\eqref{eq:currentphaserelation} and use Eq.~\eqref{eq:IH} to get the Josephson part of the Hamiltonian
\begin{equation}
    H_\mathrm J 
    = -\Delta \sum_{n=1}^N \sqrt{1-T_n\sin^2\frac{\varphi}{2}} \,.
    \label{eq:HJ_sin}
\end{equation}
Since $H_\mathrm{J}$ is symmetric and $2\pi$ periodic in $\varphi$, it can be written as a Fourier cosine series
\begin{equation}
    H_\mathrm J = - \sum_{m=1}^\infty E_{\mathrm{J}m} \cos (m\varphi)\, .
\end{equation}
To derive an analytic expression for the $E_{\mathrm{J}m}$, we apply the relations
\begin{align}
    \sqrt{1+x} &= \sum_{m=0}^\infty \binom{2m}{m}\frac{(-1)^{m+1}}{4^m(2m-1)}x^m\,, &
    \sin^{2m}\frac{x}{2} &= \frac{1}{2^{2m-1}}\sum_{j=0}^{m-1} (-1)^{m-j}\binom{2m}{j}\cos((m-j)x) \,,
\end{align}
for $m\ge 1$ to Eq.~\eqref{eq:HJ_sin}. We obtain
\begin{align}
    H_\mathrm{J} &= -\sum_{m=1}^\infty\sum_{n=1}^N\left(4\Delta  (-1)^{m+1}   \sum_{k=0}^\infty \binom{2k+2m-2}{k+m-1}\binom{2k+2m}{k}\frac{1}{k+m}\left(\frac{T_n}{16}\right)^{k+m}\right)\cos(m\varphi) \label{eq:HJ_binom}\\
    &= - \sum_{m=1}^\infty\sum_{n=1}^N\left( (-1)^{m+1} \frac{4\Delta}{m} \binom{2m-2}{m-1} {}_2F_1(m-\sfrac{1}{2},m+\sfrac{1}{2};2m+1;T_n)\left(\frac{T_n}{16}\right)^m\right)\cos(m\varphi)\,,\label{eq:HJ_hypergeometric}
\end{align}
where ${}_2F_1$ denotes the hypergeometric function which is defined by (see e.g.~\cite[Section~16.2]{DLMF})
\begin{equation}
    {}_pF_q(a_1,...,a_p;b_1,...,b_q;x) = \sum_{k=0}^\infty \frac{(a_1)_k\cdots (a_p)_k}{(b_1)_k\cdots (b_q)_k}\frac{x^k}{k!}\,,
\end{equation}
with $(a)_k = a(a+1)(a+2)\cdots (a+k-1)$ denoting the Pochhammer symbol.
Thus we find an expression for the higher-order contributions to the effective Josephson Hamiltonian,
\begin{equation}
    E_{\mathrm{J}m} = (-1)^{m+1} \frac{4\Delta}{m} \binom{2m-2}{m-1}\sum_{n=1}^N {}_2F_1(m-\sfrac{1}{2},m+\sfrac{1}{2};2m+1;T_n)\left(\frac{T_n}{16}\right)^m\,.
    \label{eq:EJm_closed_form}
\end{equation}
Note that, for small transparencies $T_n\ll 1$ where it is sufficient to consider only the leading-order term of $_2F_1$, we recover $E_{\mathrm{J}1}=E_\mathrm{J}=\Delta g/8$ and the sinusoidal current-phase relation (in this section, we identify $E_{\mathrm{J}1}\equiv E_\mathrm{J}$).

We can derive a bound for the largest possible ratio $\vert E_{\mathrm{J}m}/E_{\mathrm{J}}\vert$ when all conduction channels have a transparency $T_n\le T_{\mathrm{max}}$, independent of the particular distribution of transparencies $\{T_n\}$. Using the monotonicity of ${}_2F_1$, we have in the numerator ${}_2F_1(\cdots;T_n)\le{}_2F_1(\cdots;T_{\mathrm{max}})$ for all $n$, and in the denominator ${}_2F_1(\cdots;T_n)\ge1$ for all $n$. For the remaining quotient of sums over $n$, we use $\sum_n T_n^m/\sum_n T_n = T_{\mathrm{max}}^{m-1} [\sum_n (T_n/T_{\mathrm{max}})^m/\sum_n (T_n/T_{\mathrm{max}})] \le T_{\mathrm{max}}^{m-1} $. Thus we obtain
\begin{equation}
    \left\vert \frac{E_{\mathrm{J}m}}{E_{\mathrm{J}}}\right\vert \le \frac{1}{m} \binom{2m-2}{m-1} {}_2F_1(m-\sfrac{1}{2},m+\sfrac{1}{2};2m+1;T_{\mathrm{max}})\left(\frac{T_{\mathrm{max}}}{16}\right)^{m-1}\,.
\end{equation}
The bound is shown in Fig.~\ref{suppfig:transparency} for the leading ratios. It yields, for instance, the following statement: Given some observation for $\vert E_{\mathrm{J}2}/E_{\mathrm{J}}\vert\approx\SI{2.40}{\percent}$, there must at least be one conduction channel with a transparency $T_{\mathrm{max}}\ge 0.298$. 
We remark that the estimation of $T_{\mathrm{max}}$ in this way is based on the general current-phase relation hypothesis Eq.~\eqref{eq:currentphaserelation}.  In the presence of harmonics from a stray inductance such as $L=\SI{0.5}{\nano\henry}$ (see Section~\ref{sec:additionalinductance} below), the bound would be $T_{\mathrm{max}}\ge 0.096$.
We note that high-transparency conduction channels with energy as in Eq.~\eqref{eq:HJ_sin} have been observed experimentally, for example in break junctions (see~\cite{Janvier2015CoherentManipulationAndreevStatesHighTransmissionChannel}, in which the authors identify a channel with $T=0.99217$).

In general, the current-phase relation as a function of the Josephson energies $E_{\mathrm{J}m}$ reads
\begin{equation}
    I_S(\varphi) = \frac{2e}{\hbar} \sum_{m=1}^\infty mE_{\mathrm{J}m} \sin (m\varphi)\,.
    \label{eq:IsofphiEJm}
\end{equation}
The analysis presented in Fig.~\ref{fig:3}c of the main text shows that the corrections to the sinusoidal current-phase relation are on the level of several percent for all experiments under consideration. For given $E_{\mathrm{J}m}$, the critical current $I_\text{c}$ can be obtained by finding the maximum of Eq.~\eqref{eq:IsofphiEJm} as a function of $\varphi$ (we show three limiting cases of $I_S(\varphi)/I_\text{c}$ for each sample in the insets of Fig.~\ref{suppfig:potentialsandcprs} below; values for $I_\text{c}$ are given in Table~\ref{tab:modelparameters}).

\begin{figure*}
  \centering
  \includegraphics[width=\textwidth]{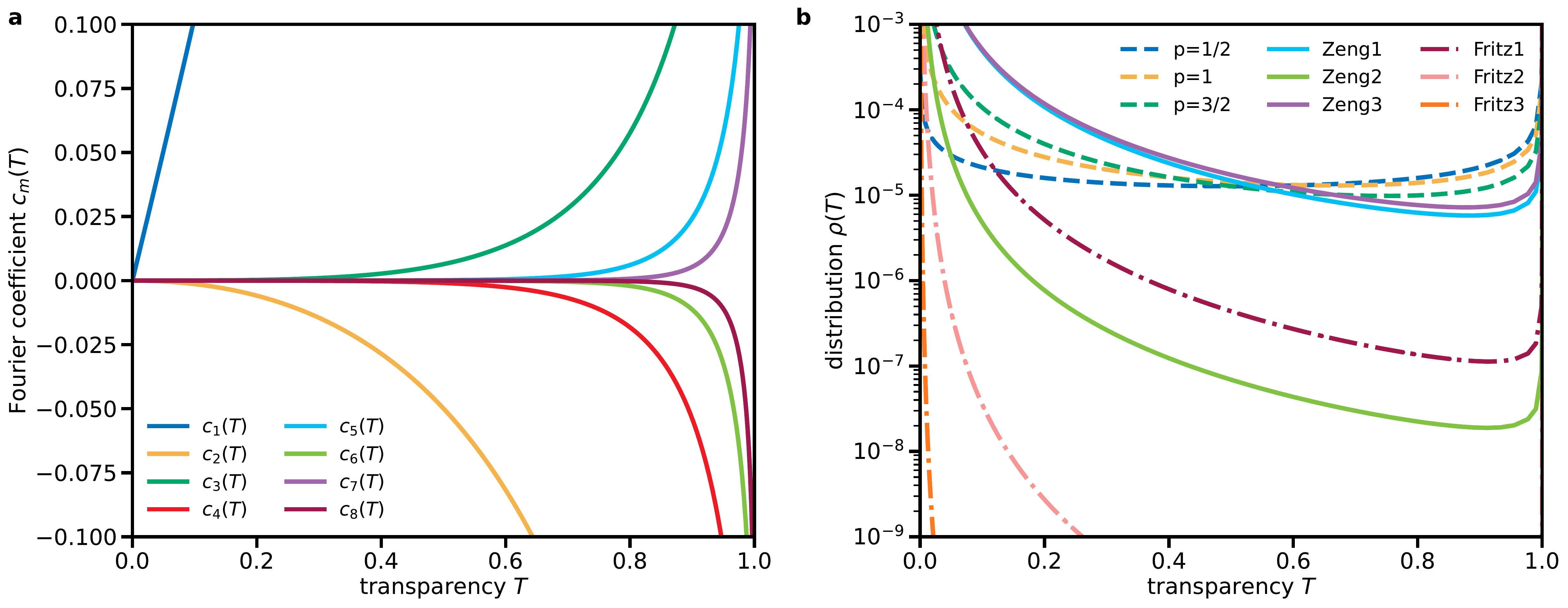}
  \caption{\textbf{Fourier coefficients $c_m(T)$ and several distributions $\rho(T)$ to describe the contribution of higher harmonics to the Josephson effect.} 
  \textbf{a} Fourier coefficients $c_m(T)$ of the series expansion of the general current-phase relation given by Eq.~\eqref{eq:IsofphiFourier}.
  \textbf{b} Distribution of transparencies $\rho(T)$ for the occasionally considered universal form Eq.~\eqref{eq:rhouni} (dashed lines), 
  and the mesoscopic model in Eq.~\eqref{eq:rhorec} using average barrier thicknesses $\bar d$ and standard deviations $\sigma$ extracted from Ref.~\cite{Zeng2015ObservationThicknessAlOxJosephsonJunctions} (Zeng, solid lines) and Ref.~\cite{Fritz2019} (Fritz, dash-dotted lines); the corresponding junction parameters are also listed in Table~\ref{tab:literaturebarriers}.
  As some of the distributions given by Eq.~\eqref{eq:rhouni} are not normalizable, the prefactors have been chosen such that $\langle T\rangle=10^{-5}$ to make them comparable (i.e., $n_{1/2}=2/\pi\times10^{-5}$, $n_1=1/2\times10^{-5}$, and $n_{3/2}=1/\pi\times10^{-5}$, cf.~\cite{Schep1997TransportThroughDirtyInterfaces}).
  Note that all distributions have two peaks, namely the bulk contribution for very small transparencies near $T=0$ and a few very high transparencies near $T=1$ responsible for non-negligible Josephson harmonics.}
  \label{suppfig:distributions}
\end{figure*}

We can use the result for $E_{\mathrm{J}m}$ in Eq.~\eqref{eq:EJm_closed_form} to obtain a closed-form expression for the Fourier coefficients $c_m(T_n)$ of the Josephson supercurrent discussed in the main text,
\begin{equation}
    I_S(\varphi) = \frac{e\Delta}{2\hbar} \sum_{m=1}^{\infty} \sum_{n=1}^N c_m(T_n) \sin (m\varphi) \,,
    \label{eq:IsofphiFourier}
\end{equation}
yielding
\begin{equation}
    c_m(T) = \binom{2m-2}{m-1}\frac{ (-1)^{m+1}T^m}{16^{m-1}} \,{}_2F_1(m-\sfrac{1}{2},m+\sfrac{1}{2};2m+1;T)\,,
\end{equation}
with $E_{\mathrm{J}m} = \frac{\Delta}{4}\frac{1}{m} \sum_{n=1}^N c_m(T_n)$.
The first eight Fourier coefficients are shown in Fig.~\ref{suppfig:distributions}a. 
For the first three of them, we give here their series expansion for small transparencies,
\begin{subequations}
\begin{align}
        c_1(T)
        & = T + \frac{T^2}{4} + \frac{15T^3}{128} 
        +  \mathcal{O}\left(T^4\right)\,,
        \label{eq:c1Texp}\\
        c_2(T)
        & = -\left[\frac{T^2}{8}+\frac{3T^3}{32}+\frac{35T^4}{512}
        +\mathcal{O}\left(T^5\right)\right]\,,
        \label{eq:c2Texp}\\
        c_3(T) 
        & = \frac{3T^3}{128} + \frac{15T^4}{512}  + \frac{945T^5}{2^{15}}
        + \mathcal{O}\left(T^6\right)\,.
\end{align}
\end{subequations}
We note that successive coefficients are always alternating in sign, and for the leading terms we have
\begin{equation}
    \frac{c_{m+1}(T)}{c_{m}(T)} \sim - \frac{m-1/2}{4m}\, T\,,
\end{equation}
so the contribution of higher harmonics becomes more relevant for larger transparencies.
Furthermore, the prefactor in this ratio is of order unity and converges to $-1/4$ for large $m$.
The properties of $c_m(T)$ propagate to the Josephson energies, for which we have
\begin{equation}
    \label{eq:EJmProperties}
    \vert E_{\mathrm{J}m}\vert = \frac{\Delta} 4 \frac 1 m \sum_{n=1}^N \underbrace{(-1)^{m+1} c_m(T_n)}_{\text{positive, $<(-1)^m c_{m-1}(T_n)$}} < \vert E_{\mathrm{J}(m-1)}\vert\,.
\end{equation}
Thus we expect that the Josephson energies $E_{\mathrm{J}m}$ for a physical model alternate in sign and decrease in magnitude for increasing order $m$.

\subsection{Transparency distributions}
\label{sec:transparencydistributions}

For a large number of channels $N$, summing over the channels corresponds to averaging over a distribution of transparencies $\rho(T)$,
\begin{equation}
        \label{eq:EJm_integral}
        E_{\mathrm{J}m} = \frac{\Delta}{4}\frac{1}{m} \sum_{n=1}^N c_m(T_n)  = \frac{\Delta}{4}\frac{N}{m} \int_0^1 \! dT \, \rho(T) c_m(T)\,.
\end{equation}
The ratio $E_{\mathrm{J}m}/E_\mathrm{J}$ can then be expressed as
\begin{equation}
    \frac{E_{\mathrm{J}m}}{E_\mathrm{J}} = \frac{1}{m} \frac{\int_0^1 dT \rho(T) c_m(T)}{\int_0^1 dT \rho(T) c_1(T)}\, .
    \label{eq:EJratios}
\end{equation}
If the transparency distribution $\rho(T)$ is not narrowly peaked around a small average value or contains significant weight for higher transparencies, it is typically not justified to neglect higher harmonics. In this section, we look at several example distributions for which this is the case.

\subsubsection{Single channel}\label{sec:single-channel-distribution}

A very elementary case is a single conduction channel with arbitrary transparency $T_1$.
In this case, the formal distribution would be $\rho(T)=\delta(T-T_1)$, but the solution can directly be obtained from Eq.~\eqref{eq:EJm_closed_form} for $N=1$,
\begin{equation}
    E_{\mathrm{J}m} = (-1)^{m+1} \frac{4\Delta}{m} \binom{2m-2}{m-1} {}_2F_1(m-\sfrac{1}{2},m+\sfrac{1}{2};2m+1;T_1)\frac{T_1^m}{16^m}\,.
    \label{eq:EJm_single_channel}
\end{equation}
For the ratio of Josephson energies, we find accordingly
\begin{equation}
    \label{eq:EJmOverEJSingleChannel}
    \frac{E_{\mathrm{J}m}}{E_\mathrm{J}} = \frac{(-1)^{m+1} }{m} \binom{2m-2}{m-1}\frac{T_1^{m-1}}{16^{m-1}} \frac{{}_2F_1(m-\sfrac{1}{2},m+\sfrac{1}{2};2m+1;T_1)}{{}_2F_1(\sfrac{1}{2},\sfrac{3}{2};3;T_1)}\,.
\end{equation}
We consider two important limits for $\vert E_{\mathrm{J}m}/E_\mathrm{J}\vert$, namely the fully open quantum point contact with $T_1 = 1$ as an upper limit, and a homogeneous barrier of very small transparency $T_1\ll 1$ as a lower limit.

For the point contact, setting $T_1=1$ in Eq.~\eqref{eq:EJmOverEJSingleChannel} yields
\begin{equation}\label{eq:pointcontact}
    \frac{E_{\mathrm{J}m}}{E_\mathrm{J}} = (-1)^{m+1}\frac{ 3}{4m^2-1}\,.
\end{equation}
This means that asymptotically, the ratios $\vert E_{\mathrm{J}m}/E_\mathrm{J}\vert$ should not scale weaker than $1/m^2$. Furthermore, in the case of a point contact, Eq.~\eqref{eq:EJm_single_channel} yields an upper limit for the first Josephson energy: For $m=1$, we have $E_{\mathrm{J}} \approx 0.4244\Delta$. Assuming an energy gap of $\Delta/h=\SI{50}{\giga\hertz}$ (see~\cite{Marchegiani2022QuasiparticlesAsymmetricJunctions}), we obtain $E_{\mathrm{J}}/h \approx \SI{21.22}{\giga\hertz}$, which is of the same order as the values of $E_{\mathrm{J}}$ given in Table~\ref{tab:modelparameters} below. We remark that the point contact $T_1=1$ also corresponds to the KO-2 model by Kulik and Omelyanchuk~\cite{Kulik1977KO2model}, which is the clean, fully ballistic counterpart~\cite{Golubov2004} of the KO-1 model~\cite{Kulik1975KO1model} discussed below.

For the lower limit with $T_1\ll1$, taking the leading-order term in Eq.~\eqref{eq:EJmOverEJSingleChannel} yields
\begin{equation}\label{eq:single_channel_low}
    \frac{E_{\mathrm{J}m}}{E_\mathrm{J}} \approx \frac{(-1)^{m+1} }{m} \binom{2m-2}{m-1}\frac{T_1^{m-1}}{16^{m-1}}\,.
\end{equation}
The binomial coefficient is bounded by
\begin{equation}
    2^{m-1} \le \binom{2m-2}{m-1} \le 4^{m-1}\,,
\end{equation}
and Stirling's approximation yields for large $m$
\begin{equation}
    \binom{2m-2}{m-1} \sim \frac{4^{m-1}}{\sqrt{(m-1)\pi}} \approx \frac{4^{m-1}}{\sqrt{m\pi}}\,.
\end{equation}
Applying this approximation to the expression in Eq.~\eqref{eq:single_channel_low} yields
\begin{equation}
    \label{eq:lowerbound}
    \frac{E_{\mathrm{J}m}}{E_\mathrm{J}} \approx (-1)^{m+1} \frac{(T_1/4)^{m-1}}{\sqrt{\pi}m^{3/2}}\,.
\end{equation}

\subsubsection{Small transparencies}

Another simple but probably only pedagogical case is a uniform distribution of transparencies below some cutoff value $T_0$.
We model this case by $\rho(T) = \nu\,\Theta(T_0-T)$, where $\Theta$ denotes the step function and $\nu$ is the normalization. We then find
\begin{equation}
    E_{\mathrm{J}m} 
    =(-1)^{m+1}\frac{\Delta}{4} \frac{N}{m}\frac{\nu}{m+1}\binom{2m-2}{m-1}\frac{T_0^{m+1}}{16^{m-1}} {}_3F_2(m-\sfrac{1}{2},m+\sfrac{1}{2},m+1;2m+1,m+2;T_0)\,,
\end{equation}
and for the ratio
\begin{equation}
    \frac{E_{\mathrm{J}m}}{E_\mathrm{J}} = (-1)^{m+1} \frac{1}{m}\frac{1}{m+1}\binom{2m-2}{m-1}\frac{T_0^{m}}{16^{m-2}}\frac{{}_3F_2(m-\sfrac{1}{2},m+\sfrac{1}{2},m+1;2m+1,m+2;T_0)}{{}_3F_2(\sfrac{1}{2},\sfrac{3}{2},2;3,3;T_0)}\,.
\end{equation}
If the cutoff value is very small, i.e., all transparencies $T\le T_0\ll1$ are very low and it is sufficient to consider only the lowest-order term, we find
\begin{equation}
    \frac{E_{\mathrm{J}m}}{E_\mathrm{J}} = \frac{(-1)^{m+1}}{m(m+1)} \binom{2m-2}{m-1}\frac{T_0^{m}}{16^{m-2}}\,.
\end{equation}
Note that this expression is similar to the case of a homogeneous barrier in Eq.~\eqref{eq:single_channel_low}, but the exponent of $T_0$ is larger by one.

\begin{figure*}
  \centering
  \includegraphics[width=\textwidth]{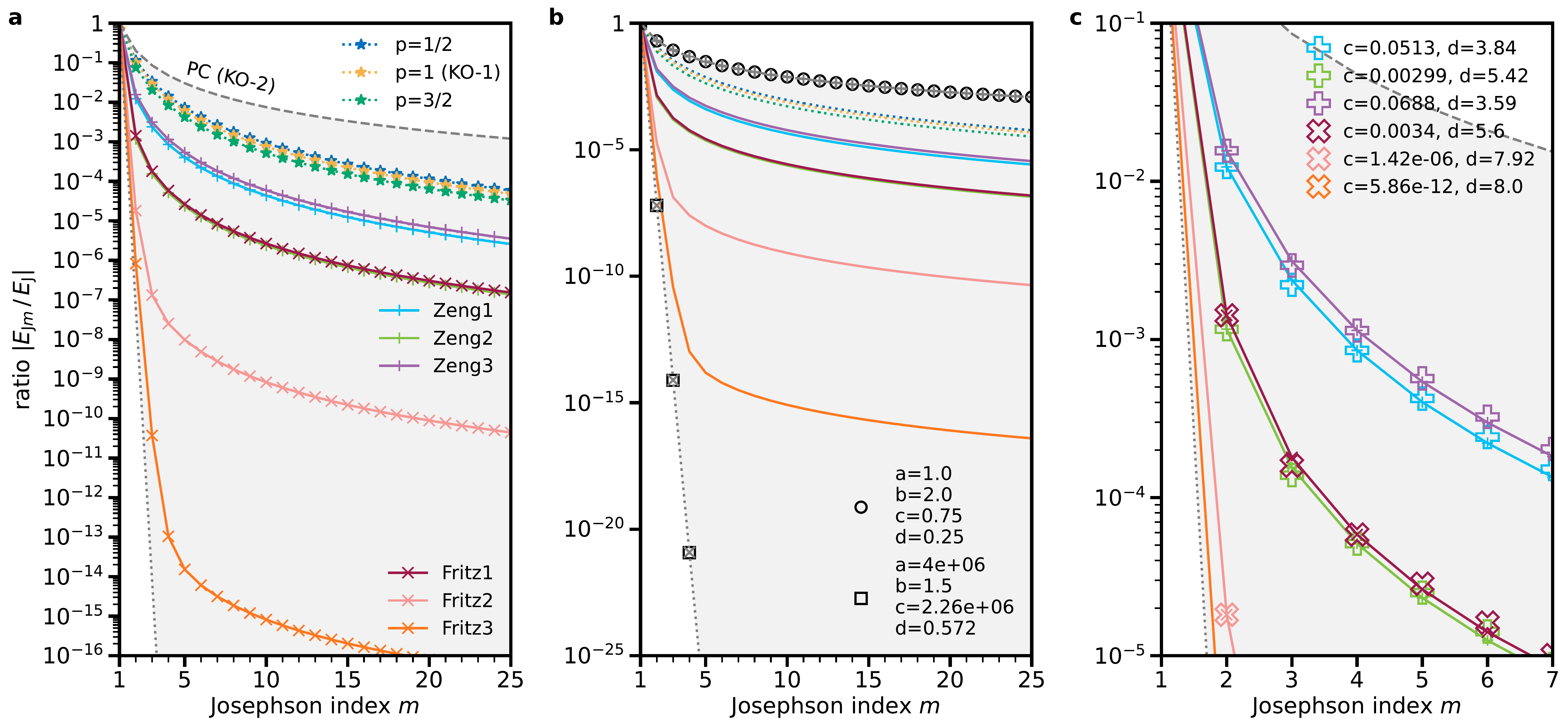}
  \caption{\textbf{Comparison of the scaling of the Josephson energy ratio $\vert E_{\mathrm{J}m}/E_\mathrm{J}\vert$ for various models.}
  \textbf{a} We show the ratios in Eq.~\eqref{eq:EJm_different_p_cases} for the universal distribution Eq.~\eqref{eq:rhouni}, as well as the ratios computed for the mesoscopic $(\bar d,\sigma)$ model using parameters for the average barrier thickness $\bar d$ and the standard deviation $\sigma$ extracted from the literature~\cite{Zeng2015ObservationThicknessAlOxJosephsonJunctions,Fritz2019} (see Table~\ref{tab:literaturebarriers}).
  \textbf{b}~The $(a,b,c,d)$ model in Eq.~\eqref{eq:EJmOverEJGeneral} can reproduce the upper and lower limits, namely the point contact (black circles, see Eq.~\eqref{eq:pointcontactModel}, where the reference result from Eq.~\eqref{eq:pointcontact} is shown with gray plusses) and the homogeneous barrier of small transparency (black squares, see Eq.~\eqref{eq:lowerboundModel}, where the reference result from Eq.~\eqref{eq:EJmOverEJSingleChannel} for $T_1=10^{-6}$ is shown with gray crosses). As a reference, results from the mesoscopic model given by Eq.~\eqref{eq:rhorec} for the samples listed in Table~\ref{tab:literaturebarriers} are shown with solid lines and results from the universal distribution are shown with dotted lines (see legend of panel a).
  \textbf{c}~The $(a,b,c,d)$ model in Eq.~\eqref{eq:EJmOverEJGeneral} can approximately reproduce the ratios predicted by the mesoscopic $(\bar d,\sigma)$ model. Since there is no visible exponential decay for the $(\bar d,\sigma)$ model (see panel a), $a=1$ is always fixed and $b=3$ is determined from an asymptotic fit, leaving only $c$ and $d$ as free parameters. 
  The resulting values are given in the legend. 
  Colors correspond to the same cases as in panel a.
  Hollow symbols lie on top of the plusses and crosses if the approximation in terms of the $(a,b,c,d)$ model is good.
  }
  \label{suppfig:EJmOverEJ}
\end{figure*}

\subsubsection{Universal distributions from the literature}\label{sec:universal_distributions}

There are regimes in which the transparency distribution $\rho(T)$ is expected to take a universal form, independent of the details of the junction (see e.g.~\cite{Likharev1985JosephonJunctionsPhasePeriodic,Schep1997TransportThroughDirtyInterfaces,Golubov2004,Kirmani2019QuasiclassicalCircuitTheoryContiguousDisordered} and references therein):
\begin{equation}\label{eq:rhouni}
    \rho(T) = \frac{n_p}{T^p\sqrt{1-T}}\,,
\end{equation}
where $n_p$ is a normalization factor, and $p=1/2,1,3/2$ enumerates three distinct cases which we discuss in the following (the corresponding distributions are shown in Fig.~\ref{suppfig:distributions}b). As we show in this work, typical qubit tunnel junctions are not in any of the universal regimes, so their microscopic and mesoscopic properties matter.

The case $p=1/2$ represents a chaotic dot 
connected via identical ballistic point contacts to two superconducting
leads~\cite{Brouwer1997ChaoticDot}. The case $p=1$, also known as the KO-1 model by Kulik and Omelyanchuk~\cite{Kulik1975KO1model}, represents a diffusive quasi-1D wire of length $L$ longer than the Fermi wavelength, $L\gg\lambda_F$, but still in the short-junction regime $L\ll \sqrt{\xi_0 l}$, where $\xi_0$ is the superconducting coherence length and $l$ the electron mean free path with $l\ll \xi_0$; the wire transverse size is $\ll L$~\cite{Golubov2004}.
It has been implemented with aluminum nanobridges~\cite{Vijay2009OptimizingAnharmonicityNanoscaleWeakLink,Vijay2010ApproachingIdealWeakLinkNanobridges,LevensonFalk2011NonlinearMicrowaveResponseWeakLink} and has been proposed to describe transmons based on superconductor-constriction-superconductor (ScS) junctions~\cite{liu2023ScSTransmonKO1}.
The case $p=3/2$ represents a disordered interface ($L \ll \lambda_F$)~\cite{Schep1997TransportThroughDirtyInterfaces} and it could be relevant to semiconductor quantum well junctions~\cite{OConnell2011EpitacialSS2D}.

Evaluating Eq.~\eqref{eq:EJm_integral} with the distribution Eq.~\eqref{eq:rhouni} yields the higher-order Josephson energies
\begin{align}
    E_{\mathrm{J}m}
    &= \frac{\Delta}{4}\frac{N}{m}n_p\frac{(-1)^{m+1}}{16^{m-1}}\binom{2m-2}{m-1}\sqrt{\pi}\frac{\Gamma(m-p+1)}{\Gamma(m-p+3/2)} {}_3F_2(m-\sfrac{1}{2},m+\sfrac{1}{2},m-p+1;2m+1,m-p+\sfrac{3}{2};1)\\[1.3em]
    &=\frac{\Delta}{4}\frac{N}{m}n_p\left\{
    \begin{array}{ll}
        \dfrac{(-1)^{m+1}}{16^{m-1}}\displaystyle\binom{2m-2}{m-1}\binom{2m}{m}\dfrac{\pi}{4^m} {}_3F_2(m-\sfrac{1}{2},m+\sfrac{1}{2},m+\sfrac{1}{2};2m+1,m+1;1)& (p=\frac1 2) \\[1.2em]
        (-1)^{m+1}\dfrac{8}{4m^2-1}& (p=1)\\[1em]
        \dfrac{(-1)^{m+1}}{16^{m-1}}\displaystyle\binom{2m-2}{m-1}\binom{2m-2}{m-1}\dfrac{\pi}{4^{m-1}} {}_3F_2(m-\sfrac{1}{2},m+\sfrac{1}{2},m-\sfrac{1}{2};2m+1,m;1)& (p=\frac3 2)
    \end{array}\right. \,.\label{eq:EJm_different_p_cases}
\end{align}
The absolute ratios $\vert E_{\mathrm{J}m}/E_\mathrm{J}\vert$ up to $m=25$ are shown in Fig.~\ref{suppfig:EJmOverEJ}a. Interestingly, in all cases the contribution of the second Josephson energy is approximately $10\,\%$ (we have $E_{\mathrm{J}2}/E_\mathrm{J} \approx -11\,\%,-10\,\%,-7.5\,\%$ for $p=1/2, 1, 3/2$).  The increase in absolute value of the ratio with decreasing $p$ is due to the relatively lower weight given to low transparencies in the respective transparency distributions (see upper limits in Fig.~\ref{suppfig:distributions}b).

The closed-form Josephson harmonics expansion for $p=1$ in Eq.~\eqref{eq:EJm_different_p_cases} also provides an easy way to construct the KO-1 Hamiltonian in the charge basis (in which the $\cos(m\varphi)$ term describes the $m^{\mathrm{th}}$ subdiagonal),
\begin{equation}
    H_\mathrm{J}^{\text{KO-1}} 
    = \sum_{m=1}^\infty E_{\mathrm{J1}} (-1)^{m}\frac{3}{m(4m^2-1)} \cos (m\varphi)\,.
\end{equation}
This form agrees with the Josephson part of the ScS Hamiltonian $E_\mathrm{J}'(2\ln(\cos^2(\varphi/2)) + 4\sin(\varphi/2)\tanh^{-1}(\sin(\varphi/2)))$ given in~\cite{liu2023ScSTransmonKO1} (after the substitution $E_\mathrm{J}'=3E_{\mathrm{J1}}/4$) where it was approximately diagonalized using finite differences in $\varphi$ space.

\subsubsection{Mesoscopic model of barrier inhomogeneity}\label{sec:mesoscopicmodel}

In this section, we derive a transparency distribution $\rho(T)$ from a model of an inhomogeneous tunnel barrier. To this end, we describe the charge transport through the insulating barrier of a JJ in terms of tunneling processes through a potential barrier with a non-uniform thickness distribution $\rho(d)$. We expect the result to be applicable to ``regular'' barriers, in the sense that the thickness $d$ of the barrier between the leads can be described by a Gaussian distribution with average thickness $\bar d$ and standard deviation $\sigma$ (cf.~Fig.~\ref{fig:3}c of the main text and the STEM images and the molecular dynamics simulations in Section~\ref{sec:moleculardynamics}). 

Tunneling through AlO$_x$ barriers has previously been studied using a trapezoidal barrier model~\cite{Hartman1964ElectronTunnelingThroughAluminumOxide,Aref2014NonuniformAlOxBarriers}. Here we use the simpler rectangular barrier model with height given by the average height of the trapezoidal barrier. This is a good approximation up to corrections quadratic in the ratio of the difference over the average height $\phi$ (as measured from the Fermi energy $E_F$). For a rectangular barrier of thickness $d$, the textbook result for the transmission probability $T$ is~\cite{LandauLifshitz}
\begin{equation}\label{eq:Td}
    T = \frac{1}{1+a^2 \sinh^2\left(d/d_0\right)}\,,
\end{equation}
with
\begin{equation}\label{eq:Tparams}
    a^2 = \frac{1}{4}\left(\sqrt{\frac{m_i \phi}{m E_F}}-\sqrt{\frac{m E_F}{m_i \phi}}\right)^2, \quad d_0 = \hbar/\sqrt{2m_i \phi}\,,
\end{equation}
and where we have assumed tunneling electrons to have the Fermi energy. For aluminum, within the free electron model, $m$ is the electron mass and $E_F = 11.67\,$eV. The parameter $m_i$ denotes the effective mass in the band of the insulator closest to the energy of the tunneling electrons; $m_i$ enters Eq.~\eqref{eq:Tparams} as a prefactor for $\phi$, so only the product of the two quantities is relevant.
From the literature, see Table~\ref{tab:barpar}, we estimate $a^2 = 2.87$ and $d_0 = 0.21\,$nm (we caution the reader that the parameters can depend on how the oxide is interfaced with the aluminum layer~\cite{Koberidze2016EffectOfGeometryOnAlOTunnelJunctions}).

\begin{table}[bt]
    \caption{Parameters for AlO$_x$ barriers taken from the indicated references. $m_i/m$ is the effective electron-mass ratio, $\phi$ is the average height of the potential barrier, and $a$ and $d_0$ are parameters for the tunnel-transmission probability in Eq.~\eqref{eq:Td}.}
    \begin{center}
        \begin{ruledtabular}
        \begin{tabular}{@{}lcccr@{}}
        Reference & $m_i/m$ & $\phi$ (eV) & $a^2$ & $d_0$ (nm) \\
        \colrule
         \cite{Gundlach1967InvestigationOfAl2O3TunnelThickness,Gundlach1971LogarithmicConductivityAl2O3TunnelJunctions} & 0.44 & 2.05 & 2.75 & 0.206 \\
         \cite{Hartman1964ElectronTunnelingThroughAluminumOxide} & 1 & 0.84 & 2.99 & 0.213
        \end{tabular}
        \end{ruledtabular}
    \label{tab:barpar}
    \end{center}
\end{table}

Since measurements of the barrier thickness $d$ have been well described by a Gaussian distribution~\cite{Zeng2015ObservationThicknessAlOxJosephsonJunctions}, we consider the distribution
\begin{equation}
\rho(d) = \frac{2\Theta(d)}{1+\mathrm{Erf}(\bar{d}/\sqrt{2}\sigma)} \frac{1}{\sqrt{2\pi} \sigma} e^{-\left(d - \bar{d}\right)^2/\left(2\sigma^2\right)}\,,
\end{equation}
where $\Theta$ is the step function and $\mathrm{Erf}$ is the error function.
The prefactor of the Gaussian ensures normalization and that $d\geq 0$. Inverting Eq.~\eqref{eq:Td} to get the thickness $d$ as a function of the transparency $T$, one can then find the probability density for the transparency,
\begin{equation}\label{eq:rhorec}
        \rho(T) = \frac{2}{1+\mathrm{Erf}(\tilde{d}/\sqrt{2}\tilde{\sigma})} \frac{1}{2 T \sqrt{1-T}\sqrt{1-T+a^2T}} \frac{1}{\sqrt{2\pi}\tilde{\sigma}} e^{-\left(f(T) + \alpha \right)^2/(2\tilde{\sigma}^2)}\,,
\end{equation}
where $\tilde{d}=\bar{d}/d_0$, $\tilde{\sigma} = \sigma/d_0$, and
\begin{equation}
    f(T) = \log \frac{\sqrt{T}}{\sqrt{1-T}+\sqrt{1-T+a^2T}} \, ,\quad \alpha = \tilde{d} + \log a\,.
\end{equation}
For given $(\tilde d,\tilde \sigma)$, the Josephson harmonics $E_{\mathrm{J}m}$ can be obtained by evaluating
Eq.~\eqref{eq:EJratios} with the distribution $\rho(T)$ of Eq.~\eqref{eq:rhorec}. Examples for the absolute ratios $\vert E_{\mathrm{J}m}/E_{\mathrm{J}}\vert$ for the samples listed in Table~\ref{tab:literaturebarriers} are shown in Fig.~\ref{suppfig:EJmOverEJ}a. Results for the KIT, ENS, and K\"oln experiments are shown in Fig.~\ref{suppfig:additionalresults} and Table~\ref{tab:modelparameters} below. Note that in this model a non-negligible ratio $\sigma/\bar{d}$ indicates that regions of smaller thickness and thus high-transparency conduction channels are likely, which is not completely unexpected given the inhomogeneity of AlO$_x$ barriers (see the STEM images in Section~\ref{sec:moleculardynamics}).

For $a\gg 1$ in Eq.~\eqref{eq:Td}, at leading order the distribution in Eq.~\eqref{eq:rhorec} reduces for $T\gg 1/(1+a^2)$ to that in Eq.~\eqref{eq:rhouni} with $p=3/2$. However, it takes log-normal form for $T\ll 1/(1+a^2)$, indicating that the channels with low transparencies are responsible for a deviation from universality; for Al/AlO$_x$/Al, $a^2$ is not very large, so one should not expect to observe any universal behavior.

We can now investigate when a junction can be considered a ``good'' tunnel junction, meaning not only $\langle T \rangle\ll 1$ but also $\langle T^2 \rangle/\langle T \rangle \ll 1$. The latter is a necessary condition to be able to neglect higher harmonics, although one should also check what happens for higher moments;
the second condition is not satisfied for the universal distributions. For the transparency distribution $\rho(T)$ in Eq.~\eqref{eq:rhorec}, we have approximately
\begin{equation}
    \langle T^n \rangle \simeq \frac{1}{1-\mathrm{Erfc}(\bar{d}/\sqrt{2}\sigma)/2} 
    \left(\frac{4}{a^2}\right)^n e^{\frac{2n\sigma}{d_0}\left(\frac{n\sigma}{d_0}-\frac{\bar{d}}{\sigma}\right)} \left[1 -\frac12 \mathrm{Erfc}\bigg(\frac{\bar{d}}{\sqrt{2}\sigma}-\frac{\sqrt{2}n\sigma}{d_0}\bigg)\right]\,,
\end{equation}
where $\mathrm{Erfc}=1-\mathrm{Erf}$ is the complementary error function.
For a sharp barrier, $\sigma \ll \sqrt{\bar{d} d_0}$, this reduces to $ \langle T^n \rangle \simeq (4/a^2)^n e^{-2n\bar{d}/d_0}$ (assuming $n$ not too large, so that the Erfc factors can be negligible), and a junction with $\bar{d}\gg d_0 $ would indeed behave as a ``good'' tunnel junction. For AlO$_x$ barriers,
since $4/a^2$ is of order unity, the condition $\langle T \rangle\ll 1$ reduces to $\bar{d}/d_0 \gg \max\{1, 2(\sigma/d_0)^2\}$, while $\langle T^2 \rangle/\langle T \rangle \ll 1$ is satisfied if $\bar{d}/d_0 \gg \max\{1, 4(\sigma/d_0)^2\}$; therefore, assuming $\sigma/d_0 \gtrsim 1$, we also need $\bar{d}/d_0 \gg 4(\sigma/d_0)^2$ (note that since these ratios are in the argument of the exponential, the large inequality sign ``$\gg$'' can mean a factor of 2 or 3, not orders of magnitude). Based on direct observations on three samples~\cite{Zeng2015ObservationThicknessAlOxJosephsonJunctions}, the latter condition is not met, which implies that $E_{\mathrm{J}2}/E_\mathrm{J}$ can be non-negligible, see Table~\ref{tab:literaturebarriers}. 
It should be noted that barriers of similar thickness and smaller standard deviation can be fabricated if proper care is taken~\cite{Fritz2019}; those junctions have $\tilde{d} > 4 \tilde\sigma^2$, see Table~\ref{tab:literaturebarriers}. They also have $\langle T \rangle\ll 1$ and $\langle T^2 \rangle/\langle T \rangle \ll 1$, but $\langle T^2 \rangle/\langle T \rangle > \langle T \rangle$, meaning that significant fluctuations in transparency are nonetheless present. 

As an additional check of our approach, we note that the distribution of transparencies could explain the observed magnitude of the subgap current in SIS and SIN junctions, see e.g.~\cite{Greibe2011PinholesRuledOut} and references therein. For a narrow distribution of transparencies, the ratio $G_{sg}/G_N$ between subgap conductance and normal-state conductance should be of the order of the average transparency, since $G_{sg} \propto T^2$ and $G_N \propto T$. For the data in Fig.~2 of \cite{Greibe2011PinholesRuledOut}, we can convert the ratio $G_N/A$ into average transparencies by multiplying it by $\lambda_F^2/G_Q \approx  \SI{1.6e-9}{\ohm\milli\meter\squared}$; this means that for the Al-AlO$_x$-Al junctions, $\langle T \rangle$ varies  roughly between $7\times 10^{-7}$ and $10^{-4}$. Over that range, the ratio $G_{sg}/G_N$ goes from about $3\times 10^{-5}$ to $5\times 10^{-3}$. At the lower end of this range, the pair of numbers is similar to that of sample Fritz3 in Table~\ref{tab:literaturebarriers}. Assuming for simplicity $G_{sg}/G_N=\langle T^2 \rangle/\langle T \rangle$, we can match the observation by taking $\bar{d}/d_0 =8.19$ and $\sigma/d_0=0.97$. Then decreasing $\bar{d}$ while keeping constant $\sigma$ as to get $\langle T \rangle\sim 10^{-4}$, we get $\langle T^2 \rangle/\langle T \rangle\sim 4\times 10^{-3}$, in fair agreement with the upper end of the observed range (keeping the standard deviation $\sigma$ constant is consistent with the finding of Ref.~\cite{Fritz2019} that it is correlated with variations in thickness of the bottom aluminum layer, whose fabrication was presumably not changed from junction to junction in Ref.~\cite{Greibe2011PinholesRuledOut}). 

\begin{table*}
  \caption{Values for the average thickness $\bar d$ and the standard deviation $\sigma$ of AlO$_x$ barriers in tunnel junctions found in the literature. Shown are the first three samples reported in Ref.~\cite{Zeng2015ObservationThicknessAlOxJosephsonJunctions} (Zeng) and Ref.~\cite{Fritz2019} (Fritz), respectively. Columns 4--7 contain the corresponding unitless quantities, $\tilde{d}=\bar{d}/d_0$ and $\tilde{\sigma} = \sigma/d_0$, in terms of $d_0=\SI{0.21}{\nano\metre}$ \cite{Koberidze2016EffectOfGeometryOnAlOTunnelJunctions}. The quantities in columns 8--10 contain the moments $\langle T \rangle$ and $\langle T^2 \rangle /\langle T \rangle$ as well as the ratio $E_{\mathrm{J}2}/E_\mathrm{J}$, which have been computed using Eq.~\eqref{eq:rhorec}.}
  \begin{center}
    \begin{ruledtabular}
      \begin{tabular}{@{}lccccccccr@{}}
      Sample & $\bar d$ (nm) & $\sigma$ (nm) & $\tilde{d}$ & $\tilde{\sigma}$ & $\tilde{d}/\tilde{\sigma}$ & $4\tilde{\sigma}^2$ & $\langle T \rangle$ & $\langle T^2 \rangle/\langle T \rangle$ & $E_{\mathrm{J}2}/E_\mathrm{J}$\\
        \colrule
 Zeng1~\cite{Zeng2015ObservationThicknessAlOxJosephsonJunctions} & 1.66 & 0.35 & 7.90 & 1.67 & 4.74 & 11.11 &  $4.6\times 10^{-5}$ & 0.098 & $-0.012$ \\
 Zeng2~\cite{Zeng2015ObservationThicknessAlOxJosephsonJunctions} & 1.88 & 0.32 & 8.95 & 1.52 & 5.87 & 9.29 & $2.4\times 10^{-6}$ & 0.011 & $-0.0012$\\
 Zeng3~\cite{Zeng2015ObservationThicknessAlOxJosephsonJunctions} & 1.73 & 0.37 & 8.24 & 1.76 & 4.68 & 12.42 & $4.3\times 10^{-5}$ & 0.120 & $-0.016$\\
Fritz1~\cite{Fritz2019} & 1.62 & 0.29 & 7.71 & 1.38 & 5.59 & 7.63 & $1.3\times 10^{-5}$ & $0.014$ & $-0.0014$\\
Fritz2~\cite{Fritz2019} & 1.65 & 0.23 & 7.86 & 1.10 & 7.17 & 4.80 & $2.3\times 10^{-6}$ & $2.8\times 10^{-4}$ & $-1.8\times 10^{-5}$\\
Fritz3~\cite{Fritz2019} & 1.73 & 0.19 & 8.24 & 0.90 & 9.11 & 3.27 & $5.0\times 10^{-7}$ & $1.3\times 10^{-5}$ & $-8.3\times 10^{-7}$\\
      \end{tabular}
      \end{ruledtabular}
    \label{tab:literaturebarriers}
  \end{center}
\end{table*}

Our mesoscopic model can describe the harmonics in the Josephson junctions of KIT, ENS and K\"oln very well (see Fig.~\ref{fig:3}b in the main text). 
For the IBM junctions, the Josephson harmonics are much larger and stay relevant until high order (cf.~Fig.~\ref{suppfig:scan} below).
They are close to the point contact limit, which asymptotically scales as $1/m^2$.
The mesoscopic model, however, shows a stronger decay (see Fig.~\ref{suppfig:EJmOverEJ}a) that cannot reproduce the large contributions required up to $E_{\mathrm{J}6}$ and thus cannot describe the observed spectra very well.
A physical reason for this may lie in the fact that the assumption of a Gaussian thickness distribution with only two parameters is too simple to model the corresponding barriers. 
We note that for the samples for which the model works well, the fit mostly constrains the ratio $\bar d/\sigma$ rather than their absolute values.

\subsection{Josephson harmonics}
\label{sec:josephsonharmonics}

In this section, we test several models motivated by the discussion of transparency distributions in the previous section. Additionally, we apply these models to previously published transmon spectra that can be found in the literature. Subsequently, we report the scan of the full range of Josephson harmonics for all samples considered in the main text, to distill the maximum information about the Josephson harmonics from the available experimental data. Finally we discuss properties of the Josephson harmonics model such as the charge dispersion, the Hamiltonian, the Josephson potential and current-phase relations, a perturbative expansion, and how engineering Josephson harmonics can both reduce the charge dispersion and increase the anharmonicity.

\subsubsection{Effective parameterizations}
\label{sec:motivationabmodel}

The results for $E_{\mathrm{J}m}$ for the transparency distributions $\rho(T)$ considered in the previous section suggest that a suitable parameterization of $E_{\mathrm{J}m}/E_\mathrm{J}$ should be (i) alternating in sign, (ii) decreasing in absolute magnitude, and (iii) should allow for both exponential and power-law decay. A suitable parameterization for $m\ge2$ that also includes the limits given by Eqs.~\eqref{eq:pointcontact} and \eqref{eq:lowerbound} is
\begin{equation}
    \frac{E_{\mathrm{J}m}}{E_\mathrm{J}} = (-1)^{m+1} \frac{c}{a^m(m^b - d)}\,,
    \label{eq:EJmOverEJGeneral}
\end{equation}
where $(a,b,c,d)$ are the model parameters to be determined. The point contact given by Eq.~\eqref{eq:pointcontact} corresponds to
\begin{equation}
\label{eq:pointcontactModel}
    a = 1\,,\quad
    b = 2\,,\quad
    c = \frac 3 4\,,\quad
    d = \frac 1 4\,,
\end{equation}
and the lower bound of a homogeneous barrier with small transparency given by Eq.~\eqref{eq:lowerbound} corresponds to
\begin{equation}
\label{eq:lowerboundModel}
    a = \frac 4 {T_1}\,,\quad
    b = \frac 3 2\,,\quad
    c = \frac 4 {\sqrt{\pi}T_1}\,,\quad
    d = 2\sqrt{2}-\frac 4 {\sqrt{\pi}}\,,
\end{equation}
where we used the remaining freedom in $d$ to match $E_{\mathrm{J}2}/E_\mathrm{J}$. Both are shown in Fig.~\ref{suppfig:EJmOverEJ}b which confirms that the $(a,b,c,d)$ model can cover the limiting cases. Furthermore, as Fig.~\ref{suppfig:EJmOverEJ}c shows, the $(a,b,c,d)$ model can approximately reproduce the mesoscopic $(\bar d,\sigma)$ model introduced in Section~\ref{sec:mesoscopicmodel}. Note that, although $a$ and $c$ are related to the inverse transparency in Eq.~\eqref{eq:lowerboundModel}, this is not necessarily the case when the model is fitted to given spectroscopy data.

In addition to the $(a,b,c,d)$ model in Eq.~\eqref{eq:EJmOverEJGeneral}, we consider the simpler $(a,b)$ model with $c=1$ and $d=0$ fixed and given by
\begin{equation}
    \frac{E_{\mathrm{J}m}}{E_\mathrm{J}} = (-1)^{m+1} \frac{1}{a^mm^b}\,.
    \label{eq:EJmOverEJ}
\end{equation}
Evidently, this choice captures the most prominent features, namely the power-law scaling of the free quantum point contact as well as an asymptotic exponential scaling towards zero as present in the lower limit of a homogeneous barrier with small transparency. In terms of fitting, we note that a data set might be describable by various combinations of $a$ and $b$ that differ by orders of magnitude. The reason is that, while the model is inspired by the asymptotic scaling as a function of $m$, for an experiment only the first $E_{\mathrm{J}m}$ coefficients may be relevant, and e.g.~the same $E_{\mathrm{J}2}$ can be obtained by various combinations of $a$ and $b$.

We emphasize in this context that, although one might want to ascribe some meaning to the model parameters (e.g.~$a$ and $b$), they are internal, unobservable variables (in the sense of the inverse problem, cf.~\cite{ChuGolub2005InverseEigenvalueProblemsBook,jaynes2003probability}). There can be many different parameterizations that describe the observed spectra similarly well. Fundamentally, the distribution of transparencies for the conduction channels of a general Josephson junction can be very complicated.
Therefore, it might not be possible to capture this complex many-body configuration in terms of only a few parameters, or in other words, reality does not necessarily fit into a simple mathematical model.

\begin{table*}
  \caption{Model parameters of all models presented in Fig.~\ref{suppfig:additionalresults}.
  The ratios $E_{\mathrm{J}m}/E_{\mathrm{J}}$ and the parameters $(a,b,c,d)$ of the models discussed in Section~\ref{sec:motivationabmodel} are unitless, $\bar d$ and $\sigma$ of the mesoscopic model discussed in Section~\ref{sec:mesoscopicmodel} are given in nanometers, $I_\text{c}$ is given in nanoamperes, and all other parameters are given in gigahertz. Gray entries denote derived model properties that are computed from the model parameters. Note that for the KIT results, the bare resonator frequencies $\Omega$ agree very well with the high-power cavity frequency given in Figs.~\ref{suppfig:KIT_CD1}a and \ref{suppfig:KIT_CD3_1}a below, see also~\cite{Lescanne2019EscapeUnconfinedStatesFloquet}).
  In the second-last row, we show results for the K\"oln sample (note that $E_{\mathrm{J}}$ is not fixed but tuned according to the magnetic field bias).
  In the last row, we show representative results for qubit 0 of the IBM Hanoi device (additional data is available in the repository~\cite{repository}). Note that for the $(a,b)$ model for IBM Q0, although the spectrum is reproduced by definition, the value of $E_{\mathrm{C}}$ comes out much larger than the reference value of $\SI{300}{\mega\hertz}$.
  The last line of the last row contains the parameters of an engineered model with reduced charge dispersion and increased anharmonicity (see Fig.~\ref{suppfig:engineering} below).
  All parameters have been obtained by solving the HamPIEP (see~Section~\ref{sec:HamPIEP}).}
  \begin{center}
    \begin{ruledtabular}
      \begin{tabular}{@{}lccccccccccccccr@{}}
      Sample & Model & $E_{\mathrm{C}}/h$ & $E_{\mathrm{J}}/h$ & $E_{\mathrm{J2}}/E_{\mathrm{J}}$ & $E_{\mathrm{J3}}/E_{\mathrm{J}}$ & $E_{\mathrm{J4}}/E_{\mathrm{J}}$ & $I_\text{c}$ & a & b & c & d & $\bar d$ & $\sigma$ & $\Omega/h$ & $G/h$ \\
        \colrule
\multirow[c]{5}{*}{KIT}
& standard & $0.197$ & $24.852$ & - & - & - & {\color{gray}$50.0$} & - & - & - & - & - & - & $7.454$ & $0.078$ \\
& $E_\mathrm{J2}$ & $0.242$ & $21.801$ & $-0.019$ & - & - & {\color{gray}$44.0$} & - & - & - & - & - & - & $7.454$ & $0.086$ \\
& $E_\mathrm{J4}$ & $0.242$ & $21.997$ & $-0.026$ & $0.004$ & $-0.001$ & {\color{gray}$43.9$} & - & - & - & - & - & - & $7.454$ & $0.086$ \\
& $a,b$ & $0.266$ & $20.983$ & {\color{gray}$-0.042$} & {\color{gray}$0.006$} & {\color{gray}$-0.002$} & {\color{gray}$42.0$} & $1.00$ & $4.58$ & - & - & - & - & $7.454$ & $0.090$ \\
& $\bar d,\sigma$ & $0.293$ & $20.186$ & {\color{gray}$-0.067$} & {\color{gray}$0.017$} & {\color{gray}$-0.007$} & {\color{gray}$40.1$} & - & - & - & - & $1.06$ & $0.45$ & $7.454$ & $-0.095$ \\
\colrule
\multirow[c]{5}{*}{KIT-CD2}
& standard & $0.206$ & $22.704$ & - & - & - & {\color{gray}$45.7$} & - & - & - & - & - & - & $7.454$ & $0.086$ \\
& $E_\mathrm{J2}$ & $0.242$ & $20.530$ & $-0.016$ & - & - & {\color{gray}$41.4$} & - & - & - & - & - & - & $7.454$ & $0.093$ \\
& $E_\mathrm{J4}$ & $0.242$ & $20.383$ & $-0.010$ & $-0.003$ & $0.001$ & {\color{gray}$41.5$} & - & - & - & - & - & - & $7.454$ & $0.093$ \\
& $a,b$ & $0.280$ & $19.337$ & {\color{gray}$-0.048$} & {\color{gray}$0.009$} & {\color{gray}$-0.002$} & {\color{gray}$38.6$} & $1.52$ & $3.16$ & - & - & - & - & $7.454$ & $0.100$ \\
& $\bar d,\sigma$ & $0.284$ & $19.361$ & {\color{gray}$-0.057$} & {\color{gray}$0.014$} & {\color{gray}$-0.005$} & {\color{gray}$38.4$} & - & - & - & - & $1.62$ & $0.50$ & $7.454$ & $0.101$ \\
\colrule
\multirow[c]{5}{*}{KIT-CD3}
& standard & $0.223$ & $13.803$ & - & - & - & {\color{gray}$27.8$} & - & - & - & - & - & - & $7.454$ & $0.080$ \\
& $E_\mathrm{J2}$ & $0.242$ & $13.164$ & $-0.009$ & - & - & {\color{gray}$26.5$} & - & - & - & - & - & - & $7.454$ & $0.083$ \\
& $E_\mathrm{J4}$ & $0.242$ & $13.225$ & $-0.013$ & $0.002$ & $-0.001$ & {\color{gray}$26.5$} & - & - & - & - & - & - & $7.454$ & $0.083$ \\
& $a,b$ & $0.247$ & $13.075$ & {\color{gray}$-0.016$} & {\color{gray}$0.002$} & {\color{gray}$-0.000$} & {\color{gray}$26.2$} & $4.07$ & $1.96$ & - & - & - & - & $7.454$ & $0.084$ \\
& $\bar d,\sigma$ & $0.254$ & $12.977$ & {\color{gray}$-0.024$} & {\color{gray}$0.005$} & {\color{gray}$-0.002$} & {\color{gray}$25.9$} & - & - & - & - & $1.62$ & $0.38$ & $7.454$ & $0.085$ \\
\colrule
\multirow[c]{6}{*}{ENS}
& standard & $0.167$ & $23.191$ & - & - & - & {\color{gray}$46.7$} & - & - & - & - & - & - & $7.739$ & $0.179$ \\
& $E_\mathrm{J2}$ & $0.181$ & $22.053$ & $-0.008$ & - & - & {\color{gray}$44.4$} & - & - & - & - & - & - & $7.739$ & $0.187$ \\
& $E_\mathrm{J4}$ & $0.186$ & $21.811$ & $-0.014$ & $0.001$ & $-0.000$ & {\color{gray}$43.8$} & - & - & - & - & - & - & $7.739$ & $0.189$ \\
& $a,b$ & $0.186$ & $21.835$ & {\color{gray}$-0.015$} & {\color{gray}$0.001$} & {\color{gray}$-0.000$} & {\color{gray}$43.9$} & $2.21$ & $3.81$ & - & - & - & - & $7.739$ & $0.189$ \\
& $a,b,c,d$ & $0.185$ & $21.876$ & {\color{gray}$-0.013$} & {\color{gray}$0.001$} & {\color{gray}$-0.000$} & {\color{gray}$44.0$} & $1.50$ & $5.49$ & $1.28$ & $0.58$ & - & - &  $7.739$ & $0.189$ \\
& $\bar d,\sigma$ & $0.195$ & $21.505$ & {\color{gray}$-0.028$} & {\color{gray}$0.006$} & {\color{gray}$-0.002$} & {\color{gray}$42.9$} & - & - & - & - & $1.63$ & $0.39$ & $7.739$ & $0.194$ \\
\colrule
\multirow[c]{3}{*}{MIT}
& standard & $0.243$ & $14.073$ & - & - & - & {\color{gray}$28.3$} & - & - & - & - & - & - & $10.975$ & $0.145$ \\
& $E_\mathrm{J2}$ & $0.248$ & $13.901$ & $-0.003$ & - & - & {\color{gray}$28.0$} & - & - & - & - & - & - & $10.975$ & $0.146$ \\
& $a,b$ & $0.250$ & $13.876$ & {\color{gray}$-0.004$} & {\color{gray}$0.000$} & {\color{gray}$-0.000$} & {\color{gray}$27.9$} & $4.09$ & $4.07$ & - & - & - & - & $10.975$ & $0.147$ \\
\colrule
\multirow[c]{2}{*}{KIT2}
& standard & $0.176$ & $17.444$ & - & - & - & {\color{gray}$35.1$} & - & - & - & - & - & - & $8.561$ & $0.194$ \\
& $a,b$ & $0.176$ & $17.443$ & {\color{gray}$-0.000$} & {\color{gray}$0.000$} & {\color{gray}$-0.000$} & {\color{gray}$35.1$} & $8.27$ & $12.47$ & - & - & - & - & $8.561$ & $0.194$ \\
\colrule
\multirow[c]{6}{*}{TUM}
& standard & $0.167$ & $44.908$ & - & - & - & {\color{gray}$90.4$} & - & - & - & - & - & - & $5.631$ & $0.127$ \\
& $E_\mathrm{J2}$ & $0.253$ & $33.884$ & $-0.031$ & - & - & {\color{gray}$68.7$} & - & - & - & - & - & - & $5.631$ & $0.156$ \\
& $E_\mathrm{J4}$ & $0.337$ & $29.023$ & $-0.078$ & $0.011$ & $-0.001$ & {\color{gray}$59.5$} & - & - & - & - & - & - & $5.631$ & $0.180$ \\
& $a,b$ & $0.252$ & $35.539$ & {\color{gray}$-0.057$} & {\color{gray}$0.011$} & {\color{gray}$-0.003$} & {\color{gray}$71.1$} & $1.00$ & $4.14$ & - & - & - & - & $5.631$ & $0.156$ \\
& $a,b,c,d$ & $0.360$ & $28.262$ & {\color{gray}$-0.098$} & {\color{gray}$0.022$} & {\color{gray}$-0.007$} & {\color{gray}$57.5$} & $1.00$ & $4.03$ & $1.94$ & $-3.48$ & - & - &  $5.631$ & $0.186$ \\
& $\bar d,\sigma$ & $0.261$ & $34.675$ & {\color{gray}$-0.062$} & {\color{gray}$0.016$} & {\color{gray}$-0.006$} & {\color{gray}$68.8$} & - & - & - & - & $1.61$ & $0.54$ & $5.628$ & $0.152$ \\
\colrule
\multirow[c]{3}{*}{K\"oln}
& standard & $0.285$ & $7.666$--$16.785$ & - & - & - & {\color{gray}$33.8$} & - & - & - & - & - & - & $7.545$ & $0.077$ \\
& $E_\mathrm{J4}$ & $0.330$ & $7.117$--$15.577$ & $-0.023$ & $0.004$ & $-0.001$ & {\color{gray}$31.1$} & - & - & - & - & - & - & $7.545$ & $0.083$ \\
& $\bar d,\sigma$ & $0.331$ & $7.144$--$15.627$ & {\color{gray}$-0.027$} & {\color{gray}$0.006$} & {\color{gray}$-0.002$} & {\color{gray}$31.1$} & - & - & - & - & $1.93$ & $0.43$ & $7.545$ & $0.083$ \\
\colrule
\multirow[c]{4}{*}{IBM Q0}
& standard & $0.302$ & $11.925$ & - & - & - & {\color{gray}$24.0$} & - & - & - & - & - & - & $7.160$ & $0.133$ \\
& $E_\mathrm{J4}$ & $0.300$ & $14.672$ & $-0.141$ & $0.083$ & $-0.027$ & {\color{gray}$34.7$} & - & - & - & - & - & - & $7.160$ & $0.133$ \\
& $a,b$ & $0.531$ & $9.217$ & {\color{gray}$-0.118$} & {\color{gray}$0.039$} & {\color{gray}$-0.014$} & {\color{gray}$18.7$} & $2.39$ & $0.56$ & - & - & - & - & $7.160$ & $0.176$ \\
& engineered & $0.300$ & $15.049$ & $-0.168$ & $0.084$ & $-0.008$ & {\color{gray}$37.0$} & - & - & - & - & - & - & $7.159$ & $0.136$ \\
      \end{tabular}
      \end{ruledtabular}
    \label{tab:modelparameters}
  \end{center}
\end{table*}

\begin{figure*}
  \centering
  \includegraphics[width=.85\textwidth]{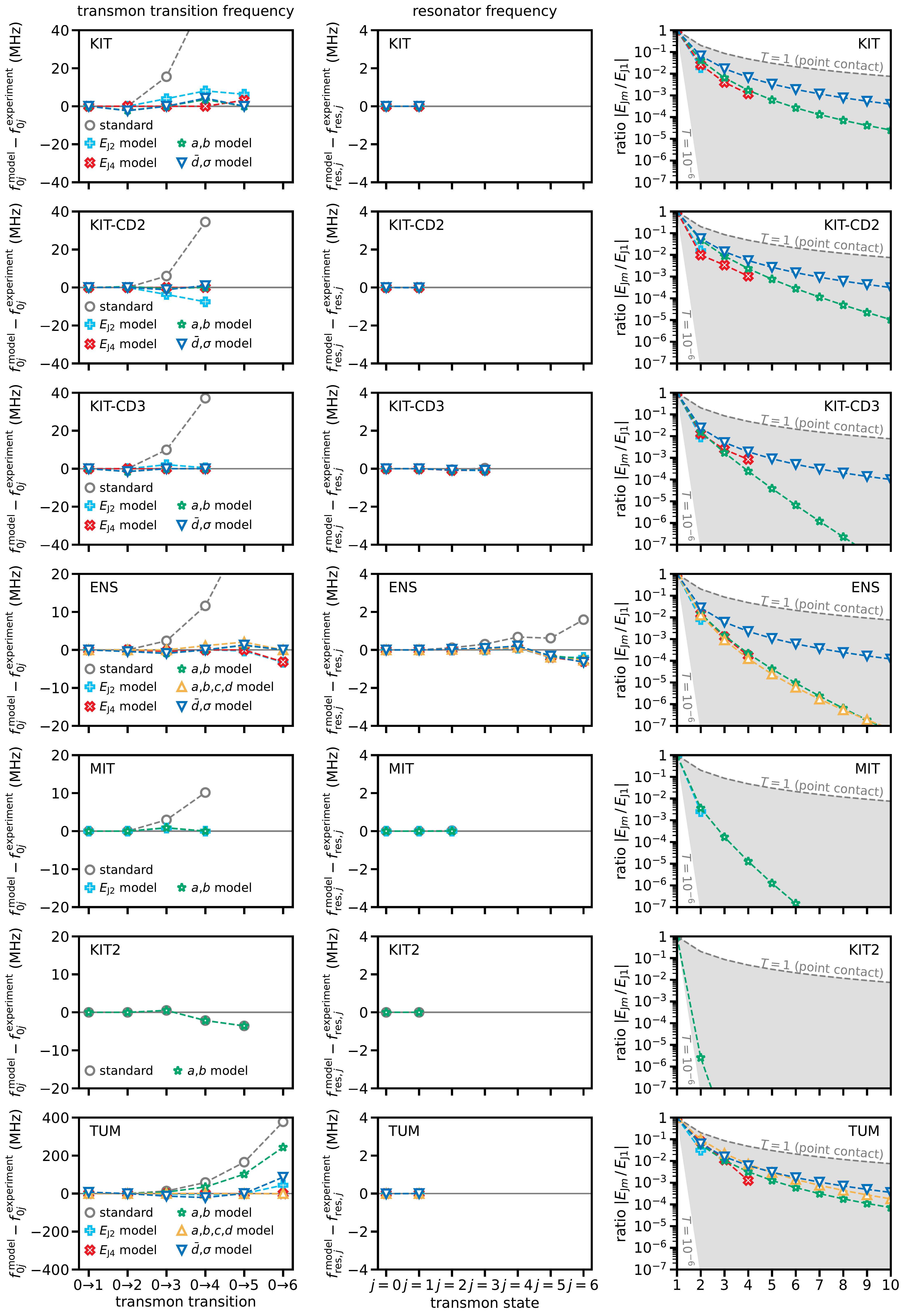}
  \caption{\textbf{Comparison of alternative models and further evidence.} Left column: Differences between the transmon transition frequencies $\smash{f_{0j}^\mathrm{model}}$ predicted by the fitted models and $\smash{f_{0j}^\mathrm{experiment}}$ for the transitions $0\to j$. Middle column: Differences between the resonator frequencies $\smash{f_{\mathrm{res},j}^\mathrm{model}}$ predicted by the fitted models and $\smash{f_{\mathrm{res}, j}^\mathrm{experiment}}$, conditional on the transmon being in state $j$. Right column: Absolute values of the Josephson energy ratios $|E_{\mathrm{J}m}/E_\mathrm{J}|$ obtained from the models. Each row shows a different experiment. Rows 1 through 4 show additional models (see text) for the KIT and ENS experiments discussed in the main text. The other rows show additional experiments with data extracted from the literature: Peterer et.~al.~\cite{Peterer2015HigherTransmonStates} (MIT), Schneider et.~al.~\cite{Schneider2018TransmonSpectroscopy,Schneider2020PhDThesisQuantumSensingExperiments} (KIT2), Xie et.~al.~\cite{Xie2018Compact3DQuantumMemory,Xie2019PhDThesis} (TUM).
  All models are identified by the same markers and colors across all panels (see legends).
  All model parameters are given in Table~\ref{tab:modelparameters}.
  }
  \label{suppfig:additionalresults}
\end{figure*}

A selection of alternative models that we have considered for various experiments is shown in Fig.~\ref{suppfig:additionalresults}. 
In addition to the $(a,b)$ model given by Eq.~\eqref{eq:EJmOverEJ}, we show results for the $(a,b,c,d)$ model given by Eq.~\eqref{eq:EJmOverEJGeneral} for the experiments at KIT and ENS (since in these cases we have sufficient experimental data to determine all model parameters), as well as for the mesoscopic $(\bar d,\sigma)$ model described in Section~\ref{sec:mesoscopicmodel}. 
The values of all model parameters are given in Table~\ref{tab:modelparameters}. Note that all parameterized models have physical ratios $E_{\mathrm{J}m}/E_{\mathrm{J}}$ (i.e., alternating, decaying, and within the two limits). This is not automatically the case for the truncated $E_{\mathrm{J}2}$ and $E_{\mathrm{J}4}$ models for which the ratios are fitted independently (e.g.~the KIT-CD2 case in Table~\ref{tab:modelparameters}).

As these results show, none of the models should be considered as the ultimate answer for the effects caused by the higher harmonics. 
It is an important task for future research to increase the accuracy in measuring fine-structure effects on the experimental side, and to find and study other alternative models and transparency distributions on the theoretical side, which could shed further light on understanding the complex structure of the Josephson effect in tunnel junctions.

\subsubsection{Additional evidence}
\label{sec:additionalevidence}

We find further support for our Josephson harmonics hypothesis by analyzing experiments with published spectroscopy results available in the literature, namely the experiments by Peterer et.~al.~\cite{Peterer2015HigherTransmonStates} (MIT), Schneider et.~al.~\cite{Schneider2018TransmonSpectroscopy} (KIT2; additional information is available in \cite{Schneider2020PhDThesisQuantumSensingExperiments}), and Xie et.~al.~\cite{Xie2018Compact3DQuantumMemory} (the spectroscopy is shown in Fig.~2.20 of \cite{Xie2019PhDThesis}). The results are shown in Fig.~\ref{suppfig:additionalresults}. In this figure, we also show evidence from the differences between predicted and measured resonator frequencies $f_{\mathrm{res},j}$ when the transmon is in state $j$ (computed from the dispersive shifts for $j\ge1$), as an addition to the evidence from the transmon transition frequencies shown in the main text (see Fig.~\ref{fig:3}). We see that a model including the higher-order Josephson harmonics can describe the data better than the standard transmon model. The only exception is KIT2, which is either an example for a highly uniform tunnel junction with very low transparencies or the data reported may not be accurate enough to reveal percent-level corrections in the Josephson harmonics. 

\subsubsection{Exhaustive scan}
\label{sec:scan}

\begin{figure*}
  \centering
  \includegraphics[width=\textwidth]{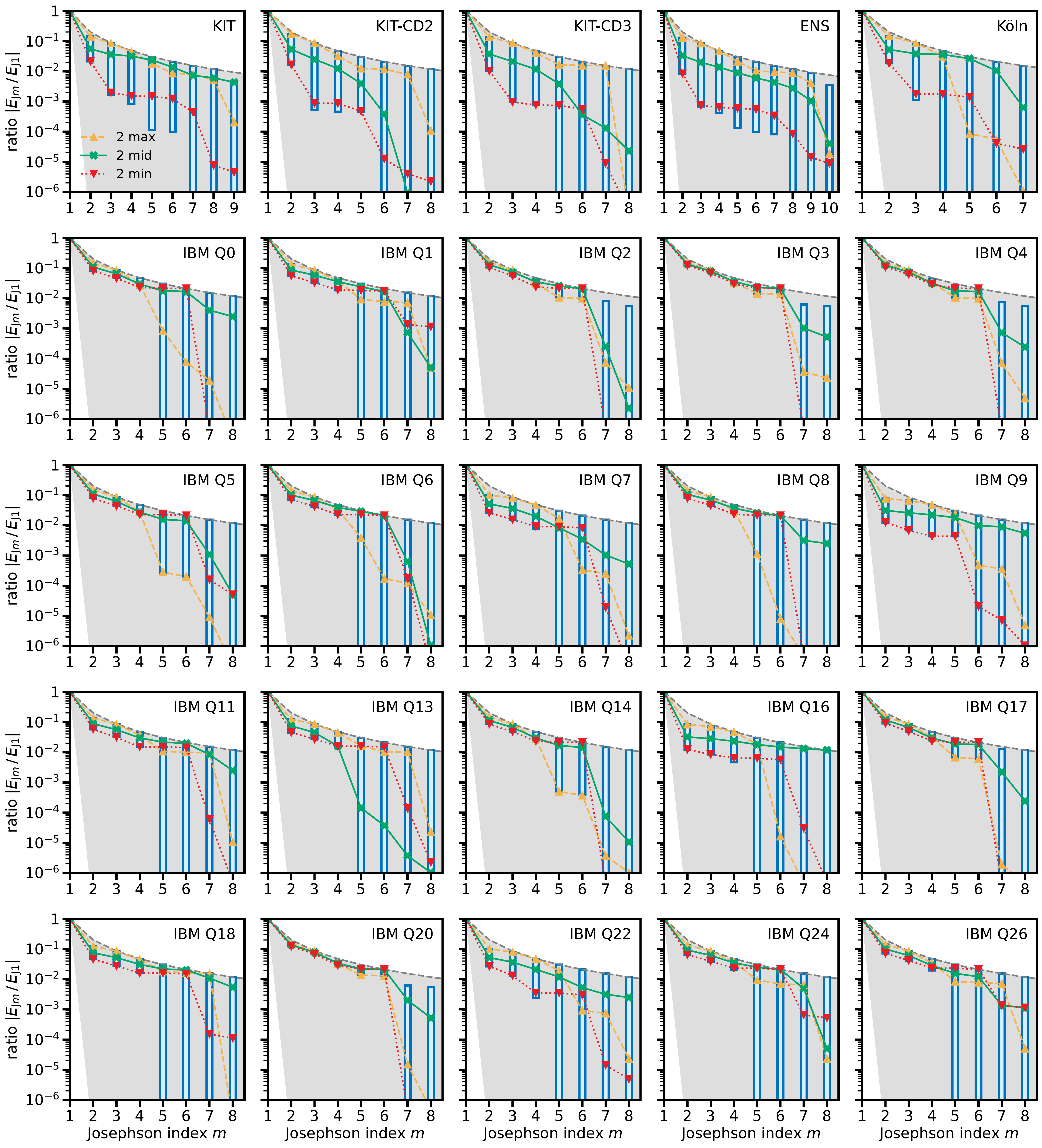}
  \caption{\textbf{Ranges of $E_{\mathrm{J}m}$ coefficients that can reproduce the observed spectra.}
  The blue bars show the ranges of possible $\vert E_{\mathrm{J}m}/E_\mathrm{J}\vert$, obtained by scanning four higher-order ratios $\mathbf y$ and for each $\mathbf y$ solving the HamPIEP (cf.~Section~\ref{sec:HamPIEP}) for the leading $N_f$ ratios (see \hyperref[sec:methods]{Methods}).
  The lines represent three particular trajectories $\mathbf{e}=(1,E_{\mathrm{J}2}/E_{\mathrm{J}1},\ldots,E_{\mathrm{J}N_f+4}/E_{\mathrm{J}1})$ corresponding to the largest $\vert E_{\mathrm{J}2}/E_{\mathrm{J}}\vert$ (yellow upward-pointing triangles, ``max''), the smallest $\vert E_{\mathrm{J}2}/E_{\mathrm{J}}\vert$ (red downward-pointing triangles, ``min''), and the geometric mean of both (green crosses, ``mid'').
  The magnitude of the corresponding $\sin(m\varphi)$ contribution to the current-phase relation $I_S(\varphi)$ is given by $m\vert E_{\mathrm{J}m}/E_{\mathrm{J}}\vert$ (see Eq.~\eqref{eq:IsofphiEJm}).
  Each panel shows the result for the transmon indicated by the label in the top right corner.
  The ranges for the K\"oln transmon (top right panel) correspond to the same dataset considered in Fig.~\ref{fig:3} of the main text (explicitly excluding further information about the charge dispersion, which would additionally reduce the extent of the bars, see Section~\ref{sec:chargedispersion}).}
  \label{suppfig:scan}
\end{figure*}

In this section, we give some additional details about the scanning procedure described in the \hyperref[sec:methods]{Methods} section. Furthermore, we show in Fig.~\ref{suppfig:scan} the detailed scanning results for all experiments considered in the main text, including the remaining transmons of the IBM Hanoi chip (cf.~Fig.~\ref{fig:3}c in the main text). Additionally, we show three specific trajectories corresponding to the minimum, maximum, and geometric mean of $\vert E_{\mathrm{J}2}/E_\mathrm{J}\vert$. 

Interestingly, when constraining $E_\mathrm{J5}$ and $E_\mathrm{J6}$ of some transmons to zero (see.~e.g.~IBM Q2), fitting only $E_\mathrm{J2}$ to $E_\mathrm{J4}$ yields unphysical results above the point-contact limit. These transmons have a very narrow range of allowed values for $E_{\mathrm{J}m}$, even for high orders such as $m=6$. Furthermore, we see that when the bars for $\vert E_{\mathrm{J}N_f+1}/E_\mathrm{J}\vert$ do not extend down towards zero, truncation at $E_{\mathrm{J}N_f}$ is not possible to describe the spectrum. This is the case for the first two cooldowns of the KIT system (note however that it is indeed possible for KIT-CD3). 

We note that by design, this procedure does not include non-perfect agreement with the experimental data. Taking into account experimental imprecision etc.~would be a complementary study and it is to be expected that the allowed ranges for the $E_{\mathrm{J}m}/E_\mathrm{J}$ become slightly larger.
Note, however, that the ranges can also be reduced by including additional information such as charge-dispersion measurements, as we discuss in the following section.

\subsubsection{Charge dispersion}
\label{sec:chargedispersion}

\begin{figure*}
  \centering
  \includegraphics[width=\textwidth]{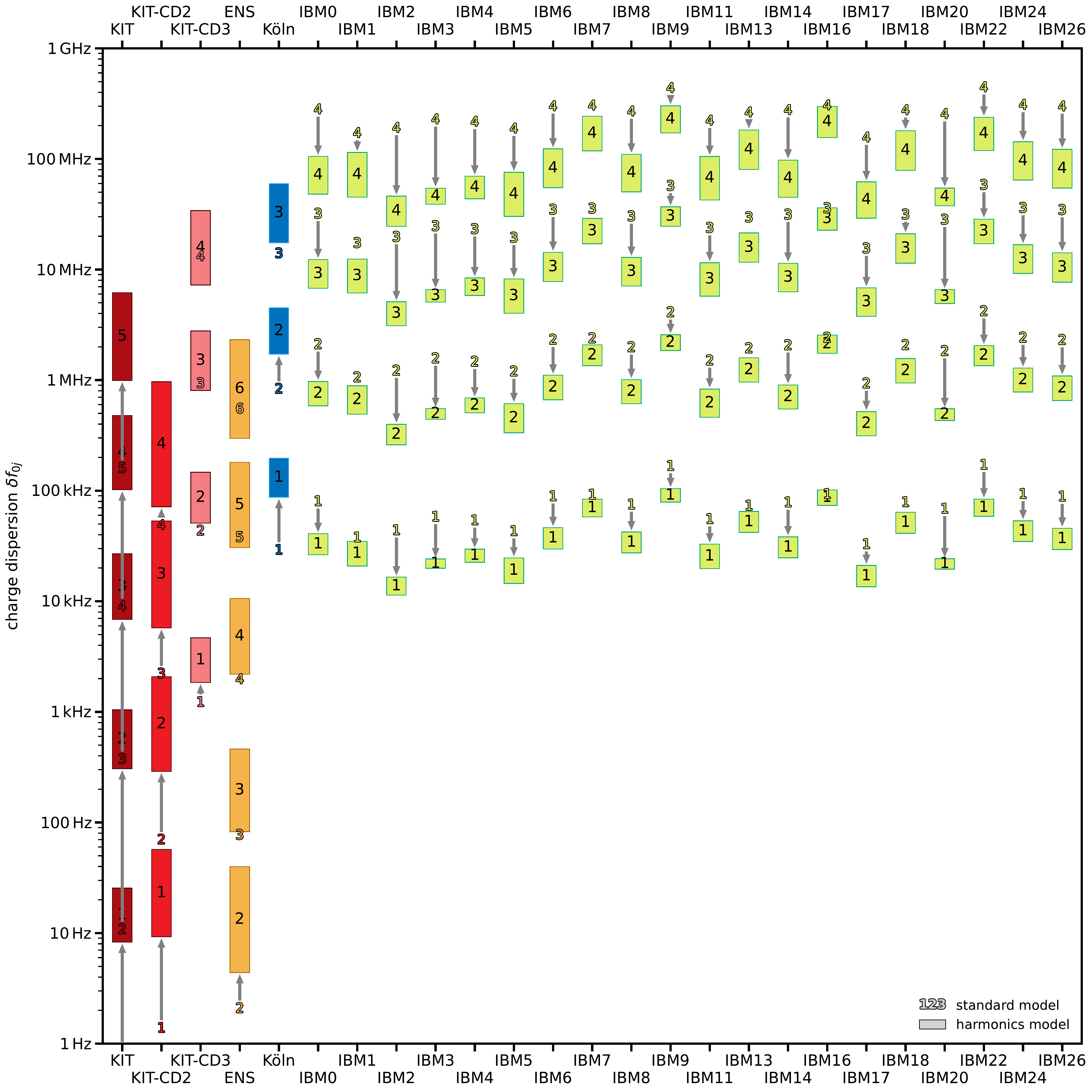}
  \caption{\textbf{Josephson harmonics can significantly increase or reduce the charge dispersion.}
  For each transmon considered in the main text, we use colored numbers as markers to represent the charge dispersions $\delta f_{0j}$ predicted by the standard transmon model and filled bars with the corresponding level indices $j$ to represent the ranges extracted from all trajectories of the harmonics model.
  Gray arrows are shown if there are significant corrections from the standard model prediction.
  The ranges for the K\"oln transmon correspond to the same dataset considered in Fig.~\ref{fig:3} of the main text.}
  \label{suppfig:chargedispersion}
\end{figure*}

For both the standard transmon model and all trajectories of the harmonics model, we evaluate the charge dispersion
\begin{equation}
    \delta f_{0j} = \lvert f_{0j}(n_{\mathrm{g}}=0) - f_{0j}(n_{\mathrm{g}}=0.5)\rvert\,,
    \label{eq:chargedispersionmodel}
\end{equation}
for the transmon transitions $0\to j$. The results are shown in Fig.~\ref{suppfig:chargedispersion} for all transmons considered in the main text.

While the standard model can underestimate the charge dispersion (sometimes dramatically, see Fig.~\ref{fig:4}a of the main text), it is also possible that the charge dispersions are overestimated (see the green bars in Fig.~\ref{suppfig:chargedispersion}). The latter can happen in the presence of high-transparency conduction channels. This agrees with the findings of \cite{Bargerbos2020PointContactChargeDispersion}, where the charge dispersion was observed to decrease with increasing junction transparency, which happens because the height of the potential increases with transparency. However, we emphasize that when constraining both standard and harmonics models to describe the same measured frequencies, whether the charge dispersion actually decreases or increases (compared to the standard model) is a combination of where the transmon levels lie in the potential well and how the potential is affected by the Josephson harmonics (see the discussion after Eq.~\eqref{eq:josephsonharmonicspotentialheight}).

If additional measurements $\delta f_{0j}^{\mathrm{experiment}}$ of the charge dispersion are available, this information can also be used to reduce the uncertainty of suitable Josephson harmonics models. One would then filter out, from all suitable trajectories $\mathbf{e}=(1,E_{\mathrm{J}2}/E_{\mathrm{J}1},\ldots,E_{\mathrm{J}N_f+4}/E_{\mathrm{J}1})$, only those trajectories that agree with the measured charge dispersions.
For the ENS sample, for instance, one can extract from Fig.~6b(6) of \cite{Lescanne2019EscapeUnconfinedStatesFloquet} that the charge dispersion of level $j=6$ is on the order of a few MHz. This is incompatible with the standard model (see the yellow ``6'' in Fig.~\ref{suppfig:chargedispersion}) and would restrict the parameter range in the harmonics model.
The charge dispersion is thus a sensitive probe for deviations from the standard model (cf.~also Fig.~\ref{fig:4} in the main text for the K\"oln experiment).

We note that both the standard model and the harmonics model give an exponential \emph{scaling} of the charge dispersion with $E_\mathrm{J}/E_\mathrm{C}$~\cite{koch2007transmon}. For this reason, it is always possible to use the two parameters $(E_\mathrm{C},E_\mathrm{J})$ of the standard model to fit either (i) $f_{01}$ and $f_{02}$ or (ii) $f_{01}$ and $\delta f_{01}$. The latter case has been successfully demonstrated for qubit 2 in \cite{Schreier2008SuppressingChargeNoise}. For the K\"oln sample, one could analogously increase $E_\mathrm{C}$ (and decrease $E_\mathrm{J}$) such that the gray dashed lines in Fig.~\ref{fig:4}a rise towards the measured data. However, in this case, the predictions for the higher level transitions $f_{02}, f_{03}, \ldots$ would be even further away from the measured frequencies.

\subsubsection{Hamiltonian}
\label{sec:Hamiltonian}
In this section, we state the transmon-resonator Hamiltonian including the higher-order Josephson harmonics and discuss some properties of this Hamiltonian. The Hamiltonian of the transmon-resonator system is given by
\begin{equation}
    H = 4 E_\mathrm{C} (n - n_{\mathrm{g}})^2 - \sum_{m\ge1} E_{\mathrm{J}m} \cos(m\varphi)
    + \Omega a^\dagger a + G n(a + a^\dagger)\,,
    \label{eq:Hamiltonian}
\end{equation}
where, in the charge basis $\{\ket n\}$, the operator $n=\sum_n n \ket{n}\!\bra{n}$ is diagonal and the operator $\cos(m\varphi)=\sum_n 1/2\,(\ket{n}\!\bra{n+m}+\ket{n+m}\!\bra{n})$ has constant entries on the $m^{\mathrm{th}}$ subdiagonal (which represent correlated $m$-Cooper pair tunneling), $a^\dagger a$ is the number operator of the harmonic oscillator with $a=\sum_k\sqrt{k+1}\ket{k}\!\bra{k+1}$ the bosonic annihilation operator given in the harmonic oscillator's eigenbasis. $E_\mathrm{C}$ denotes the transmon's charging energy, $E_{\mathrm{J}m}$ the $m^{\mathrm{th}}$-harmonic Josephson energy, $\Omega$ is the resonator frequency and $G$ denotes the transmon-resonator coupling strength. 

\begin{figure*}
  \centering
  \includegraphics[width=\textwidth]{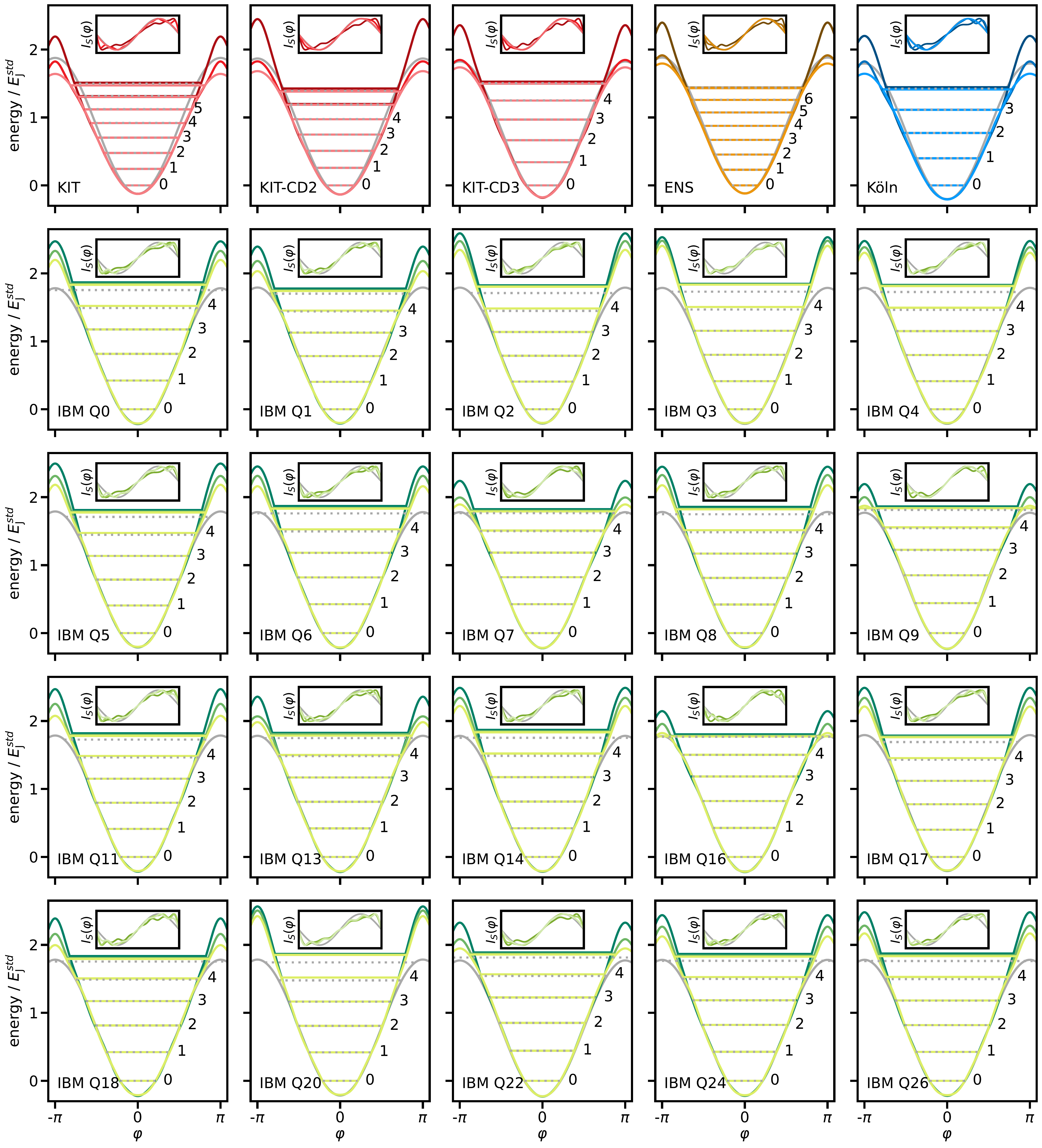}
  \caption{\textbf{Energy spectra and phenomenological current-phase relations for each sample considered in the main text.}
  The standard model is shown in gray.
  Three harmonics models given by the three trajectories ``max'', ``mid'', and ``min'' in Fig.~\ref{suppfig:scan} are shown in color (darker colors represent larger $\vert E_{\mathrm{J}2}/E_{\mathrm{J}}\vert$).
  The energy levels of the dressed transmon states $\ket{\overline{0j}}$ for $j=0,1,2,\ldots$ are computed by diagonalizing the corresponding transmon-resonator Hamiltonian in Eq.~\eqref{eq:Hamiltonian}.
  Each potential $V(\varphi)$ is shifted such that the energy corresponding to $j=0$ is at zero.
  Labels are shown for all measured transition frequencies.
  The insets show the respective current-phase relations $I_S(\varphi)$, in units of the critical current $I_\text{c}$ obtained by maximizing over Eq.~\eqref{eq:IsofphiEJm} (cf.~Table~\ref{tab:modelparameters}).
  We note that the more extreme deviations from the sinusoidal current-phase relations (darker lines) might be easily detectable in DC measurements~\cite{DellaRocca2007MeasurementCurrentPhaseRelation,Thompson2023NonsinusoidalCPR,Thompson2023NonsinusoidalCPRThesis}.
  The results for the K\"oln transmon correspond to the same dataset considered in Fig.~\ref{fig:3} of the main text.}
  \label{suppfig:potentialsandcprs}
\end{figure*}

The energy levels of $H$ corresponding to the dressed transmon states $\ket{\overline{0j}}$ for $j=0,1,2,\ldots$ together with the $\varphi$-dependent Josephson potentials 
\begin{equation}
    V(\varphi) = - \sum_{m\ge1} E_{\mathrm{J}m}\cos(m\varphi)
    \label{eq:josephsonharmonicspotential}
\end{equation}
and the current-phase relations $I_S(\varphi)$ are shown in Fig.~\ref{suppfig:potentialsandcprs} for all samples. Note that the energy levels of all harmonics models are equal to all measured transitions and only differ for even higher levels. Depending on the height of the Josephson potentials, they can be either above or below the standard model.

We can compute the height of the potential (a.k.a.~the depth of the potential well) as
\begin{equation}
    V(\pi) - V(0) 
    = 2(E_{\mathrm{J}1} + E_{\mathrm{J}3} + E_{\mathrm{J}5} + \cdots)\,.
    \label{eq:josephsonharmonicspotentialheight}
\end{equation}
Here we see that, compared to the height of the standard model $2E_{\mathrm{J}}^{\mathrm{std}}$, a large contribution of odd higher harmonics (which are all positive) can increase the height of the potential significantly.

This helps to understand why in Fig.~\ref{fig:3}a, the IBM transmons are the only devices for which most standard model predictions are lower than the measured transitions: Since the IBM transmons have the largest contributions from odd higher harmonics (cf.~Fig.~\ref{suppfig:scan}), the Josephson potentials are much deeper than expected from the standard model. Therefore, the energy levels of $j=4$ (and often also $j=3$) lie much higher than the standard model suggests (see Fig.~\ref{suppfig:potentialsandcprs}.)
Furthermore, the increase of the height of the IBM potentials due to strong odd harmonics in all trajectories explains the decrease of the charge dispersions, while for all other devices the dominating $E_{\mathrm{J}2}$ term (cf.~Fig.~\ref{suppfig:scan}) effectively flattens the potentials such that the charge dispersion increases (cf.~Fig.~\ref{suppfig:chargedispersion}).

\subsubsection{Perturbative expansion}

To obtain the leading-order Josephson harmonics corrections to the qubit frequency $\omega=2\pi f_{01}$ and the anharmonicity $\alpha=2\pi(f_{12} - f_{01})$, we expand the $\cos(m\varphi)$ terms in $\varphi$. 
We emphasize that this step, which is frequently done to obtain an anharmonic ``Duffing'' oscillator approximation of the transmon (see e.g.~\cite{koch2007transmon,gambetta2013controlIFF,DidierRigetti2017AnalyticalParametericTransmon,Blais2021}), is not suitable for an accurate study of higher transmon states (and neither the charge dispersion).
We also note that this approximation is not applicable to all of the expressions for $E_{\mathrm{J}m}$ obtained for certain transparency distributions in Section~\ref{sec:transparencydistributions}, as the double series
\begin{equation}
    \label{eq:HamiltonianArgumentDoubleSeries}
    \sum_{m=1}^\infty \sum_{k=0}^\infty (-1)^k E_{\mathrm{J}m} \frac{(m\varphi)^{2k}}{(2k)!}
\end{equation}
may not converge absolutely (as in the case of e.g.~the point contact Eq.~\eqref{eq:pointcontact}) and rearranging the terms as well as neglecting higher orders may lead to inconsistent results.
The second-order expression of the Josephson contribution $H_\mathrm{J}=-\sum_m E_{\mathrm{J}m} \cos(m\varphi)$ to $H$ is given by
\begin{equation}\label{eq:HJ_approx_quadratic}
   H_\mathrm{J} \approx \left( \sum_{m=1}^\infty m^2E_{\mathrm{J}m} \right)\frac{\varphi^2}{2}\,.
\end{equation}
Equation~\eqref{eq:HJ_approx_quadratic} implies that the second-order correction to the standard model  corresponds to the substitution rule
\begin{equation}
    E_\mathrm{J} \to E_\mathrm{J}\left( 1 + \sum_{m=2}^\infty m^2\frac{E_{\mathrm{J}m}}{E_\mathrm{J}}\right)\,.
\end{equation}
Considering terms up to fourth order in $\varphi$ (see~\cite{gambetta2013controlIFF}), we find the analogous correction to the often-used approximation $\alpha\approx-E_\mathrm{C}$ (using units with $\hbar=1$),
\begin{equation}
    \alpha \approx -E_\mathrm{C}\frac{1+\sum_{m\ge 2} m^4E_{\mathrm{J}m}/E_\mathrm{J}}{1+\sum_{m\ge 2} m^2E_{\mathrm{J}m}/E_\mathrm{J}}\,.
\end{equation}
Similarly, we obtain for the qubit frequency,
\begin{equation}
    \omega \approx \sqrt{8E_\mathrm{C}E_\mathrm{J}\left( 1 + \sum_{m=2}^\infty m^2\frac{E_{\mathrm{J}m}}{E_\mathrm{J}} \right)} +\alpha\,.
\end{equation}
Considering the expansion up to $E_{\mathrm{J}2}$ only yields
\begin{align}
    \omega &\approx \sqrt{8E_\mathrm{C}E_\mathrm{J}\left(1+4\frac{E_{\mathrm{J}2}}{E_\mathrm{J}}\right)} + \alpha\,,\label{eq:approx_omega_EJ2}\\
    \alpha &\approx - E_\mathrm{C}\frac{1+16E_{\mathrm{J}2}/E_{\mathrm{J}}}{1+4E_{\mathrm{J}2}/E_{\mathrm{J}}}\,,\label{eq:approx_alpha_EJ2}
\end{align}
for which we can make the following observation:
Since $E_{\mathrm{J}2} < 0 < E_\mathrm{J}$ (see Eq.~\eqref{eq:EJmProperties}), it follows that $1+16E_{\mathrm{J}2}/E_\mathrm{J} < 1+4E_{\mathrm{J}2}/E_\mathrm{J} < 1$ and thus the fraction in Eq.~\eqref{eq:approx_alpha_EJ2} is smaller than 1.
This implies that, for the same measured values of $\alpha$ and $\omega$, the value of $E_\mathrm{C}$ resulting from Eq.~\eqref{eq:approx_alpha_EJ2} has to be larger than the one obtained from the analogous standard model relation $\alpha\approx-E_\mathrm{C}$.
Similarly, the value of $E_\mathrm{J}$ resulting from Eq.~\eqref{eq:approx_omega_EJ2} has to be smaller than the one obtained from the analogous standard model relation $\omega\approx\sqrt{8E_\mathrm{C}E_\mathrm{J}}-\alpha$.
Hence, it is reasonable to expect that the ratio $E_\mathrm{J}/E_\mathrm{C}$ is too large when using the standard model.
This expectation is confirmed by several of the observed shifts of $E_\mathrm{J}/E_\mathrm{C}$ presented in Fig.~\ref{fig:4}c in the main text.

We caution the reader, however, that this argument relies on crude approximations and does not apply in general. For instance, if the ratios of Josephson energies do not decay fast enough and are relevant to high order, the value of $E_\mathrm{J}/E_\mathrm{C}$ computed from the standard model can actually be too small rather than too large (and the charge dispersion may be overestimated by the standard model). This is supported by the IBM data shown in Fig.~\ref{fig:4}c in the main text.

\subsubsection{Engineering $E_{\mathrm{J}m}$ coefficients}
\label{sec:engineering}

\begin{figure*}
  \centering
  \includegraphics[width=\textwidth]{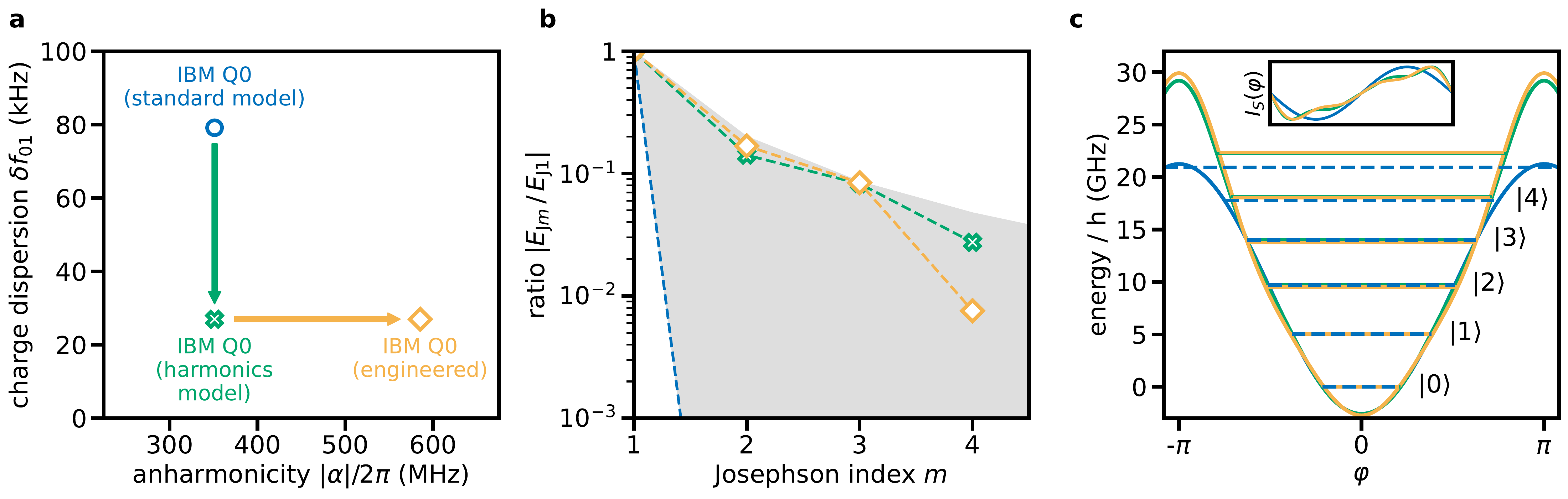}
  \caption{\textbf{Engineering $E_{\mathrm{J}m}$ coefficients can both reduce charge dispersion and increase anharmonicity.}
  \textbf{a}~Map of the charge dispersion $\delta f_{01}$ and the absolute anharmonicity $\vert\alpha\vert$ for the standard model for IBM Q0 (blue circle), the $E_{\mathrm{J}4}$ harmonics model (green cross), and an engineered $E_{\mathrm{J}4}$ model with increased anharmonicity (yellow diamond).
  The model parameters for the latter were chosen such that the qubit frequency $f_{01}$, the resonator frequency $f_{\mathrm{res},0}$, and the charge dispersion $\delta f_{01}$ stay constant (they are listed in the last row of Table~\ref{tab:modelparameters}).
  \textbf{b}~The corresponding Josephson energy ratios $\vert E_{\mathrm{J}m}/E_{\mathrm{J}1}\vert$, using the same coloring.
  \textbf{c}~The corresponding potentials and current-phase relations (cf.~Fig.~\ref{suppfig:potentialsandcprs}), using the same coloring.}
  \label{suppfig:engineering}
\end{figure*}

As discussed in the main text (see Fig.~\ref{fig:4}b),
the IBM transmons give evidence that one can use strong contribution from higher harmonics $E_{\mathrm{J}2}, E_{\mathrm{J}3}, E_{\mathrm{J}4}$ (cf.~Fig.~\ref{suppfig:scan}) to engineer devices with a reduced charge dispersion $\delta f$ while keeping the anharmonicity $\vert\alpha\vert=2\pi\vert f_{12}-f_{01}\vert$ constant. A natural question is now whether one can also increase the anharmonicity (e.g.~to mitigate leakage~\cite{motzoi2009drag,chow2010dragexperiment,gambetta2010dragtheory}) by further engineering the $E_{\mathrm{J}m}$ contributions. We note that in practice it is not straightforward to tune the values of coefficients $(E_{\mathrm{J}2}, E_{\mathrm{J}3}, E_{\mathrm{J}4})$ to arbitrary values. To achieve this, one probably requires a combination of various strategies, such as shaping the channels' transparencies, adding inductive elements in series, flux bias, etc. (see also~\cite{bozkurt2023doublefourier}).

In Fig.~\ref{suppfig:engineering}a, we show that there exist sets of $E_{\mathrm{J}m}$ coefficients which give a reduced charge dispersion and increased anharmonicity (yellow arrow). The corresponding Josephson energy ratios are given in Fig.~\ref{suppfig:engineering}b. We see that the main reason for the increased anharmonicity is a reduction of $E_{\mathrm{J}4}$, while $E_{\mathrm{J}2}$ and $E_{\mathrm{J}3}$ stay roughly the same. This makes sense as $E_{\mathrm{J}3}$ must stay large to maintain the height of the potential for the reduced charge dispersion (see Eq.~\eqref{eq:josephsonharmonicspotentialheight}). The reduced $E_{\mathrm{J}4}$ then makes the potential slightly wider around level $\vert2\rangle$, which draws the level towards the bottom of the potential well and thus increases the anharmonicity (see Fig.~\ref{suppfig:engineering}c). We emphasize that whether it is possible or not to engineer such a set of $E_{\mathrm{J}m}$ coefficients in a real device is currently an open question.

\subsection{Alternative corrections}
\label{sec:alternativecorrections}

After observing that the standard transmon model cannot describe the measured transition frequencies, it might seem natural to consider alternative possible modifications to the model, such as an additional stray inductance present in the circuit, hidden electromagnetic modes, the coupling to other qubits as present on the IBM device, or an asymmetry in the superconducting electrodes. Although the theoretical discussion given above suggests that higher harmonics are the \emph{expected} correction to describe the measurements, from a phenomenological perspective, such alternative modifications should not be ruled out a priori.

For this reason, we here discuss alternative corrections to the model, and why we have come to the conclusion that these alternatives do not provide the same universal quality to solve the disagreement between standard model and experiment for the considered samples. It would be an interesting direction of research to study how the inclusion of Josephson harmonics would influence advanced circuit quantization and measurement techniques \cite{Reed2010HighFidelityReadoutCQED,Nigg2012BlackBoxCircuitQuantization,vooldevoret2017circuitqed,kafri2016tunableinductivecoupling,Ansari2019BlackBoxCircuitQuantization,Naghiloo2019QuantumMeasurementSuperconductingQubits,Riwar2022CircuitQuantizationTimeDependentMagneticField,Blais2021,Miano2023HamiltonianFluxBiasedJJ}.

\subsubsection{Series inductance}
\label{sec:additionalinductance}

\begin{figure*}
  \centering
  \captionsetup[subfigure]{position=top,labelfont=bf,textfont=normalfont,singlelinecheck=off}
  \textbf{a}
  \subfloat{\includegraphics[width=.31\textwidth]{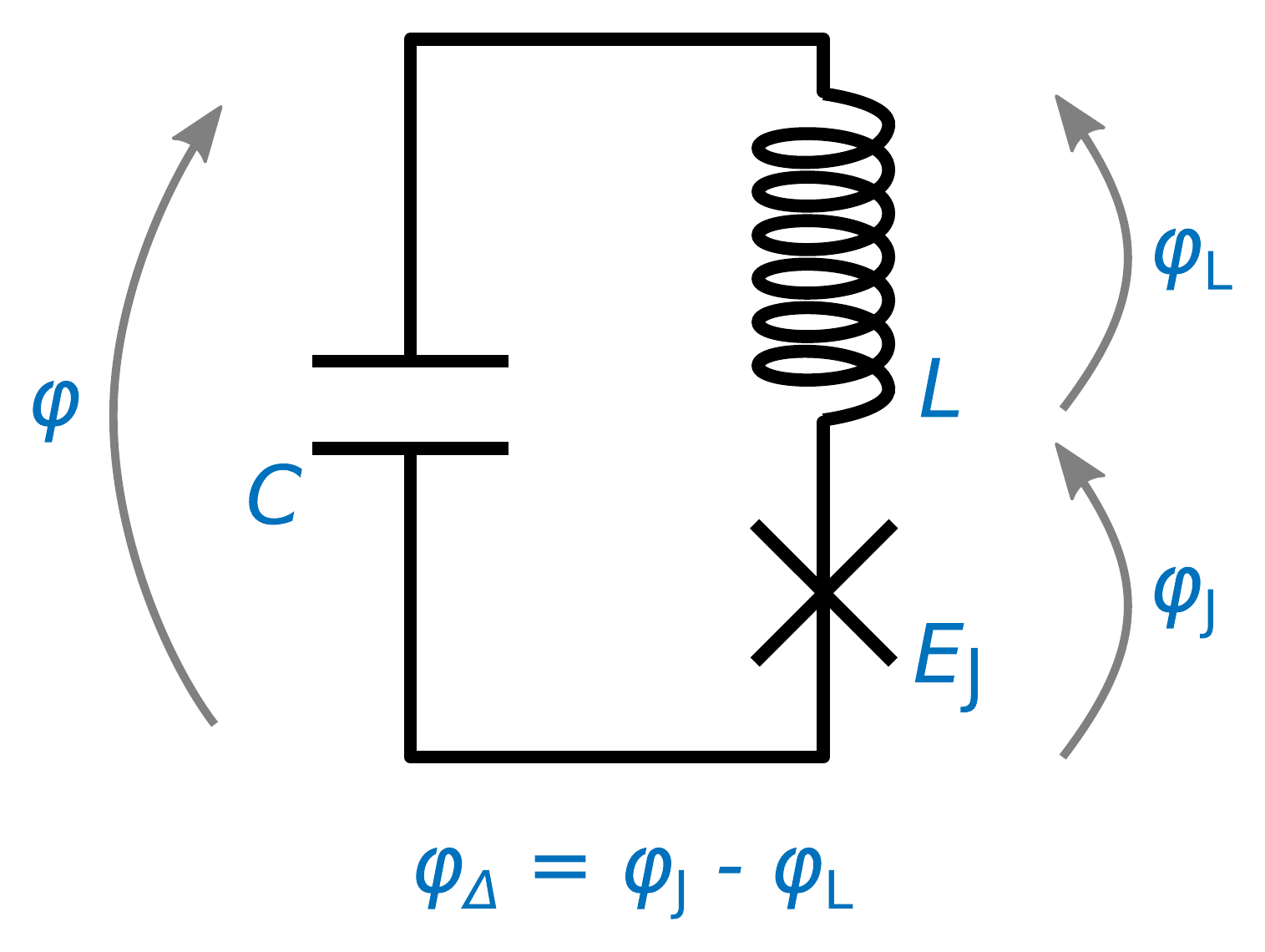}}
  \hfill
  \textbf{b}
  \subfloat{\includegraphics[width=.3\textwidth]{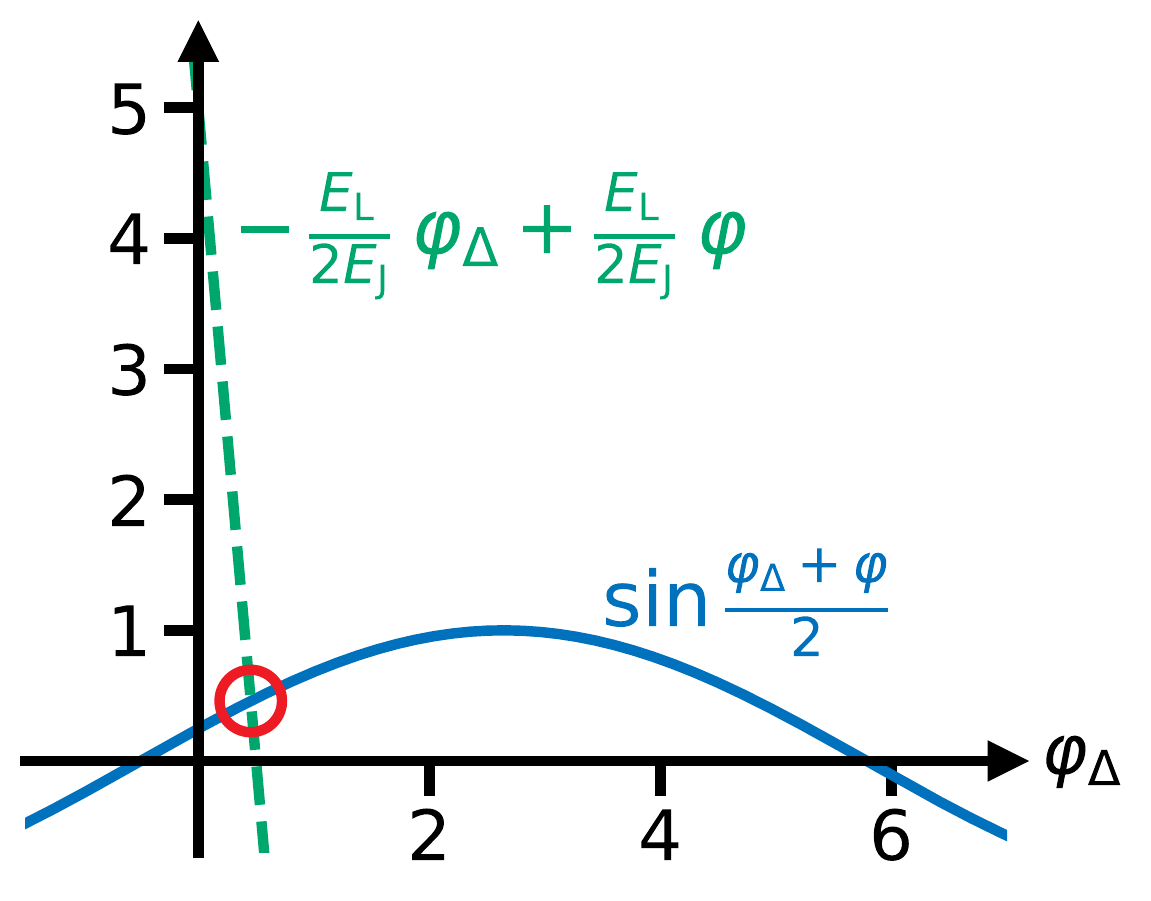}}
  \hfill
  \textbf{c}
  \subfloat{\includegraphics[width=.3\textwidth]{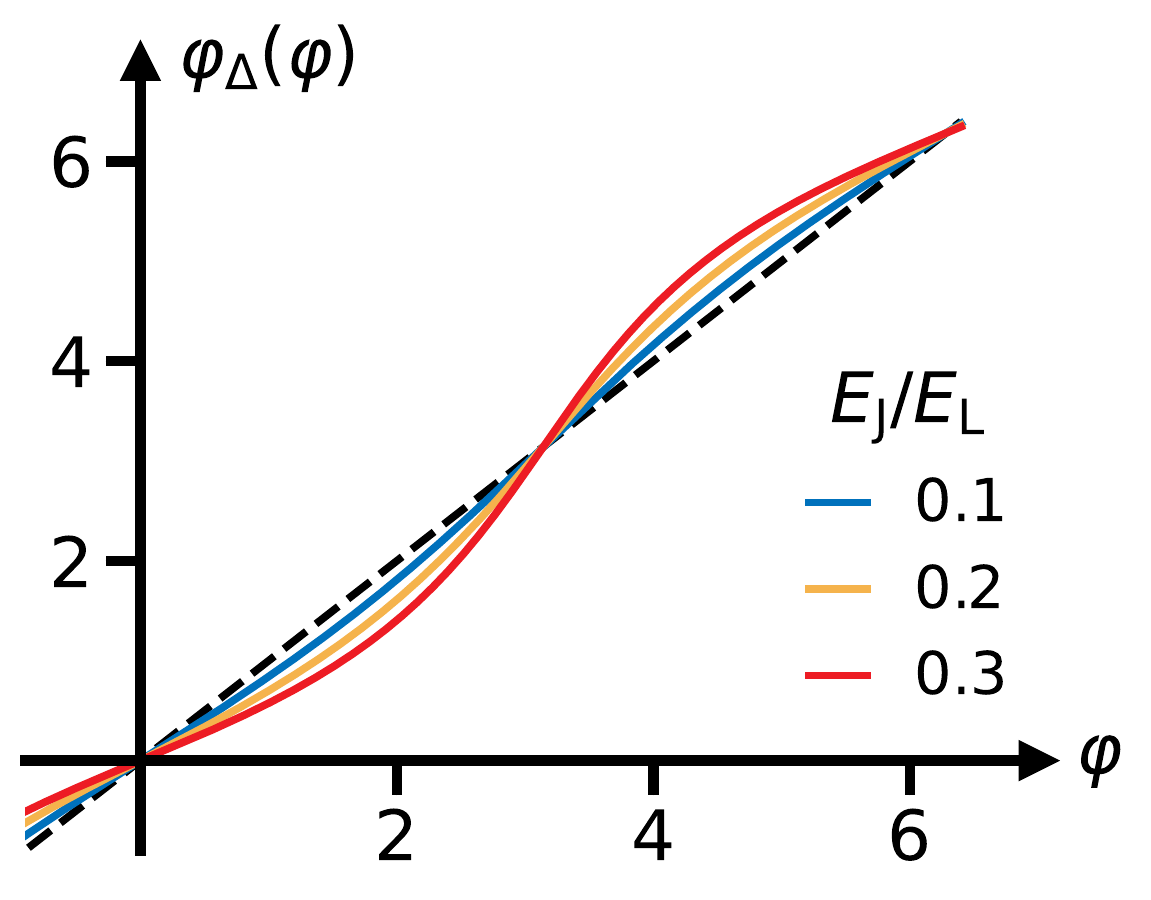}}
  \caption{\textbf{Circuit quantization with an additional linear stray inductance in series.} 
  \textbf{a} Lumped-element circuit diagram of a JJ with an additional linear stray inductance. The transmon is described in terms of the total capacitance $C$, the Josephson energy $E_\text{J}$ characterizing the non-linear Josephson inductance, and an additional finite stray inductance $L$, in series with the JJ.
  \textbf{b} The solution of Eq.~\eqref{eq:phideltaofphi} is determined as the intersection of a sine function (blue) and a straight line (green). Since the largest slope of the sine function is $\max\vert\partial/\partial{\varphi_\Delta}\sin((\varphi_\Delta+\varphi)/2)\vert=1/2$, this intersection is unique if $E_\text{L}/2E_\text{J}>1/2$, i.e., $E_\text{J}/E_\text{L}<1$.
  The example solution (red circle) corresponds to $\varphi_\Delta(\varphi=0.5)\approx0.45$ for $E_\text{J}/E_\text{L}=0.05$.
  \textbf{c} Characteristic form of the solution $\varphi_\Delta(\varphi)$ of Eq.~\eqref{eq:phideltaofphi} for several values of $E_\text{J}/E_\text{L}$ (see legend). $\varphi_\Delta(\varphi)$ is an odd function with a $2\pi$ translation invariance that oscillates closely around $\varphi$ (dashed diagonal).
  }
  \label{suppfig:additionalinductancecircuit}
\end{figure*}

In this section, we show that a small linear stray inductance $L$ in series with the JJ also induces higher harmonic corrections. In a transmon qubit, such a linear stray inductance can arise from the leads that connect the Josephson junction to the shunt capacitance. In an equivalent circuit diagram, this inductance would appear in series with the Josephson junction as shown in Fig.~\ref{suppfig:additionalinductancecircuit}a.

To get a feeling for the size of $L$, we consider the KIT system as an example (cf.~Table~\ref{tab:modelparameters} and Section~\ref{suppsec:samplesKIT}).
From electromagnetic simulation (see Fig.~\ref{suppfig:HFSS}e), we obtain a value for a linear stray inductance of $L\approx\SI{0.380}{\nano\henry}$. In terms of energies, using the relation $E_\text{L}=(\Phi_0/2\pi)^2/L$ with the magnetic flux quantum $\Phi_0$, we have $E_\text{L}/h\approx\SI{430}{\giga\hertz}$,
so $E_\text{J}/E_\text{L}\approx0.058$. 
The characteristic energy ratio $E_\text{J}/E_\text{L}$ is also called the \emph{screening parameter} in SQUID terminology~\cite{clarke2006squidhandbook}.

\begin{figure*}
  \centering
  \begin{minipage}{.49\textwidth}
  \includegraphics[width=\columnwidth]{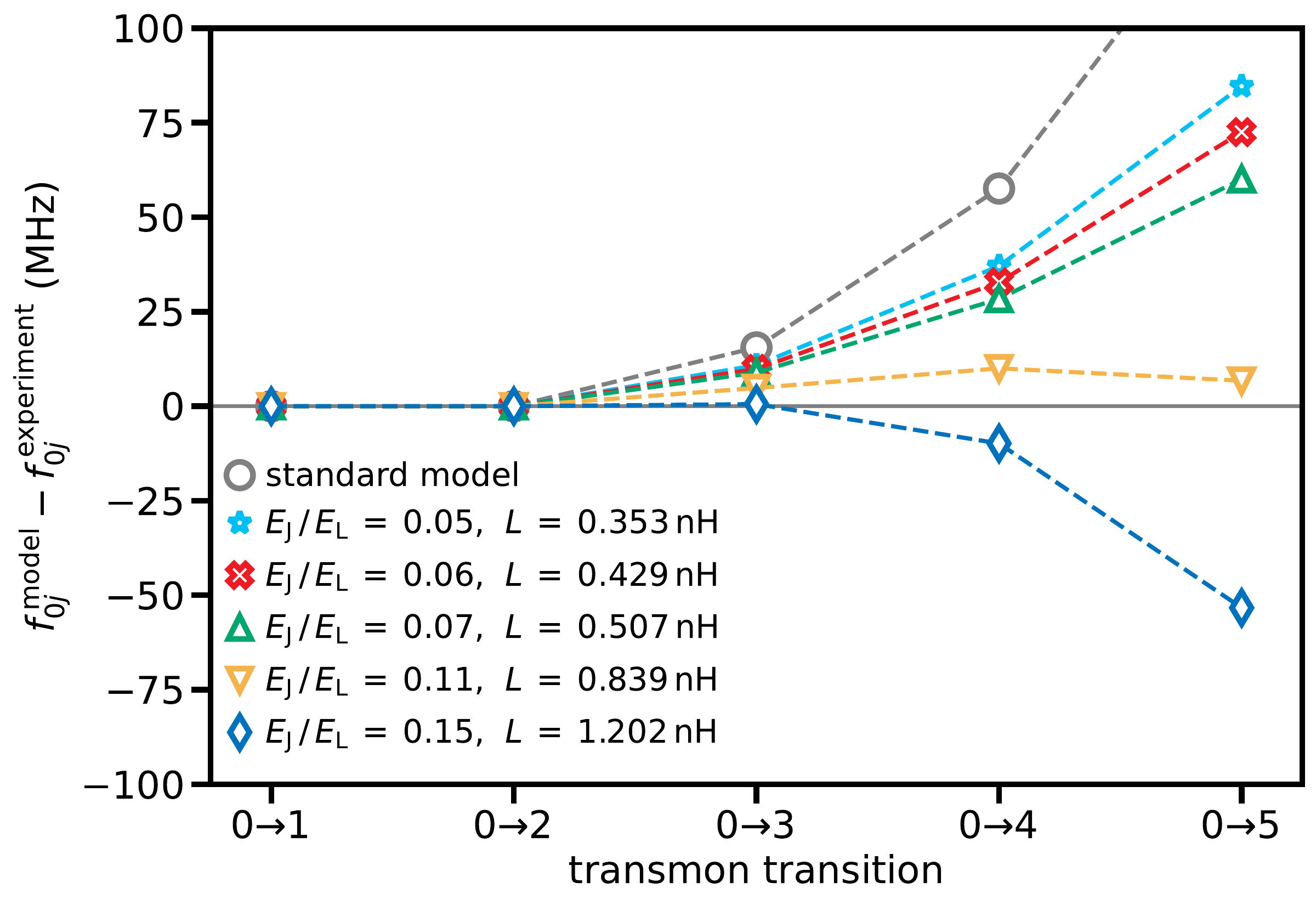}
\end{minipage}
\hfill
\begin{minipage}{.49\textwidth}
\caption{\textbf{Harmonics from additional series inductance alone cannot correct the standard transmon model.}
  Markers show the difference between the frequency $f_{0j}^{\mathrm{model}}$ predicted by the Hamiltonian Eq.~\eqref{eq:Hinductance_cos_mphi} (including the resonator $H_\mathrm{res}$ from Eq.~\eqref{eq:standardmodel} in the main text) and the measured transition frequency $f_{0j}^{\mathrm{experiment}}$ from $\ket0$ into a higher transmon state $\ket j$ for the KIT experiment.
  Different colors represent different values of $E_\mathrm{J}/E_\mathrm{L}$, corresponding to a different series inductance $L$ (see legend).
  In particular, the result for the KIT inductance extracted from electromagnetic simulation, $L=\SI{0.380}{\nano\henry}$ (cf.~Fig.~\ref{suppfig:HFSS}) corresponding to $E_\text{J}/E_\text{L}\approx0.058$, would fall between the blue stars and the red crosses; including the contribution from kinetic inductance would increase $L$ to about 0.5~nH (cf.~Section~\ref{suppsec:samplesKIT}), corresponding approximately to the upward pointing green triangles.
  Gray circles denote the standard transmon model, corresponding to the limit $L=0$.
  For each case, the given $E_\mathrm{J}/E_\mathrm{L}$ only fixes the ratios $E_{\mathrm{J}m}/E_\mathrm{L}$ (cf.~Table~\ref{tab:additionalinductanceEJm}), and the four standard model parameters $(E_\mathrm{C},E_\mathrm{J},\Omega,G)$ are adjusted slightly such that the first two transmon transition frequencies and the first two resonator frequencies match the observed values, in order to make the cases easily comparable.
  Dashed lines are guides to the eye.  
  }
  \label{suppfig:additionalinductance}
\end{minipage}
\end{figure*}

The circuit in Fig.~\ref{suppfig:additionalinductancecircuit}a is described by the Lagrangian
\begin{equation}
    \label{eq:lagrangianind}
    \mathcal{L} = \frac 1 2 C (\dot\Phi_\text{J} + \dot\Phi_\text{L})^2 - \frac{1}{2L}\Phi_\text{L}^2 + E_\text{J}\cos\varphi_\text{J}\,,
\end{equation}
where $\Phi_\text{J}$ ($\Phi_\text{L}$) is the flux across the junction (inductance). Furthermore, we define $\varphi_\text{J}\equiv2\pi\Phi_\text{J}/\Phi_0$, and we implicitly assume this relation between all lower-case ($\varphi$) and upper-case ($\Phi$) flux variables in what follows. We perform a change of variables $\varphi=\varphi_\text{J}+\varphi_\text{L}$ and $\varphi_\Delta=\varphi_\text{J}-\varphi_\text{L}$. Then, as $\dot\varphi_\Delta$ does not occur in the Lagrangian, we have $\mathrm d/\mathrm dt\:(\partial\mathcal L/\partial\dot\varphi_\Delta)=0$ and thus $\partial\mathcal L/\partial\varphi_\Delta=0$. This yields an equation that we can use to eliminate $\varphi_\Delta$, which is justified if $E_\text{L}/E_\text{C}\gg1$ and the stray capacitance in parallel to the JJ is much smaller than $C$ (roughly~\SI{1}{\femto\farad} vs.~\SI{80}{\femto\farad} for the KIT sample) such that the Born-Oppenheimer approximation is applicable (see~\cite{kafri2016tunableinductivecoupling,Rymarz2022SingularQuantization} for more information).
The equation reads
\begin{equation}
    \label{eq:phideltaofphi}
    \sin\frac{\varphi_\Delta+\varphi}{2} = -\frac{E_\text{L}}{2E_\text{J}} \varphi_\Delta + \frac{E_\text{L}}{2E_\text{J}}\varphi\,,
\end{equation}
and is related to Kepler's transcendental equation~\cite{Hall1883KeplersProblem,Minev2021EnergyParticipationQuantization,Rymarz2022SingularQuantization}.
For $0<E_\text{J}<E_\text{L}$, Eq.~\eqref{eq:phideltaofphi} has a unique solution for $\varphi_\Delta$, defined by the intersection of a sine function and a straight line (see Fig.~\ref{suppfig:additionalinductancecircuit}b). Although we cannot solve for $\varphi_\Delta$ analytically, Eq.~\eqref{eq:phideltaofphi} uniquely determines $\varphi_\Delta$ as a function of $\varphi$, and we write $\varphi_\Delta(\varphi)$ to denote this solution. It is shown in Fig.~\ref{suppfig:additionalinductancecircuit}c for several values of $E_\text{J}/E_\text{L}$.

The function $\varphi_\Delta(\varphi)$ has two symmetry properties that can be extracted from Eq.~\eqref{eq:phideltaofphi}. The first is that it is odd, which can be shown by replacing $\varphi\mapsto-\varphi$ in Eq.~\eqref{eq:phideltaofphi}. Using the symmetry of the sine function, we obtain the same equation with $\varphi_\Delta\mapsto-\varphi_\Delta$, and since Eq.~\eqref{eq:phideltaofphi} is the uniquely defining equation for $\varphi_\Delta(\varphi)$, we have $\varphi_\Delta(-\varphi)=-\varphi_\Delta(\varphi)$.
The second symmetry property of $\varphi_\Delta(\varphi)$ is a $2\pi$ translation invariance, which can be shown in the same manner, yielding $\varphi_\Delta(\varphi+2\pi)=\varphi_\Delta(\varphi)+2\pi$.

\begin{table}
  \begin{minipage}{0.49\textwidth}
  \begin{center}
    \begin{ruledtabular}
      \begin{tabular}{@{}lllll@{}}
      $E_\text{J}/E_\text{L}$ & $E_{\text{J}1}/E_\text{J}$ & $E_{\text{J}2}/E_\text{J}$ & $E_{\text{J}3}/E_\text{J}$ & $E_{\text{J}4}/E_\text{J}$ \\
        \colrule
        0.01 & $0.99999$ & $-0.00250$ & $0.00001$ & $-0.00000$\\
        0.05 & $0.99969$ & $-0.01249$ & $0.00031$ & $-0.00001$\\
        0.07 & $0.99939$ & $-0.01747$ & $0.00061$ & $-0.00003$\\
        0.1 & $0.99875$ & $-0.02492$ & $0.00124$ & $-0.00008$\\
        0.2 & $0.99501$ & $-0.04934$ & $0.00489$ & $-0.00065$\\
        0.5 & $0.96907$ & $-0.11490$ & $0.02710$ & $-0.00850$\\
         &  {\color{gray}$0.96908$} &  {\color{gray}$-0.11491$} &  {\color{gray}$0.02686$} &  {\color{gray}$-0.00833$}\\
      \end{tabular}
      \end{ruledtabular}
  \end{center}
  \end{minipage}
  \hfill
  \begin{minipage}{0.49\textwidth}
  \caption{\label{tab:additionalinductanceEJm}
    Values of $E_{\text{J}m}/E_\text{J}$ for a linear stray inductance in series with the JJ (cf.~Fig.~\ref{suppfig:additionalinductancecircuit}a), for different values of $E_\text{J}/E_\text{L}$ (i.e.~the screening parameter~\cite{clarke2006squidhandbook}). Small values of $E_\text{J}/E_\text{L}$ correspond to a small series inductance $L$.
  The ratios have been computed from Eq.~\eqref{eq:Hinductance_EJm} using $K=10000$ (which gives accurate results up to machine precision). 
  Gray entries in a separate row correspond to the approximations in Eq.~\eqref{eq:Hinductance_EJm_approximation} and are only given if the approximation is not equal to the numerically exact result within five decimal digits.}
  \end{minipage}
\end{table}

Given $\varphi_\Delta(\varphi)$, we perform the Legendre transformation of Eq.~\eqref{eq:lagrangianind} to obtain the Hamiltonian
\begin{equation}
    \label{eq:Hinductance}
    H=4E_\text{C}n^2 - E_\text{J}\cos\frac{\varphi+\varphi_\Delta(\varphi)}{2} + \frac{E_\text{L}}2 \left(\frac{\varphi-\varphi_\Delta(\varphi)}{2}\right)^2\,.
\end{equation}
Note that the series inductance does \emph{not} break the $2\pi$ periodicity of the Hamiltonian, despite the occurrence of $\varphi^2$. 
This is in contrast to a shunt inductance, for which the flux $\varphi$ and the charge $n$ would need to be treated as non-compact operators with continuous spectrum $\mathbb R$, see~\cite{Koch2009chargespectrumfluxqubit,Smith2016QuantizationInductivelyShuntedCircuits,Le2020TransmonFluxoniumPeriodicPhaseZakBasis,Hassani2022InductivelyShuntedTransmonExperiment}. 
For a series inductance, though, due to the two symmetry properties of $\varphi_\Delta(\varphi)$, the two $\varphi$-dependent terms of Eq.~\eqref{eq:Hinductance} are even and $2\pi$-periodic. Therefore, we can write the two $\varphi$-dependent terms of Eq.~\eqref{eq:Hinductance} as Fourier cosine series,
\begin{subequations}
\begin{align}
    \cos\frac{\varphi+\varphi_\Delta(\varphi)}{2} &= \sum_m c_m\cos(m\varphi)\,,\\
    \left(\frac{\varphi-\varphi_\Delta(\varphi)}{2}\right)^2 &= \sum_m s_m\cos(m\varphi)\,.
\end{align}
\end{subequations}
The coefficients $c_m$ and $s_m$ can be obtained numerically by (i) solving Eq.~\eqref{eq:phideltaofphi} for $\varphi_\Delta(\varphi_k)$, where $\varphi_k=\pi(k+1/2)/K$ with $k=0,\ldots,K-1$ and $K\gg1$ controls the accuracy, and (ii) using the discrete cosine transform (DCT). To see this, we note that any even $2\pi$-periodic function $g(\varphi)$ can be written as a Fourier cosine series $g(\varphi)=\sum_m g_m\cos(m\varphi)$, where
\begin{equation}
    g_m = \frac 2 \pi \int_0^\pi g(\varphi)\cos(m\varphi)\,\mathrm{d}\varphi
    \approx \frac 2 \pi \sum_{k=0}^{K-1} g(\varphi_k) \cos(m\varphi_k) \frac\pi K
    = \frac 1 K \mathrm{DCT}_{m} (g)
\end{equation}
for $m\ge1$ and $\mathrm{DCT}_{m} (g) = 2\sum_k g(\varphi_k) \cos(m\varphi_k)$ denotes the type-II DCT, and $g_0=1/\pi\int_0^\pi g(\varphi)\,\mathrm{d}\varphi\approx \sum_{k=0}^{K-1}g(\varphi_k)$. Neglecting the constants $c_0$ and $s_0$, we obtain the Hamiltonian
\begin{equation}
    \label{eq:Hinductance_cos_mphi}
    H=4E_\text{C}n^2 - \sum_{m\ge1} E_{\text{J}m}\cos(m\varphi)\,,
\end{equation}
with the higher harmonic contributions given by
\begin{equation}
    \label{eq:Hinductance_EJm}
    E_{\text{J}m} = E_\text{J}\left(c_m - \frac{E_\text{L}}{2E_\text{J}} s_m\right)\,. 
\end{equation}
The ratios $E_{\text{J}m}/E_\text{J}$ are shown in Table~\ref{tab:additionalinductanceEJm} for several values of $E_\text{J}/E_\text{L}$. Note that here, we typically have $E_{\text{J}1}\neq E_\text{J}$, since $E_\text{J}$ characterizes the JJ while $E_{\text{J}1}$ includes the series inductance as a separate circuit element. We remark that the ratios alternate in sign and decay in magnitude with increasing order $m$, similar to the Josephson harmonics arising from the conduction channel transparencies (cf.~Section~\ref{sec:transparencydistributions}).

It is possible to obtain closed-form approximations for the leading-order ratios $E_{\text{J}m}/E_\mathrm{J}$ by expanding the small quantity in Eq.~\eqref{eq:phideltaofphi}, $(\varphi-\varphi_\Delta(\varphi))/2$, in powers of $E_\mathrm{J}/E_\mathrm{L}$. To this end, we rewrite Eq.~\eqref{eq:phideltaofphi} as
\begin{equation}
    \frac{\varphi-\varphi_\Delta(\varphi)}2 = \frac{E_\mathrm{J}}{E_\mathrm{L}}\left(\sin\varphi\cos\frac{\varphi-\varphi_\Delta(\varphi)}2 - \cos\varphi\sin\frac{\varphi-\varphi_\Delta(\varphi)}2\right)\,.
\end{equation}
After substituting $(\varphi-\varphi_\Delta(\varphi))/2=\sum_{k\ge1} a_k (E_\mathrm{J}/E_\mathrm{L})^k$ and expanding the sine and cosine functions, we iteratively obtain $a_1,\ldots,a_5$ by comparing coefficients. Performing the same expansion in the Hamiltonian Eq.~\eqref{eq:Hinductance} and inserting the expression for $a_1,\ldots,a_5$ yields the fifth-order approximations
\begin{subequations}
\label{eq:Hinductance_EJm_approximation}
\begin{align}
    \frac{E_{\mathrm{J}1}}{E_\mathrm{J}} &\simeq
    1 - \frac{1}{8}\bigg(\frac{E_\mathrm{J}}{E_\mathrm{L}}\bigg)^2 + \frac{1}{192}\bigg(\frac{E_\mathrm{J}}{E_\mathrm{L}}\bigg)^4\,,\\
    \frac{E_{\mathrm{J}2}}{E_\mathrm{J}} &\simeq
    - \frac{1}{4}\bigg(\frac{E_\mathrm{J}}{E_\mathrm{L}}\bigg) + \frac{1}{12}\bigg(\frac{E_\mathrm{J}}{E_\mathrm{L}}\bigg)^3 - \frac{1}{96}\bigg(\frac{E_\mathrm{J}}{E_\mathrm{L}}\bigg)^5\,,\\
    \frac{E_{\mathrm{J}3}}{E_\mathrm{J}} &\simeq \frac{1}{8}\bigg(\frac{E_\mathrm{J}}{E_\mathrm{L}}\bigg)^2 - \frac{9}{128}\bigg(\frac{E_\mathrm{J}}{E_\mathrm{L}}\bigg)^4\,,\\
    \frac{E_{\mathrm{J}4}}{E_\mathrm{J}} &\simeq -\frac{1}{12}\bigg(\frac{E_\mathrm{J}}{E_\mathrm{L}}\bigg)^3 + \frac{1}{15}\bigg(\frac{E_\mathrm{J}}{E_\mathrm{L}}\bigg)^5\,.
\end{align}
\end{subequations}
As Table~\ref{tab:additionalinductanceEJm} shows, these expressions are a good approximation to the numerically exact result for the leading-order harmonics if the series inductance and thus $E_\text{J}/E_\text{L}$ is not too large. 

Finally, to see whether the higher harmonics from the linear stray inductance can explain the deviations between computed and measured spectra (see Fig.~2a in the main text), we perform the same test when using $H$ in Eq.~\eqref{eq:Hinductance_EJm} as a model. The result is shown in Fig.~\ref{suppfig:additionalinductance} for the KIT system. However, the correction corresponding to $L=\SI{0.380}{\nano\henry}$ ($L=\SI{0.5}{\nano\henry}$) as extracted from electromagnetic simulation, between the blue stars and the red crosses (including kinetic inductance effects, green triangles) shows a similar deviation from the experiment as the standard transmon model. Also, when the inductance $L$ is increased so much that the next transmon transition $0\to3$ matches the experiment (blue diamonds), the deviation for higher transmon transitions systematically bends into the other direction. Thus we conclude that, although the linear stray inductance does result in an appreciable correction (and it is probably part of the ranges shown in Fig.~\ref{fig:3}c in the main text and Fig.~\ref{suppfig:scan}), it is not the main solution to revise the standard transmon model. We note that it would be interesting to work out the consequences of a series inductance on circuits with multiple JJs.

\subsubsection{Additional hidden modes}

The Hamiltonian in Eq.~\eqref{eq:Hamiltonian} includes only a single-mode resonator contribution, namely the readout resonator with bare frequency $\Omega$ and coupling strength $G$. However, the electromagnetic environment of the device may contain additional modes. A valid hypothesis for a correction to the Hamiltonian is therefore the existence of spurious, hidden electromagnetic modes (so-called ``dark'' modes) that couple weakly to the transmon. 

For this reason, we consider the addition of a second hidden mode to the Hamiltonian,
\begin{equation}
    \label{eq:Hdarkmodes}
    H' = H + \Omega_{\mathrm{dark}} b^\dagger b + G_{\mathrm{dark}} n(b + b^\dagger)\,,
\end{equation}
where $b^\dagger$ ($b$) are the hidden mode's bosonic creation (annihilation) operators, $\Omega_{\mathrm{dark}}$ is the frequency, $G_{\mathrm{dark}}$ is the coupling strength to the transmon, and $H$ is given by Eq.~\eqref{eq:standardmodel}. We consider a coupling $G_{\mathrm{dark}}/h=\SI{0.009}{\giga\hertz}$ that is ten times weaker than the coupling $G$ between the transmon and the resonator in the standard model (see~Table~\ref{tab:modelparameters}).

\begin{figure*}
\begin{minipage}[t]{0.48\textwidth}
  \centering
  \includegraphics[width=\columnwidth]{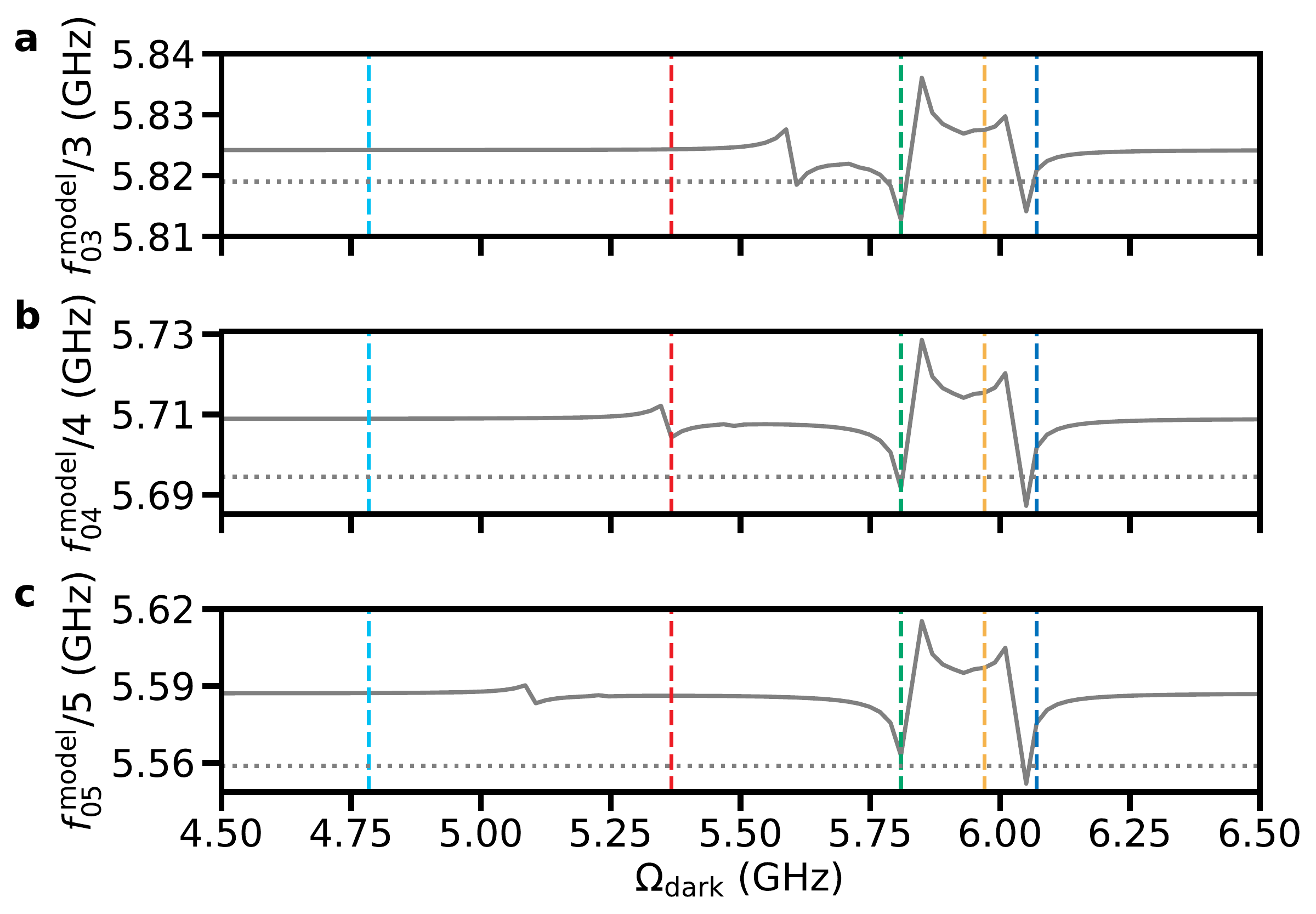}
  \caption{\textbf{Additional hidden modes coupling weakly to the qubit can affect the spectrum.}
  Transition frequencies $f^\mathrm{model}_{0j}/j$ as a function of the hidden mode frequency $\Omega_\mathrm{dark}$ for \textbf{a} $j=3$, \textbf{b} $j=4$ and \textbf{c} $j=5$ (gray lines).
  The kinks in the gray lines indicate singularities (which are not fully resolved by the spacing of the scanned $\Omega_\mathrm{dark}$ and thus connected by a straight line) when the dark mode's frequency $\Omega_\mathrm{dark}$ is close to particular transition frequencies in the spectrum such as $f_{01}=\SI{6.039}{\giga\hertz}$.
  Dashed vertical lines correspond to the specific dark modes considered in Fig.~\ref{suppfig:additionalmodes} using the same colors. 
  In each panel, a dotted horizontal line indicates the measured frequencies $f_{0j}^{\mathrm{experiment}}/j$.
  In particular, there is no single $\Omega_\mathrm{dark}$ for which each model frequency (solid line) matches the measured frequency (dotted line).
  }
  \label{suppfig:additionalmodesfrequencies}
\end{minipage}
\hfill
\begin{minipage}[t]{0.48\textwidth}
  \centering
  \includegraphics[width=\columnwidth]{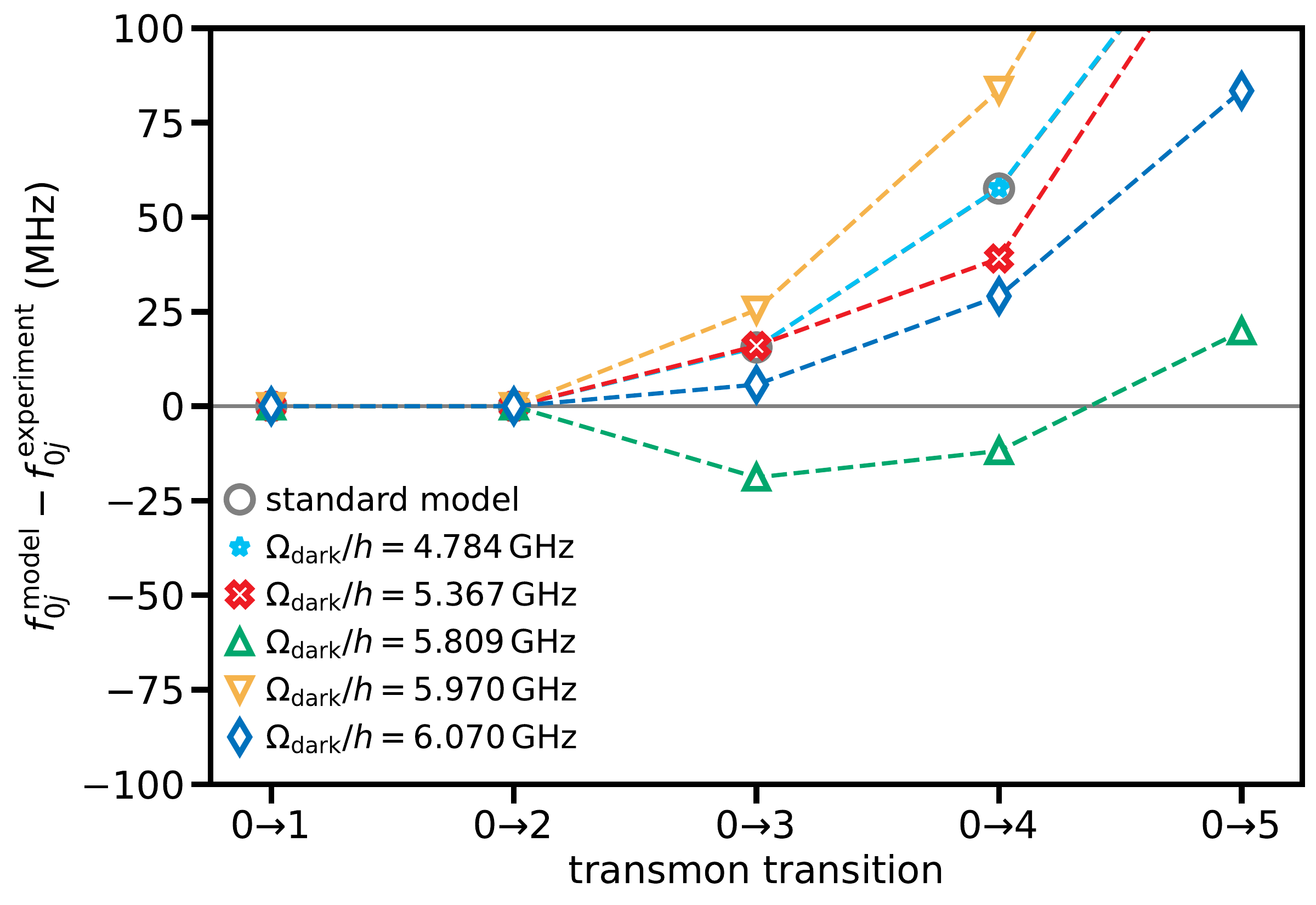}
  \caption{\textbf{Additional hidden  modes cannot easily rescue the standard transmon model.}
  Markers show the difference between the frequency $f_{0j}^{\mathrm{model}}$ predicted by the Hamiltonian Eq.~\eqref{eq:Hdarkmodes} and the measured transition frequency $f_{0j}^{\mathrm{experiment}}$ from $\ket0$ into a higher transmon state $\ket j$ for the KIT experiment.
  Gray circles denote the standard model from the main text without any dark modes.
  Different colors represent different values of $\Omega_\mathrm{dark}$ (see~Fig.~\ref{suppfig:additionalmodesfrequencies} and legend).
  For each case, the four standard model parameters $(E_\mathrm{C},E_\mathrm{J},\Omega,G)$ are adjusted slightly such that the first two transmon transition frequencies and the first two resonator frequencies match the observed values, in order to make the cases easily comparable.
  Dashed lines are guides to the eye.}
  \label{suppfig:additionalmodes}
  \end{minipage}
\end{figure*}

We study the KIT system as an example. The effect of additional modes is shown in Fig.~\ref{suppfig:additionalmodesfrequencies} as a function of $\Omega_{\mathrm{dark}}$. Around certain frequencies of $\Omega_\mathrm{dark}$, the transmon transition frequencies can deviate a lot from the predicted frequency when no hidden modes are considered. 
Thus we see that additional modes can indeed have a strong effect on the spectrum, despite the weak coupling strength. 
However, Fig.~\ref{suppfig:additionalmodesfrequencies} also shows that there is no single frequency $\Omega_\mathrm{dark}$ that can bring all model frequencies (solid lines) to the experimental values (dotted lines).

In Fig.~\ref{suppfig:additionalmodes}, we show the difference between model and experiment in the same way as in Fig.~\ref{fig:3} of the main text, for the particular dark mode frequencies indicated in Fig.~\ref{suppfig:additionalmodesfrequencies}.
These results show that additional hidden modes do not correct for the systematic deviation of the higher states, i.e., the curvature in the differences between the standard model prediction and the experimental values is still there if hidden modes are included.

Finally, we remark that no evidence for spurious hidden modes has been found in a 2D coplanar waveguide system experiment~\cite{Sheldon2017characterizationhiddenmodes}. Also for the KIT sample, we have not seen any evidence for such dark modes from $T_2$ measurements. Therefore, we do not consider the addition of spurious modes as the solution to the mismatch between the standard model and the experimental data.

\subsubsection{Multi-qubit coupling}
\label{sec:additionalqubits}

\begin{figure*}
  \centering
  \includegraphics[width=\textwidth]{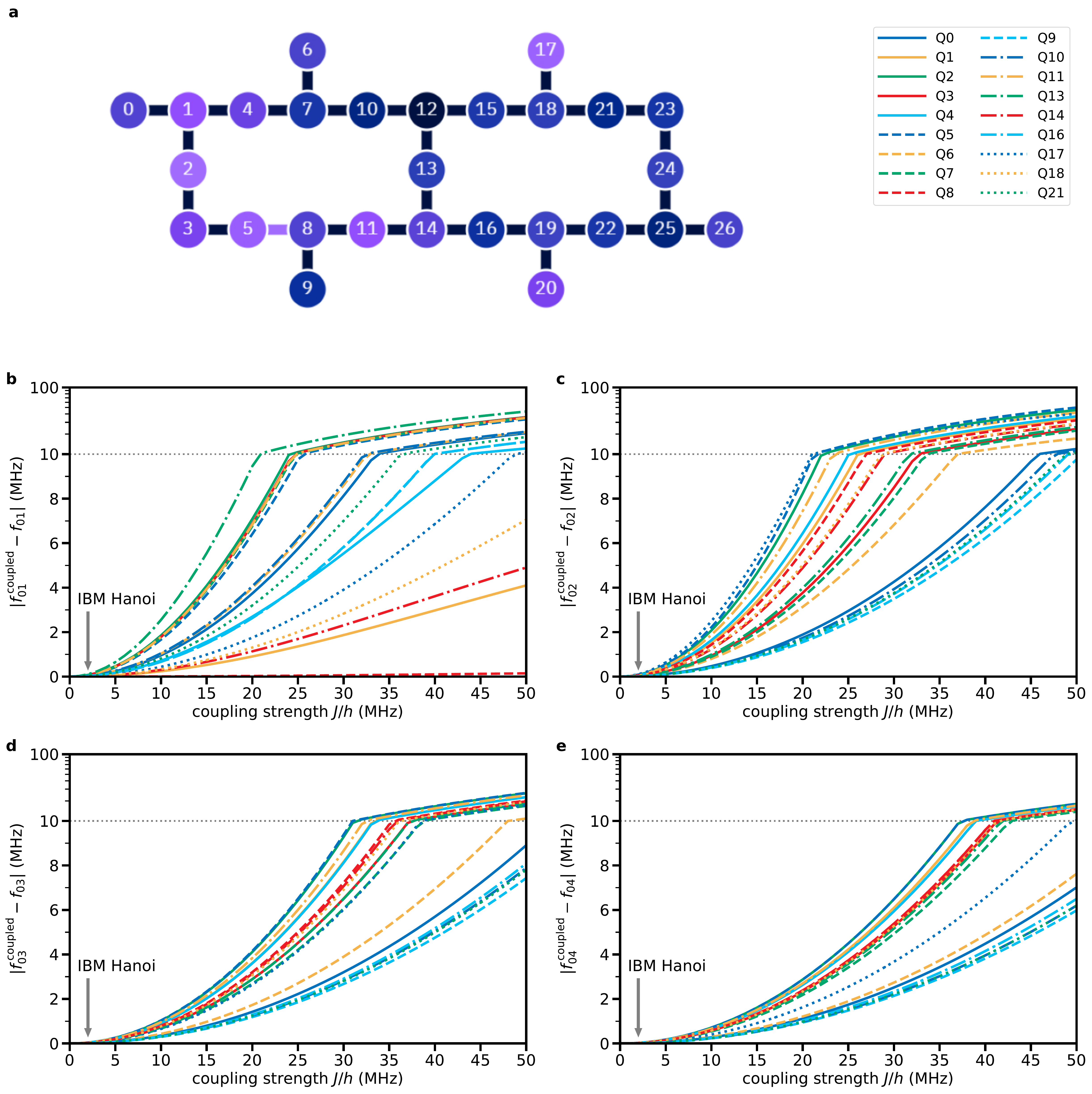}
  \caption{\textbf{Effects of the capacitive transmon-transmon coupling on the spectrum are negligible for the IBM multi-qubit device. a} The connectivity between the transmons on the IBM Hanoi device. 
  The color of each qubit represents the qubit frequency from $f_{01}=\SI{4.7190}{\giga\hertz}$ (dark blue, Q12) to $f_{01}=\SI{5.2562}{\giga\hertz}$ (light purple, Q2).
  \textbf{b--e} Absolute difference of the frequencies $f^\mathrm{coupled}_{0j}$ obtained from the diagonalization of the capacitively coupled multi-transmon system (including up to 3 transmons and their respective readout resonators) and the frequencies $f_{0j}$ obtained from the diagonalization of the single-transmon-resonator system for the \textbf{b} $0\to 1$ transition \textbf{c} $0\to 2$ transition \textbf{d} $0\to 3$ transition and \textbf{e} $0\to 4$ transition. The gray arrow indicates the design value of $2\,\mathrm{MHz}$ for the IBM Hanoi device. Although the transmon-transmon coupling can have a strong effect on the spectrum for large values of the coupling strength $J$ (note the change from linear to log-scale indicated by the dotted gray line), for values around $2\,\mathrm{MHz}$, the effects on the spectrum are negligible with less than $1\,\mathrm{MHz}$ deviation.
  }
  \label{suppfig:additionalqubits}
\end{figure*}

Since the IBM Hanoi device is a multi-qubit device, an obvious question is whether the coupling between the qubits on the chip has an effect on the transmon transition frequencies, and whether this effect may provide an alternative path to rescue the standard transmon model. Here we show that for the IBM Hanoi device, this effect is much too small to significantly shift the transmon frequencies.

The transmons on the IBM Hanoi device are coupled by very short coplanar waveguide resonators (cf.~\cite{Goeppl2008CoplanarWaveguideResonator}). Due to the small size of the resonators, their frequencies are much larger than the transmon frequencies, so the coupling can be treated as capacitive transmon-transmon coupling (cf.~\cite{Berke2022TransmonCapacitiveChaoticFluctuations}). The coupled Hamiltonian takes the form
\begin{equation}
    \label{eq:capacitivecoupling}
    H = \sum_i \left(4 E_{\mathrm{C}_i} (n_i-n_{\mathrm{g}i})^2 - E_{\mathrm{J}_i}\cos\varphi_i
    + \Omega_i a_i^\dagger a_i + G_i n_i(a_i + a_i^\dagger)\right) +
    \sum_{\langle i,j\rangle} J n_i n_j\,,
\end{equation}
where the first term represents the standard model Hamiltonians of each transmon $i$ including its readout resonator, and $J$ is the capacitive coupling strength between all connected transmon pairs $\langle i,j\rangle$.
The particular connectivity of the IBM Hanoi device is shown in Fig.~\ref{suppfig:additionalqubits}a.

To examine whether the coupling can explain the deviations in the spectra shown in Fig.~\ref{fig:3}a of the main text, we diagonalize $H$ for the joint system of up to three neighboring transmons and their respective readout resonators. In Fig.~\ref{suppfig:additionalqubits}b--e, we show the absolute difference $\vert f^\mathrm{coupled}_{0j}-f_{0j}\vert$ between the coupled and the uncoupled systems as a function of the transmon-transmon coupling strength $J/h$. Shown are only the qubits that could be measured up to level $\ket 4$ and that are also coupled to a qubit that could be measured. We scan the coupling strength $J/h$ from $0$ to $50\,\mathrm{MHz}$. Most experiments have $J/h\lesssim\SI{5}{\mega\hertz}$ although experiments beyond $J/h\approx\SI{50}{\mega\hertz}$ are possible (see~\cite{Berke2022TransmonCapacitiveChaoticFluctuations}). At this end of the scale, the coupling can have a significant impact on the spectrum, in the sense that deviations go up to $\SI{100}{\mega\hertz}$ (see Fig.~\ref{suppfig:additionalqubits}b). However, the transmon qubits of the IBM Hanoi device are around $J/h\approx\SI{2}{\mega\hertz}$ by design (see the arrows in Fig.~\ref{suppfig:additionalqubits}b--e).
For values of $J/h$ around $\SI{2}{\mega\hertz}$, the effect of the coupling on the spectrum is very small and cannot correct for the observed deviations from the experimental data.
For this reason, we conclude that also the coupling to the other qubits cannot explain the failure of the standard transmon model shown in Fig.~\ref{fig:3}a of the main text.

\subsubsection{Asymmetry in the superconducting gaps}
\label{sec:gapasymmetry}

In tunnel junctions with aluminum leads, the dependence of the superconducting gap on film thickness can cause a slight difference in the gaps of the two electrodes, $\Delta_1\neq\Delta_2$, because in general the top layer has to be thicker than the bottom layer for fabrication reasons. This difference has recently been shown to play a role in the context of quasiparticle tunneling~\cite{Diamond2022DistinguishingParityMechanisms, Marchegiani2022QuasiparticlesAsymmetricJunctions}.
In this section, we estimate the influence of gap asymmetry on the ratio between second and first Josephson harmonics. We show that the effect is negligible at typical gap asymmetries of around $10\,\%$, and even the presence of larger gap asymmetries would only slightly suppress the effect of higher harmonics.

We start from Zaitsev's treatment of asymmetric S$_1$cS$_2$ junctions~\cite{Zaitsev1984AsymmetricSuperconductingMicrocontacts}, in which the single-channel current-phase relation is expressed as
\begin{equation}
    I_S(\varphi) \propto 
    \sum_{\omega>0} \left\langle \frac{T(
    \alpha)f_1f_2\sin\varphi}{2+T(\alpha) (g_1g_2 + f_1f_2\cos\varphi-1)}\right\rangle_\alpha\,,
\end{equation}
where the angular brackets average over an angle-dependent transparency $T(\alpha)$, the sum is over Matsubara frequencies $\omega>0$, and $f_i = \Delta_i/\sqrt{\omega^2+\Delta_i^2}$ and $g_i=\omega/\sqrt{\omega^2+\Delta_i^2}$ 
depend on the two gaps.
We consider the zero-temperature limit in which the sum over $\omega$ becomes an integral and assume the transparency to be 
independent of $\alpha$; then for equal gaps $\Delta_i \equiv \Delta$ we recover Eq.~\eqref{eq:ChannelCurrent}. 

For generic values of the gaps, keeping the first two terms in the Fourier expansion of $I_S(\varphi)$ we find
\begin{equation}\label{eq:ISFourier}
      I_S(\varphi) \propto \sum_{m=1}^\infty \tilde{c}_m(T) \sin(m\varphi) \propto \int\limits_0^\infty d\omega\,\left[\frac{a-\sqrt{a^2-b^2}}{b}\sin\varphi +\frac{b^2-2a^2+2a\sqrt{a^2-b^2}}{b^2}\sin (2\varphi) +\ldots\right]
\end{equation}
with $a=1+\frac{T}{2}\left(g_1 g_2-1\right)$ and $b=\frac{T}{2}f_1 f_2$. Explicit expressions for the Fourier coefficients can be found in the low transparency limit $T\to0$:
\begin{equation}
    \tilde{c}_1 \simeq \frac{T}{4}\int\limits_0^\infty d\omega \, f_1 f_2=\frac{T}{2}\frac{\Delta_1\Delta_2}{\Delta_1+\Delta_2} K \left(\frac{|\Delta_1-\Delta_2|}{\Delta_1+\Delta_2}\right)\,,\qquad \tilde{c}_2 \simeq -\frac{T^2}{16}\int\limits_0^\infty d\omega \, f_1^2 f_2^2 = -\frac{T^2}{16}\frac{\pi}{2}\frac{\Delta_1\Delta_2}{\Delta_1+\Delta_2}\,.
\end{equation}
where $K(k)=\int_0^{\pi/2}d\theta/\sqrt{1-k^2\sin^2\theta}$ is the complete elliptic integral of the first kind and we used Eq.~(19.8.12) from \cite{DLMF}. Note that $\tilde{c}_1$ correctly displays the known dependence of the tunnel junction critical current on the two gaps, see~\cite[Eq.~(3.2.5)]{BaronePaterno1982}. Since $K(0)=\pi/2$, for small gap asymmetry, $\Delta_1\approx\Delta_2$, we have $E_{\mathrm{J}2}/E_{\mathrm{J}1} = \tilde{c}_2/2\tilde{c}_1 \approx -T/16$, cf.~Eqs.~\eqref{eq:c1Texp} and \eqref{eq:c2Texp}. In the opposite limit in which one of the two gaps goes to zero, the ratio would vanish; this shows that the gap asymmetry suppresses the impact of the higher harmonics. However, this suppression can in practice be neglected: for the realistic value $\Delta_1/\Delta_2=0.9$ we find $E_{\mathrm{J}2}/E_{\mathrm{J}1} \approx -T/16(1-0.0007)$, a value extremely close to that for the symmetric case. Even for $\Delta_1/\Delta_2=0.5$ the deviation from the symmetric value is small,  $E_{\mathrm{J}2}/E_{\mathrm{J}1} \approx -T/16(1-0.029)$. Furthermore, numerical integration of the formulas for the Fourier coefficients in Eq.~\eqref{eq:ISFourier} shows that the suppression is largest at small transparency. Hence we conclude that the effect of gap asymmetry is negligible.

\clearpage

\section{Numerical methods}

In this section, we detail the numerical methods used to obtain the parameters of the Hamiltonian Eq.~\eqref{eq:Hamiltonian} to describe the experimental data. This problem belongs to the class of \emph{inverse eigenvalue problems} (IEPs) that have been covered in a large body of scientific literature \cite{Friedland1977InverseEigenvalueProblems,Friedland1987InverseEigenvalueProblemNumericalMethods,Chu1998InverseEigenvalueProblems,ChuGolub2005InverseEigenvalueProblemsBook}. We formalize the particular IEP under consideration, i.e.~the Hamiltonian parameterized IEP (HamPIEP), and outline the numerical methods that we have used to find suitable solutions. 
Finally, we discuss several details specific to solving the HamPIEP for transmon systems.

\subsection{Inverse eigenvalue problem}

The IEP is an instance of the famous \emph{inverse problem} \cite{ChuGolub2005InverseEigenvalueProblemsBook}, which is one of the most important problems that a scientist may face: Given some data observed in an experiment, find a model that describes the data. The model is usually characterized by a set of parameters that are themselves not observable.

In quantum physics, the model is typically a Hamiltonian, and the experimental data are frequencies measured in spectroscopy experiments. An early application of the IEP in that form was the study and the description of atomic or molecular spectra, which goes back to the year 1955 \cite{DowningHouseholder1955InverseEigenvalueProblemMolecules,Toman1966MolecularSpectroscopyInverseEigenvalueProblem,Brussaard1977ShellModelApplicationsNuclearSpectroscopyIEP}.

\subsubsection{LiPIEP}

One of the simplest types of IEPs is the linear parameterized IEP (LiPIEP). Here the task is to find a set of parameters $\mathbf x=(x_1,\ldots,x_n)$ such that an $n\times n$ matrix of the form
\begin{equation}
    A(\mathbf x) = A_0 + x_1 A_1 + \cdots + x_n A_n\,,
\end{equation}
where $A_0,A_1,\cdots,A_n$ are fixed $n\times n$ matrices, has eigenvalues $\mathbf f(\mathbf x)=(\lambda_1(\mathbf x),\ldots,\lambda_n(\mathbf x))$ equal to a given set of numbers $\mathbf f^*=(\lambda_1^*,\ldots,\lambda_n^*)$. In theory, there exists a solution for almost all $A_i$ that is unique (up to the $n!$ permutations of the eigenvalues)~\cite{ChuGolub2005InverseEigenvalueProblemsBook}. In practice, this solution can be found by applying Newton's root-finding method to the function
\begin{equation}
    \mathbf f(\mathbf x) - \mathbf f^* = \begin{pmatrix}
        \lambda_1(\mathbf x) - \lambda_1^*\\
        \vdots\\
        \lambda_n(\mathbf x) - \lambda_n^*\\
    \end{pmatrix} = 0\,,
\end{equation}
and the fact that the derivative of an eigenvalue $\lambda_i(\mathbf x)$ with respect to a parameter $x_k$ is given by
\begin{equation}
\label{eq:ieplinearderivative}
    J_{ik} = \left(\frac{\partial\mathbf f}{\partial \mathbf x}\right)_{ik} = \frac{\partial\lambda_i}{\partial x_k} = q_i^T A_k q_i\,,
\end{equation}
where $q_i$ is the eigenvector corresponding to $\lambda_i$. This means that one can iteratively find a solution by (i) diagonalizing $A(\mathbf x)$ for a given set of parameters $\mathbf x$ and (ii) computing the Jacobian $J$ in Eq.~\eqref{eq:ieplinearderivative} to obtain an update $\Delta\mathbf x$ for the next iteration (see below).

\subsubsection{HamPIEP}
\label{sec:HamPIEP}

The goal of the IEP solved in this work is to find a parameterized Hamiltonian that describes certain measured transition frequencies. This problem, which we call the HamPIEP, extends the simple LiPIEP in several ways:
\begin{enumerate}
    \item It requires the diagonalization of multiple Hamiltonians $H(\mathbf x;\boldsymbol\theta)$ for a fixed set of constants $\boldsymbol\theta$ (e.g.~$\boldsymbol\theta=n_\mathrm{g}$ for $n_\mathrm{g}=0$ and $n_\mathrm{g}=1/2$).
    \item The Hamiltonians $H(\mathbf x;\boldsymbol\theta)$ may depend non-linearly on the parameters $\mathbf x$ (e.g.~the dependence on $(a,b)$ in Eq.~\eqref{eq:EJmOverEJ} for $\mathbf x=(E_\mathrm{C},E_\mathrm{J1},\Omega,G,a,b)$)
    \item The eigenvalues of $H(\mathbf x;\boldsymbol\theta)$ require a specific labeling procedure (e.g.~the assignment of photon labels $k$ and transmon labels $j$ to the eigenvalues $E_{\overline{kj}}(n_\mathrm{g})$, see \hyperref[sec:methods]{Methods})
    \item We only want specific eigenvalue combinations $\mathbf f$ to match the measured
    data $\mathbf f^*$ (e.g.~$f_{\mathrm{res},j}^{\mathrm{model}}=\sum_{n_\mathrm{g}}(E_{\overline{1j}}(n_\mathrm{g})-E_{\overline{0j}}(n_\mathrm{g}))/4\pi$
    is the average of differences between eigenvalues of two Hamiltonians, see \hyperref[sec:methods]{Methods}).
    \item The number of parameters $\#\mathbf x$ and the number of eigenvalue combinations $\#\mathbf f$ may be much smaller than the size of the Hamiltonians (i.e.~$\#\mathbf x,\#\mathbf f\ll\mathrm{dim}(H)$).
    \item The number of parameters is smaller than or equal to the number of eigenvalue combinations (i.e.~$\#\mathbf x\le\#\mathbf f$).
\end{enumerate}
To solve the HamPIEP, we need the Jacobian $J=\partial \mathbf{f}/\partial\mathbf{x}$. However, points 1--4 usually make it impossible to find a closed-form expression for $J$ such as Eq.~\eqref{eq:ieplinearderivative} for the LiPIEP. Therefore, we obtain $J$ by using automatic differentiation with TensorFlow~\cite{TensorFlow} (only for the $(\bar d,\sigma)$ model, which involves integrals over the distribution $\rho(T)$ in Eq.~\eqref{eq:rhorec}, we pre-compute a dense grid of Eq.~\eqref{eq:EJratios} using Mathematica~\cite{mathematica13}). This circumvents the need to approximate the gradients numerically using finite differences.

We first consider the case $\#\mathbf x=\#\mathbf f$, in which one can use a root-finding algorithm to find the solution to the HamPIEP. Here we use the globally convergent Newton root-finding method with line search and backtracking~\cite{numericalrecipes} (Newton-LB). If $\#\mathbf x=\#\mathbf f$, the Jacobian $J=\partial \mathbf{f}/\partial\mathbf{x}$ is a square matrix. This matrix is usually invertible, so we can use LU-decomposition of $J$ to compute the Newton step,
\begin{equation}
  \Delta\mathbf{x} = -J^{-1}\mathbf (\mathbf f(\mathbf x) - \mathbf f^*)\,,
\end{equation}
to iteratively update $\mathbf x\leftarrow\mathbf x+\Delta\mathbf x$. Note that the Newton step automatically points in a direction that decreases the squared sum of differences, 
\begin{equation}
    F(\mathbf x) = \frac 1 2 (\mathbf f(\mathbf x) - \mathbf f^*)^2 \,,
\end{equation}
because $\nabla F\cdot\Delta\mathbf{x}=-(\mathbf f(\mathbf x) - \mathbf f^*)^2<0$. The strategy to solve the HamPIEP is thus to follow the Newton step $\mathbf x\leftarrow\mathbf x+\Delta\mathbf x$ as long as $F$ decreases, to obtain quadratic convergence near the minimum. If the Newton step does not significantly reduce $F$, we backtrack by performing a line search along $\Delta\mathbf x$ to find a step $\mu\Delta\mathbf x$ with $\mu\in(0,1)$ that reduces $F$. For this, we evaluate the function $g(\mu) = F(\mathbf x+\mu\Delta\mathbf x)$ that we successively model to cubic order in $\mu$
(see~\cite[Section~9.7.1]{numericalrecipes} for more information). This procedure ensures a globally convergent method~\cite{Dennis1996NumericalMethodsUnconstrainedOptimization}.

In the case where $\#\mathbf x<\#\mathbf f$ and an exact solution to the HamPIEP may not exist, we use the Broyden-Fletcher-Goldfarb-Shanno (BFGS) optimization algorithm~\cite{NumericalOptimization} to minimize the weighted sum of absolute differences
\begin{equation}
    \label{eq:weightedsumgeneral}
    W(\mathbf x) = \lvert\mathbf w\cdot(\mathbf f(\mathbf x)-\mathbf f^*)\rvert\,,
\end{equation}
where $\vert\cdot\vert$ denotes the L1-norm and $\mathbf w$ represents the weights  (see the following section for how the weights are chosen).
In practice, we use the implementation of the BFGS algorithm from TensorFlow Probability~\cite{TensorFlow282} running on the NVIDIA A100 GPUs of JUWELS Booster~\cite{JuwelsClusterBooster}. If necessary, a suitable initial value for the BFGS algorithm is obtained from Newton-LB applied to the case in which only some elements of $\#\mathbf f$ are used to ensure $\#\mathbf x=\#\mathbf f$. 

\subsection{Choosing appropriate weights}
\label{sec:weights}

\begin{figure*}
  \centering
  \includegraphics[width=\textwidth]{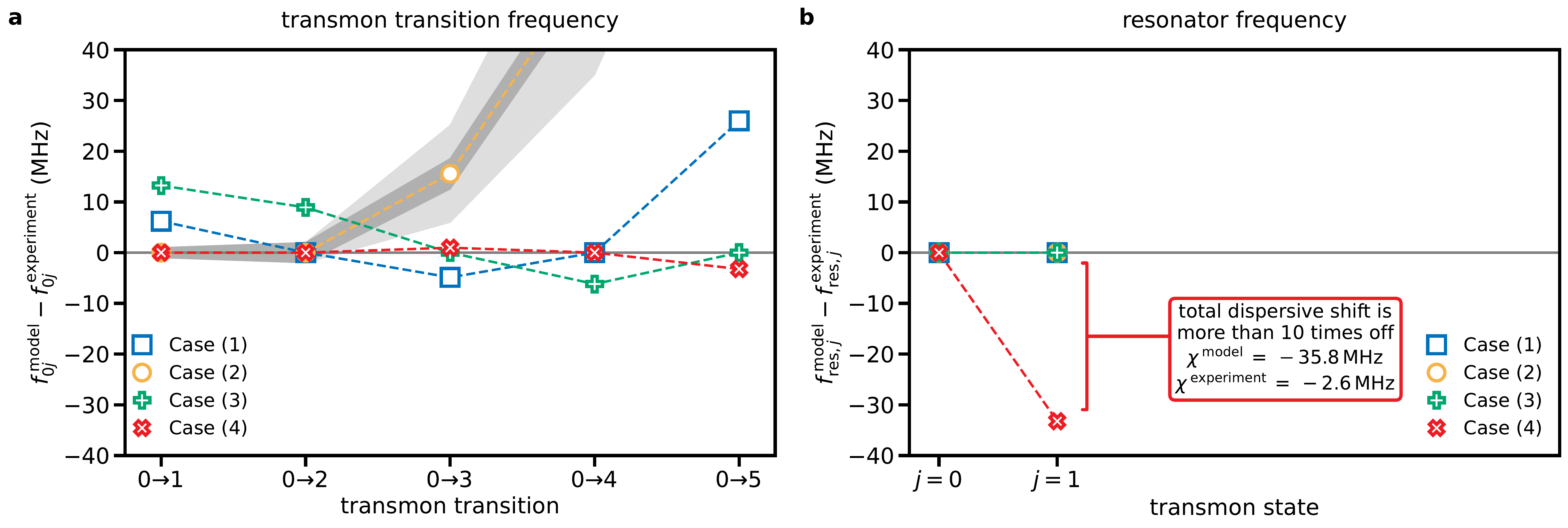}
  \caption{\textbf{Choosing appropriate weights to fit the measured transition frequencies.}
  \textbf{a} Results of fitting the parameters of the standard transmon model using different weights $w_{0j}$ and $w_{\mathrm{res},j}$ (see Eq.~\eqref{eq:fitcostfunction}). Markers show the difference between the model predictions $f_{0j}^\mathrm{model}$ and the measured frequencies $f_{0j}^\mathrm{experiment}$ for the cases (1)--(4). The corresponding weights are given in Table~\ref{tab:weights} together with the model parameters $(E_\mathrm{C},E_\mathrm{J},\Omega,G)$. The dark gray error band indicates the deviation for solving the HamPIEP unambiguously for the leading two transmon transition frequencies if one adds $\pm \SI{1}{\mega\hertz}$ to both $f_{01}^\mathrm{experiment}$ and $f_{02}^\mathrm{experiment}/2$. The light gray error band indicates this deviation if one adds $\pm \SI{1}{\mega\hertz}$ to $f_{01}^\mathrm{experiment}$ and $\mp \SI{1}{\mega\hertz}$ to $f_{02}^\mathrm{experiment}/2$.
  \textbf{b} Difference between the predicted resonator frequencies $f_{\mathrm{res},j}^\mathrm{model}$ and the measured values $f_{\mathrm{res},j}^\mathrm{experiment}$, conditional on the transmon being in state $j$. The fitted models are the same as in panel a and indicated by the same colors. All models match the two resonator frequencies exactly, except in case (4) where the predicted total dispersive shift $\chi^\mathrm{model}=f_{\mathrm{res},1}^\mathrm{model}-f_{\mathrm{res},0}^\mathrm{model}$ differs from $\chi^\mathrm{experiment}$ by more than a factor of 10.
  Dashed lines are guides to the eye.
  }
  \label{suppfig:weights}
\end{figure*}

\begin{table*}[b]
  \caption{
  Choices of weights and resulting standard model parameters for the KIT sample for four different cases.
  All model parameters are given in GHz and the weights $w_{0j}$ and $w_{\mathrm{res},j}$ are unitless.}
  \begin{center}
    \begin{ruledtabular}
      \begin{tabular}{@{}lcccclllllll@{}}
      Case &
      $E_{\mathrm{C}}/h$ & $E_{\mathrm{J}}/h$ & $\Omega/h$ & $G/h$ & $w_{01}$ & $w_{02}$ & $w_{03}$ & $w_{04}$ & $w_{05}$ & $w_{\mathrm{res},0}$ & $w_{\mathrm{res},1}$\\
        \colrule
    (1) & 0.207 & 23.720 & 7.454 & 0.078 & $1$ & $1/2$ & $1/3$ & $1/4$ & $1/5$ & $1$ & $1$ \\
    (2) & 0.197 & 24.852 & 7.454 & 0.078 & $1$ & $1$ & $1/30$ & $1/40$ & $1/50$ & $1$ & $1$ \\
    (3) & 0.211 & 23.319 & 7.454 & 0.077 & $1/50$ & $1/50$ & $1/50$ & $1/50$ & $1/50$ & $1$ & $1$ \\
    (4) & 0.227 & 22.496 & 7.364 & 0.289 & $1$ & $1$ & $1/50$ & $1$ & $1/50$ & $1$ & $0$ \\
      \end{tabular}
      \end{ruledtabular}
    \label{tab:weights}
  \end{center}
\end{table*}

For the additional physical models for Josephson harmonics considered in Fig.~\ref{suppfig:additionalresults} and Table~\ref{tab:modelparameters}, the number of model parameters $\#\mathbf x$ is often smaller than the number of measured transition frequencies $\#\mathbf f$. To obtain the model parameters $\mathbf x$ by solving the HamPIEP, we then minimize the weighted sum of absolute differences Eq.~\eqref{eq:weightedsumgeneral}. Note that this does not apply to the results presented in the main text, where e.g.~the HamPIEP for the $E_{\mathrm{J}2}$ model was solved unambiguously (see~\hyperref[sec:methods]{Methods}). However, for the additional models, the explicit form of the objective function $W(\mathbf x)$ in Eq.~\eqref{eq:weightedsumgeneral} is given by
\begin{equation}
    \label{eq:fitcostfunction}
    W(\mathbf x) 
    = \sum_{j=1}^{N_\mathrm{tr}} 
    w_{0j} \lvert f_{0j}^{\mathrm{model}} - f_{0j}^{\mathrm{experiment}} \rvert
    +
    \sum_{j=0}^{N_\mathrm{res}} 
    w_{\mathrm{res},j} \lvert f_{\mathrm{res},j}^{\mathrm{model}} - f_{\mathrm{res},j}^{\mathrm{experiment}} \rvert\,,
\end{equation}
where $w_{0j}$ are weights for the transmon frequencies for the transition $0\to j$, and $w_{\mathrm{res},j}$ are weights for the resonator frequencies when the transmon is in state $j$. Note that fitting is a complicated procedure that always requires one to make some potentially subjective choices, i.e.~choosing initial values and weights. In all cases where a certain model cannot describe the data, this will influence the final result. It is the purpose of this section to argue that the choices that we have made in this regard are justified.

When fitting the models to the measured spectra, we always put larger weights on the first two transmon frequencies and the first two resonator frequencies, to ensure that 
\begin{subequations}
\label{eq:fitcostfunctiongoal}
\begin{align}
    f_{01}^{\mathrm{model}} &\stackrel{!}{\approx} f_{01}^{\mathrm{experiment}}& &\text{(dressed qubit frequency)}\,,\\
    f_{02}^{\mathrm{model}} &\stackrel{!}{\approx} f_{02}^{\mathrm{experiment}}& &\text{(from dressed anharmonicity $\alpha$)}\,,\\
    f_{\mathrm{res},0}^{\mathrm{model}} &\stackrel{!}{\approx} f_{\mathrm{res},0}^{\mathrm{experiment}}& &\text{(dressed resonator frequency)}\,,\\
    f_{\mathrm{res},1}^{\mathrm{model}} &\stackrel{!}{\approx} f_{\mathrm{res},1}^{\mathrm{experiment}}& &\text{(from dispersive shift $\chi$)}\,,
\end{align}
\end{subequations}
where the symbol $\stackrel{!}{\approx}$ expresses that the fits shall try to match these values with priority.
The argument for this is that the transition frequencies $f_{0j}/j$ can be measured to roughly the same accuracy (say within \SI{1}{\mega\hertz}), so lower-index transitions like the qubit frequency $f_{01}$ can be measured to better accuracy than the higher transition frequencies $f_{03}, f_{04}, \ldots$. Furthermore, the offset charge dispersion of higher energy levels $j$ is much larger (cf.~Fig.~\ref{fig:4} in the main text), which additionally increases the measurement imprecision.

However, one may argue that the choice given by Eq.~\eqref{eq:fitcostfunctiongoal} results in a ``whiplash effect'', in the sense that $\pm\SI{1}{\mega\hertz}$ variations in $f_{01}$ and $f_{02}/2$ might cause drastically different model predictions for the higher levels.
In Fig.~\ref{suppfig:weights}, we show that this is not the case (shaded gray areas). 
Hence, a model that is based on the standard transmon Hamiltonian Eq.~\eqref{eq:standardmodel} and that reproduces---within the experimental imprecision---the first two transition frequencies and the resonator frequencies can also not match the higher transmon transition frequencies.
 
Additionally, we test four different cases for choices of weights (markers) when fitting to the whole spectrum (see Table~\ref{tab:weights}):
Case (1) with weights $w_{0j}=1/j$ is the most obvious choice from a theoretical point of view, since equal weights are put on $f_{0j}/j$ which are all of similar magnitude and measurable to about $\SI{1}{\mega\hertz}$ accuracy. Indeed, it works very well if the model that is fitted can describe the data (see bottom panel of Fig.~\ref{fig:3}b in the main text). However, if the model being fitted cannot describe the data, such as the standard model (see the blue squares in Fig.~\ref{suppfig:weights}), the fit tries to match all frequencies on average (with a hit at $f_{02}$ and $f_{05}$) but the systematic error in the curvature stays. This causes the qubit frequency $f_{01}$, which is the most accurately measurable transmon property, to be off way beyond the experimental imprecision.

Case (2) resembles similar weights that still enforce the constraint Eq.~\eqref{eq:fitcostfunctiongoal}, even if the model being fitted cannot describe the data. This is shown by the yellow circles in Fig.~\ref{suppfig:weights}.

Case (3) corresponds to equal weights on all transmon transition frequencies and larger weights on the resonator frequencies (which are measurable more accurately). However, here the fit similarly tries to match the transmon frequencies on average, at the cost of greatly overestimating the qubit frequency and the anharmonicity.

At first glance, the model corresponding to case (4) might look promising (red crosses in Fig.~\ref{suppfig:weights}). However, the price for matching the transmon transition frequencies is a huge error in the resonator frequency $f_{\mathrm{res},1}$ when the transmon is in state $j=1$. This frequency is determined from the dispersive shift $\chi=f_{\mathrm{res},1}-f_{\mathrm{res},0}$, and is measurable to an accuracy of less than $\SI{1}{\mega\hertz}$. The model prediction in this case, however, is off by more than $\SI{30}{\mega\hertz}$ (see Fig.~\ref{suppfig:weights}b). Furthermore, the bare resonator frequency $\Omega$ and the coupling strength $G$ (see Table~\ref{tab:weights}) are very different for this model. Therefore, we have to reject this model as a good alternative to describe the experiment. It might be an interesting idea, though, to analyze whether alternative forms of the transmon-resonator coupling $Gn(a+a^\dagger)$ in the Hamiltonian Eq.~\eqref{eq:Hamiltonian} could remedy this shortcoming.

\subsection{Identification of dressed states}
\label{sec:dressedstates}

\begin{figure*}
\begin{minipage}{0.49\textwidth}
  \centering
  \includegraphics[width=\columnwidth]{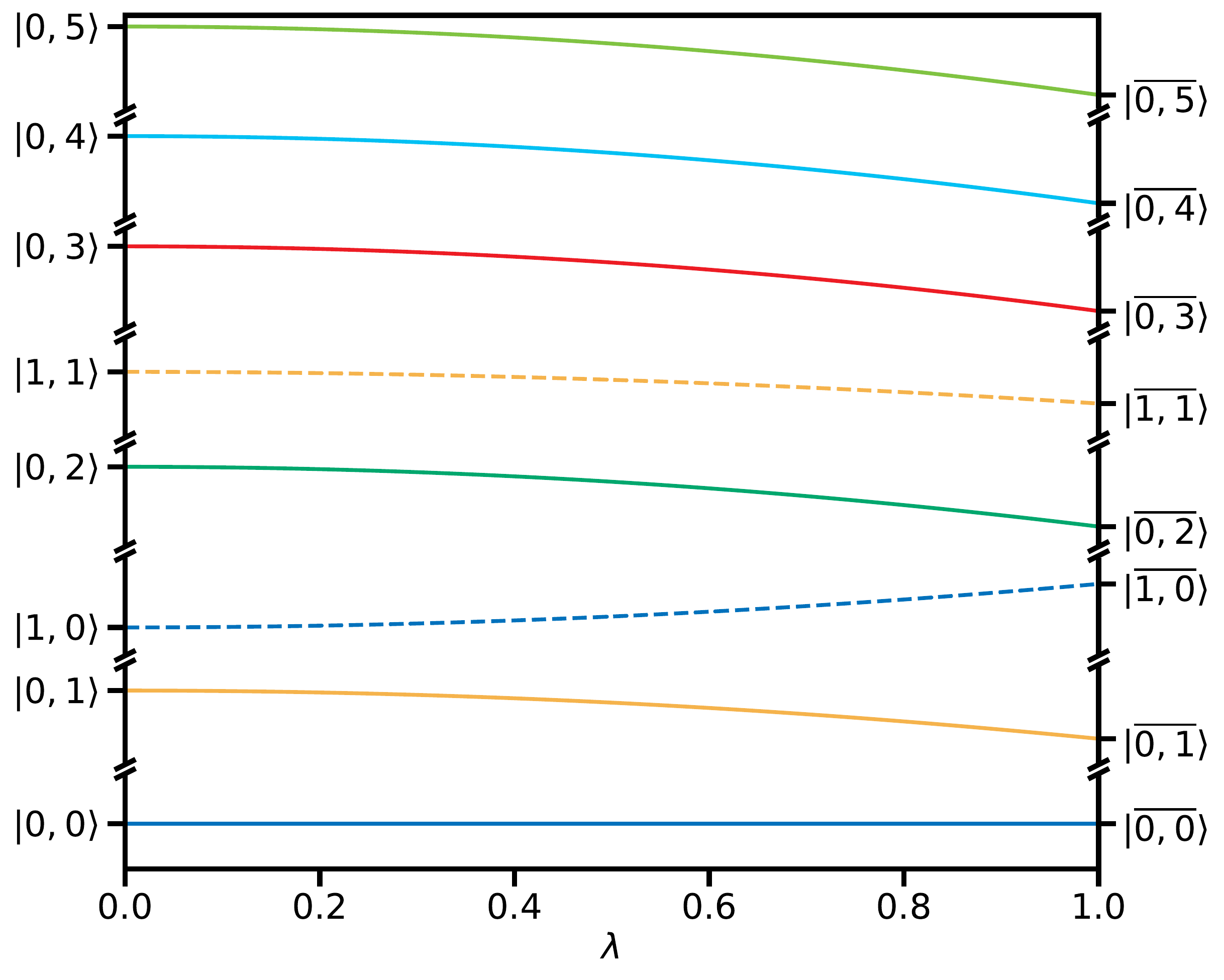}
  \end{minipage}
  \hfill
  \begin{minipage}{0.49\textwidth}
  \caption{\textbf{Evolution of the energy spectrum from bare states $\ket{k,j}$ to dressed states $\ket{\overline{k,j}}$.}
  We show the evolution of the eigenenergies $E_{\overline{l}}(\lambda)$ of $H(\lambda)$ in Eq.~\eqref{eq:jointtransmonresonatorH} as a function of the normalized coupling strength $\lambda$ for the standard model of the KIT sample (cf.~Table~\ref{tab:modelparameters}). 
  Note that the vertical axes between the cuts do not use the same scale; the eigenenergies $E_{\overline{l}}(\lambda)$ differ by several GHz, whereas the variation between $E_{\overline{l}}(\lambda=0)$ and $E_{\overline{l}}(\lambda=1)$ is less than \SI{25}{\mega\hertz}.
  In this part of the spectrum, there are no avoided crossings.
  }
  \label{suppfig:spectrum}
  \end{minipage}
\end{figure*}

\begin{figure*}
  \centering
  \includegraphics[width=\textwidth]{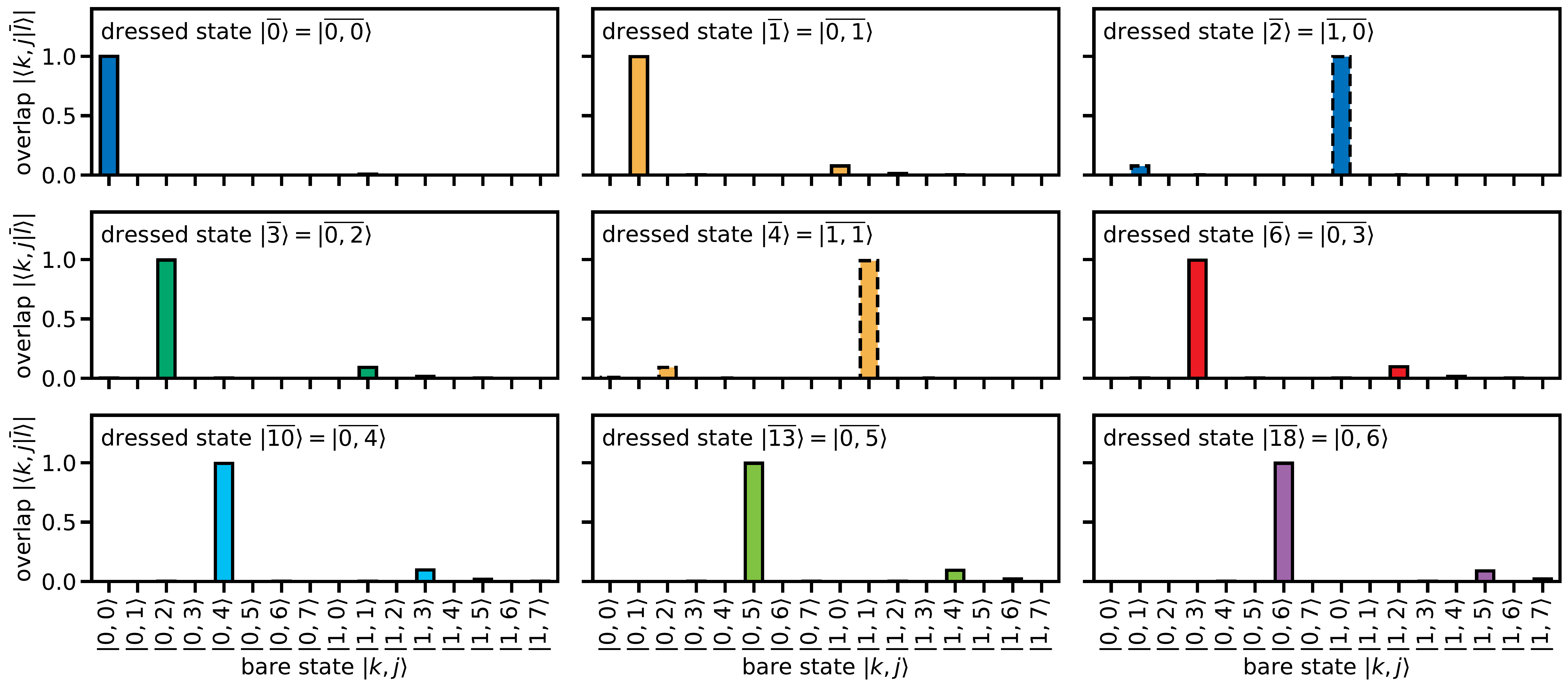}
  \caption{\textbf{Identification of dressed states within the low-energy subspace.}
  The overlap is given by the absolute values of the entries of the unitary change-of-basis matrix $U$ resulting from the diagonalization. Specifically, each column $\overline{l}$ of $U$ contains the coefficient $\braket{k,j|\overline{l}}$ in the row identified by $k$ and $j$.
  Colors correspond to the respective eigenenergy curves shown in Fig.~\ref{suppfig:spectrum}.
  }
  \label{suppfig:overlap}
\end{figure*}

In both the standard and the harmonics model, the joint transmon-resonator Hamiltonian couples the \emph{bare states} $\ket{k,j}$, given by the product of the transmon state $\ket j$ and the bare Fock state $\ket k$ of the resonator, through the transmon-resonator coupling $G n (a+a^\dagger)$ (see \hyperref[sec:methods]{Methods}). The eigenstates of this Hamiltonian are the \emph{dressed states} $\ket{\overline{k,j}}$, named because the hybridization of the two systems dresses each eigenenergy and eigenstate by a contribution from the other system. However, after numerical diagonalization, the eigenstates $\ket{\overline{l}}$ are typically ordered by increasing eigenenergies $E_{\overline{l}}$ for $\overline{l}=0,1,\ldots$, and it is not immediately obvious how to uniquely assign labels $k$ and $j$ to every eigenstate, i.e., how to identify $\ket{\overline{k,j}}\leftrightarrow\ket{\overline{l}}$, especially for large photon numbers $k$. 

In this case, one has to make a potentially subjective choice. In the literature, one can find assignments based on iteratively constructing transmon \emph{ladders} or \emph{branches} using, for instance, the projection onto the bare transmon manifolds~\cite{Peterer2015HigherTransmonStates} or the overlap with the state $a^\dagger\ket{\overline{k-1,j}}$ obtained from applying the bare raising operator of the resonator to the previously assigned state~\cite{Shillito2022DynamicsTransmonIonization,Cohen2022ClassicalChaosDrivenTransmons}.

An alternative approach is to trace the evolution from bare to dressed states through the hybridization, by considering the Hamiltonian 
\begin{equation}
    H(\lambda) = \sum_j E_j\ket{j}\!\bra{j} + \Omega a^\dagger a
    + \lambda G n (a+a^\dagger)\,,
    \label{eq:jointtransmonresonatorH}
\end{equation}
which interpolates between bare and dressed states using the parameter $0\le\lambda\le1$. For $\lambda=0$, the assignment of labels $(k,j)$ to the energies $E_{\overline{l}}(\lambda=0)$ is clear by definition of the bare states. Then for increasing $\lambda$, whenever two eigenenergies $E_{\overline{l}}(\lambda)$ and $E_{\overline{l}'}(\lambda)$ are about to cross, one needs to \emph{swap} the labels $(k,j)$ and $(k',j')$. After the crossing, the labels in the ordered list of eigenenergies have thus changed.

We remark that all such crossings are \emph{avoided} crossings, since
$H(\lambda)$ in Eq.~\eqref{eq:jointtransmonresonatorH} is a one-parameter matrix flow, and by the Hund-von Neumann-Wigner theorem
\cite{Hund1927DeutungMolekuelspektren, VonNeumann1929EigenvaluesInAdiabaticProcesses} (see also~\cite{Uhlig2020CoalescingEigenvaluesAndCrossingEigencurves}), the eigenvalue curves do not intersect. Still, the eigenstates hybridize (such that they form a superposition of the previous eigenstates), and by swapping the labels we can ensure that the eigenenergies leaving the crossing continue to carry the energy dependence that would cause them to be characterized as one or the other
state in an experiment.

The relevant low-energy spectrum of $H(\lambda)$ in Eq.~\eqref{eq:jointtransmonresonatorH} is shown in Fig.~\ref{suppfig:spectrum} as a function of $\lambda$ for the KIT sample. Here the eigenenergies are still well separated and no avoided crossings occur. We can thus use the overlap of the dressed states with the bare states to assign the proper labels $(k,j)$, as shown in Fig.~\ref{suppfig:overlap}. However, it is important to realize that this procedure does not always yield the correct assignment (as one could test by investigating higher parts of the spectra for $k>1$ or by further increasing $\lambda$).

\subsection{Residuals for the K\"oln data}
\label{sec:iepkoeln}

\begin{figure*}
  \centering
  \includegraphics[width=\textwidth]{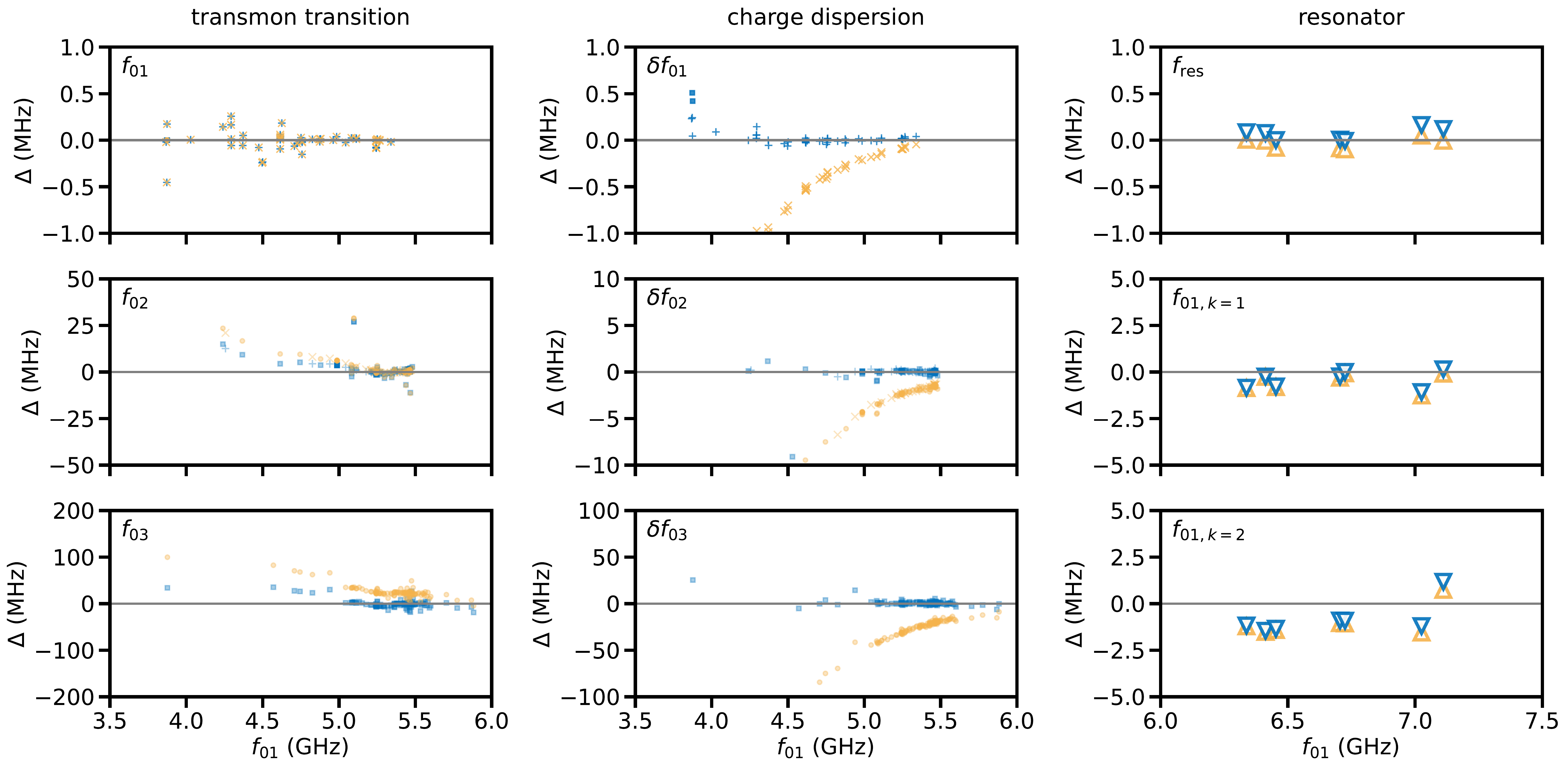}
  \caption{\textbf{Residuals for the K\"oln data.}
  All panels show the difference $\Delta$ between the frequencies obtained from the standard (yellow) or the harmonics (blue) model and the measured frequencies as a function of the transmon frequency $f_{01}$.
  The left column shows the differences in transmon transition frequencies $f_{01}$ (top), $f_{02}$ (middle), $f_{03}$ (bottom). The top left panel basically represents the calibration of $f_{01}$ depending on the in-plane magnetic field strength $B_{\parallel}$ (note the small scale on the vertical axis).
  The middle column shows the differences in the charge dispersions $\delta f_{01}$ (top), $\delta f_{02}$ (middle), $\delta f_{03}$ (bottom).
  The right column represents resonator properties: the top right panel shows the difference in the (dressed) resonator frequencies, and the middle and bottom panels show the difference in the qubit frequencies $f_{01,k}$ when the resonator is not in the ground state but in the $k$-photon state with $k=1$ (middle) or $k=2$ (bottom). The frequency $f_{01,k=1}$ is shifted from $f_{01}$ by the total dispersive shift $\chi$ (i.e.~$f_{01,k=1}=f_{01}-\chi$), and $f_{01,k=2}$ is approximately shifted by $2\chi$ from $f_{01}$.
  Data in the left and middle columns is obtained from spectroscopy (yellow squares and blue circles) and Ramsey (yellow plusses and blue crosses) experiments.
  Data in the right column is obtained from photon-number-splitting experiments (triangles).
  }
  \label{suppfig:koelnresiduals}
\end{figure*}

In contrast to the other transmon experiments considered in this work, the K\"oln data is special in the sense that it consists of 288 data points for the same frequency-tunable transmon sample, comprised of spectroscopy, Ramsey, and photon-number-splitting experiments. Furthermore, the data points correspond to many different qubit frequencies $f_{01}$ tuned by an in-plane magnetic field $B_{\parallel}$. We cover this dependence by varying only the Josephson energy $E_\mathrm{J}$ while keeping all other model parameters $\mathbf{x}^{\mathrm{std}}=(E_\mathrm{C},\Omega,G)$ of the standard model and $\mathbf{x}^{\mathrm{har}}=(E_\mathrm{C},E_\mathrm{J2}/E_\mathrm{J},E_\mathrm{J3}/E_\mathrm{J},E_\mathrm{J4}/E_\mathrm{J},\Omega,G)$ of the harmonics model constant. Technically, this is done by diagonalizing the Hamiltonians for 100 linearly spaced values of $E_\mathrm{J}/h$ between $\SI{2}{\giga\hertz}$ and $\SI{60}{\giga\hertz}$ and using cubic interpolation on the resulting data to obtain $(f_{02}(f_{01}), f_{03}(f_{01}), \delta f_{01}(f_{01}), \delta f_{02}(f_{01}), \delta f_{03}(f_{01}), f_{\mathrm{res}}(f_{01}), \chi(f_{01}))$, where the dispersive shift is computed from $\chi(f_{01}) = f_{01,k=1}(f_{01}) - f_{01}$. The model parameters are obtained by fitting each of these frequencies to two measured frequencies that are closest to medians in the sets of all frequencies. We obtain $\mathbf{x}^{\mathrm{std}}=(h\times\SI{0.2848}{\giga\hertz},h\times\SI{7.54498}{\giga\hertz},h\times\SI{0.0772}{\giga\hertz})$ and $\mathbf{x}^{\mathrm{har}}=(h\times\SI{0.3299}{\giga\hertz},-0.02298,0.00382,-0.00128,h\times\SI{7.54504}{\giga\hertz},h\times\SI{0.0832}{\giga\hertz})$ (see also Table~\ref{tab:modelparameters}). The residuals of this procedure are shown in Fig.~\ref{suppfig:koelnresiduals}.
As expected from the results shown in the main text, the Josephson harmonics model can describe the full dataset much better than the standard model. There are a few out of the 288 data points that still show deviations in $f_{02}$ and $f_{03}$ (left column for $f_{01}<\SI{5}{\giga\hertz}$). These data points correspond to very strong magnetic fields (the strongest $B_{||}$ is about $\SI{0.4}{\tesla}$). We assume that for these outliers, other higher-order, strong-field effects, not taken into account in the current models, start to play a role in this regime. However, the bulk of all data points is well described by the Josephson harmonics model.

\clearpage
\section{Samples description}
\label{suppsec:samples}

\subsection{KIT}
\label{suppsec:samplesKIT}
\begin{figure*}
  \centering
  \includegraphics[width=0.9\textwidth]{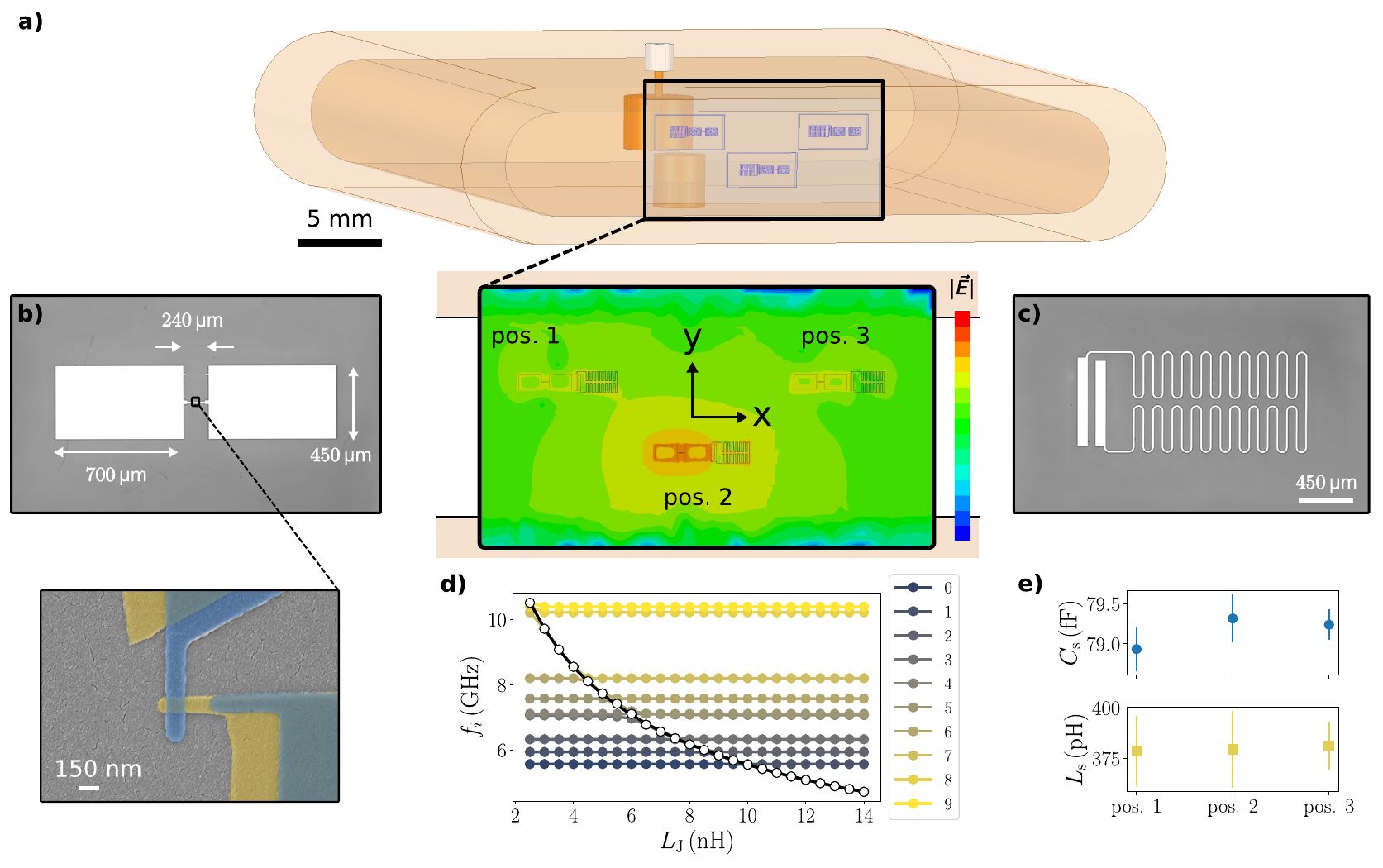}
  \caption{\textbf{Finite-element method (FEM) simulation of the KIT transmon pad geometry to obtain an estimate of the shunt capacitance and geometric stray inductance.} \textbf{a)} High-frequency structure simulator (HFSS) geometry including the copper waveguide sample holder (light orange), the input port with impedance matching (top dark orange cylinder), the tuning screw (bottom cylinder), and the sapphire substrate hosting three transmon samples. A zoom-in shows the three samples labeled according to their position on the substrate, together with the simulation result of the electric field magnitude for the transmon mode at position 2 in logarithmic scale, ranging from low (blue) to high (red) electric field strength. \textbf{b)} Optical microscope image of a transmon sample showing the aluminum thin film in white and the sapphire substrate in gray, indicating the dimensions of the electrodes. The zoom-in shows a scanning electron microscope (SEM) image of a JJ fabricated with the Dolan shadow-angle technique. The top and bottom electrodes are false-color coded in ocher and blue, respectively. \textbf{c)} Optical microscope image of the resonator geometry, with the in-plane capacitor on the left-hand side, and the meandered inductor on the right-hand side. \textbf{d)} FEM simulation results showing the frequencies of the first 10 eigenmodes above $5\,\mathrm{GHz}$ as a function of the lumped-element Josephson inductance $L_{\mathrm{J}}$ associated to the transmon at position 2. Error bars stemming from simulation inaccuracies on the data points are too small to be visible. The identified transmon frequencies are highlighted with white markers. The shunt capacitance $C_{\mathrm{s}} \approx 80\,\mathrm{fF}$ ($E_\text{C}/h \approx \SI{242}{\mega\hertz}$) and the geometric stray inductance $L_{\mathrm{s}} \approx 380\,\mathrm{pH}$ ($E_\text{L}/h\approx\SI{430}{\giga\hertz}$), which is an independent contribution in series with the Josepshon inductance (cf.~Section~\ref{sec:additionalinductance}), are extracted from a fit (black solid line) to the highlighted transmon frequencies using Eq.~\eqref{eq:transmon_freq}. \textbf{e)} Simulation results for the shunt capacitance and stray inductance as a function of the sample position following the same procedure. The errorbars are given by the fit uncertainty.}
  \label{suppfig:HFSS}
\end{figure*}

The KIT transmon qubit contains a single JJ shunted by an in-plane plate capacitor with rectangular pads, and is capacitively coupled to a lumped-element readout resonator (Fig.~\ref{suppfig:HFSS}). We determine the value of the charging energy $E_\text{C}/\text{h}=\SI{242\pm 1}{\mega\hertz}$ from finite-element method (FEM) simulations. The simulated geometry includes the 3D-waveguide sample holder hosting three qubit samples and their readout resonators, as shown in Fig.~\ref{suppfig:HFSS} a) together with the electric field distribution of the eigenmode associated with the central qubit. While the transmons have identical capacitor electrodes (see Fig.~\ref{suppfig:HFSS}b) but different JJ overlap areas, the three readout resonators differ in the number of meanders (Fig.~\ref{suppfig:HFSS}c) to ensure a frequency detuning of around $800\,\mathrm{MHz}$. The samples are located on a single c-plane sapphire substrate ($t = 330\,\si{\micro\metre}$). The dielectric permittivity of the sapphire substrate is anisotropic and defined via a tensor, where the value in parallel to the c-axis (perpendicular to the surface) is $\epsilon_{\mathrm{r},\parallel} = 11.5$ and the value perpendicular to the c-axis is $\epsilon_{\mathrm{r},\perp} = 9.3$. The shunt capacitance $C_{\mathrm{s}}$, from which we calculate the charging energy $E_{\mathrm{C}} = e^2 / (2 C_{\mathrm{s}})$, and the linear stray inductance $L_{\mathrm{s}}$ arising from the leads are extracted from the simulated eigenfrequencies by varying the lumped-element inductance $L_{\mathrm{J}}$ (see Fig.\,\ref{suppfig:HFSS}d), and using the fit function
\begin{equation}
    f = \frac{1}{2 \pi \sqrt{(L_{\mathrm{J}} + L_{\mathrm{s}}) C_{\mathrm{s}}}}\,.
    \label{eq:transmon_freq}
\end{equation}
The obtained stray inductance is $L_{\mathrm{s}} \approx 380\,\mathrm{pH}$ for all samples (see Fig.~\ref{suppfig:HFSS}e). In addition to the geometric contribution to the stray inductance extracted from the FEM simulations, we expect a small contribution arising from the kinetic inductance of the pure Al thin film. We can estimate this contribution from the normal-state resistivity of the Al film $\rho_{\mathrm{n}} = 4.3\,\si{\micro\ohm\centi\metre}$ \cite{Friedrich_2019}, which we have measured for a sample fabricated in the same evaporator using the same recipe. For a film thickness $t = 70\,\si{\nano\metre}$, we arrive at a kinetic sheet inductance $L_{\mathrm{k},\square} = 0.6\,\si{\pico\henry}/\square$ from which we can estimate the contribution of the kinetic inductance $L_{\mathrm{k}}\approx 60-120\,\si{\pico\henry}$ to the stray inductance in the sample. Hence, we expect a total stray inductance $L_{\mathrm{s,tot}} = L_{\mathrm{s}} + L_{\mathrm{k}} \approx 500\,\mathrm{pH}$.   

\subsubsection*{Fabrication}
The sample is fabricated on c-plane, double side polished sapphire substrate. A bi-layer resist stack of \SI{700}{\nano\meter} MMA EL-13 and 300 nm PMMA A4 and a $\SI{10}{\nano\meter}$ gold conduction layer is used for writing with a 50 keV e-beam writer. The structures are developed in a beaker with an isopropanol-water mixture with volume ratio 3:1 at \SI{6}{\celsius}. Before the metal deposition in a Plassys evaporation system, the substrate is cleaned with a Kaufmann ion source in an Ar/O2 descum process and the vacuum is improved using titanium gettering. The target film thicknesses of the first and second aluminum layer are \SI{30}{\nano\meter} and \SI{40}{\nano\meter}, respectively, and the evaporation angles are $\SI{0}{\degree}$ and $\SI{20}{\degree}$, respectively. The $(\SI{100}{\nano\meter})^2$ sized JJ is fabricated in Dolan-style \cite{Dolan1977JunctionTechnique} and the insulating aluminum oxide barrier is grown under static oxidation at an oxygen pressure of \SI{10}{\milli\bar} for \SI{2.5}{\minute}.

\subsubsection*{Cooldown 1}
\begin{figure*}
  \centering
  \includegraphics[width=\textwidth]{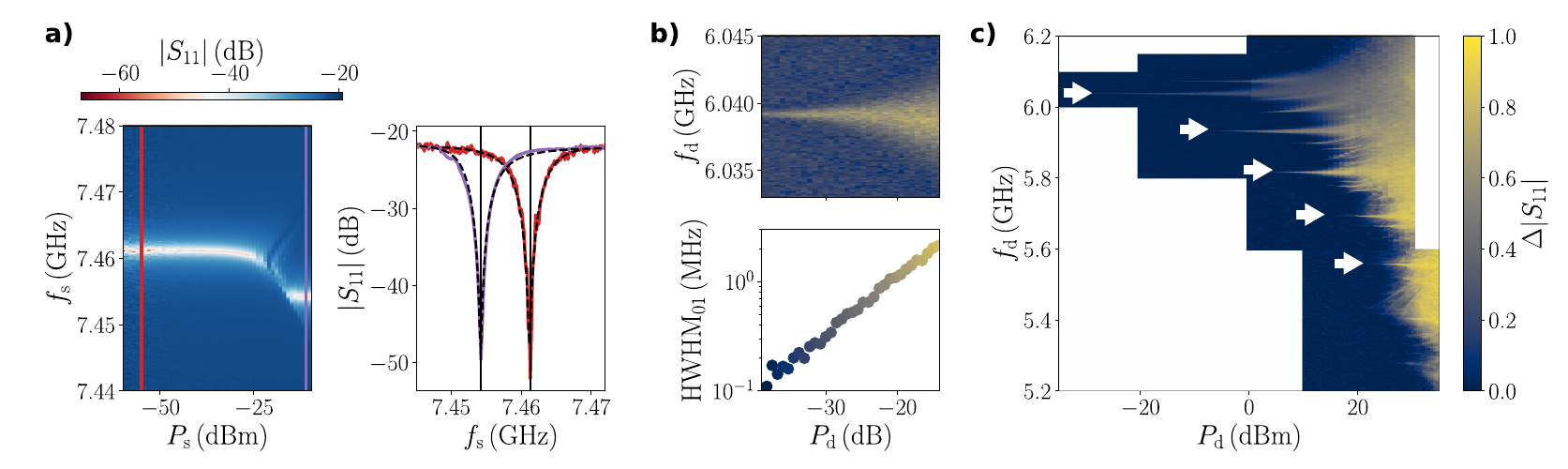}
  \caption{\textbf{Spectroscopic measurements of the KIT sample (Cooldown 1).} \textbf{a)}~Amplitude of the reflection coefficient $S_{11}$ in log-scale as a function of signal frequency $f_{\mathrm{s}}$ and signal power $P_{\mathrm{s}}$ at room temperature (left panel). At low readout power, the readout resonator is found at $f_{\mathrm{r}} = 7.4613 \, \mathrm{GHz}$. With increasing power the resonator shifts in frequency due to the inherited non-linearity until it eventually reaches a constant frequency at $f_{\mathrm{r}} = 7.4543 \, \mathrm{GHz}$ when the transmon is in a highly excited state. The right-hand panel shows two traces at low (red) and high power (violet), respectively, as indicated by the vertical lines in the 2D sweep. From the difference in transition frequency we extract the Lamb shift $\Delta \omega_{\mathrm{r}} = - 2 \pi \times 7 \, \mathrm{MHz}$. The dashed black lines indicate the results of a circle fit to the complex reflection coefficient, from which we determine the internal and external decay rates. \textbf{b)}~Two-tone spectroscopy of the fundamental transition frequency of the transmon qubit. Due to the dispersive interaction, the readout resonator shifts in frequency when the additional drive tone excites the transmon, resulting in a change in the reflected signal amplitude from low (dark) to high (yellow). With increasing drive power $P_{\mathrm{d}}$, the transition broadens, as indicated in the bottom panel. \textbf{c)}~Two-tone spectroscopy of the transmon spectrum measured by sweeping the drive frequency and drive power. The white arrows indicate the features identified as transitions in the transmon spectrum. Many other features are visible in the spectrum which are drive power dependent. The feature above the fundamental transition frequency around $f_{01} = 6.0392\,\mathrm{GHz}$ is the transition frequency of another qubit, which is orders of magnitude more weakly coupled to the readout resonator.}
  \label{suppfig:KIT_CD1}
\end{figure*}

The sample is measured in reflection in a 3D copper waveguide sample holder similar to Refs.~\cite{Winkel2020TransmonImplementationSuperconductingGranularAluminum} and \cite{Winkel_2020_DJJAA}. The amplitude of the reflection coefficient is shown in Fig.~\ref{suppfig:KIT_CD1} as a function of the signal frequency $f_{\mathrm{s}}$ and signal power $P_{\mathrm{s}}$ around the resonance frequency $f_{\mathrm{r}} = 7.4613\,\mathrm{GHz}$ of the readout resonator. With increasing signal power, the readout resonator shifts in frequency due to the non-linearity inherited from the linear coupling to the transmon, until it becomes independent of power when the transmon is in a highly excited state \cite{Lescanne2019EscapeUnconfinedStatesFloquet}. From the difference in the resonator frequency we extract the Lamb shift $\Delta \omega_{\mathrm{r}} = - 2 \pi \times 7 \, \mathrm{MHz}$, which is the difference between the bare and the dressed resonator frequency. In the simple transmon Hamiltonian and using a Schrieffer-Wolff transformation, the Lamb shift is identified with the partial dispersive shift $\tilde{\chi}_{12}/2$ \cite{koch2007transmon}, where $\tilde{\chi}_{ij} = g_{ij}^2/(\omega_{ij} - \omega_{\mathrm{r}})$, from which we estimate the coupling rate $g_{12} = 2\pi\times 151\,\mathrm{MHz}$. This value is in agreement with our FEM simulations for $g_{01} = 2\pi\times 127\,\mathrm{MHz}$ when accounting for the numerical factor arising from the transition matrix element. For the fit to our extended transmon model including the higher harmonics in the Josephson effect, we do not rely on this indirect determination of the total dispersive shift $\chi_{01}$, which under the Schrieffer-Wolff transformation is given by $\chi_{01} = 2 \tilde{\chi}_{01} - \tilde{\chi}_{12}$. Instead we use the value $\chi_{01} = - 2 \pi \times 2.6\,\mathrm{MHz}$ measured in CD2 from the pointer states in the IQ-plane associated to the transmon ground and first excited state (see Fig.\,\ref{suppfig:KIT_CD2}a). The frequencies of the individual transitions in the transmon spectrum are extracted from two-tone spectroscopy, as shown in Fig.~\ref{suppfig:KIT_CD1}b and c, and are listed in Tab.~\ref{supptab:measured_frequencies} together with the readout frequencies estimated from the Lamb shift. The linewidth of the fundamental transition Fig.~\ref{suppfig:KIT_CD1}b is extracted from Lorentzian fits to the response in the reflection coefficient. As can be seen from the spectroscopic data in Fig.~\ref{suppfig:KIT_CD1}c, there are many additional features visible in the spectrum, especially at high drive powers, since we are probing the dressed energy spectrum of the transmon and the readout resonator. We argue that the frequencies associated with transitions in the transmon spectrum are independent of drive power, and only show power broadening. The transition frequencies of the multi-photon transitions indicated by the white arrows are extracted from individual measurements in the frequency vicinity of each transition, similar to Fig.~\ref{suppfig:KIT_CD1}b.

\subsubsection*{Cooldown 2}
\begin{figure*}
  \centering
  \includegraphics[width=\textwidth]{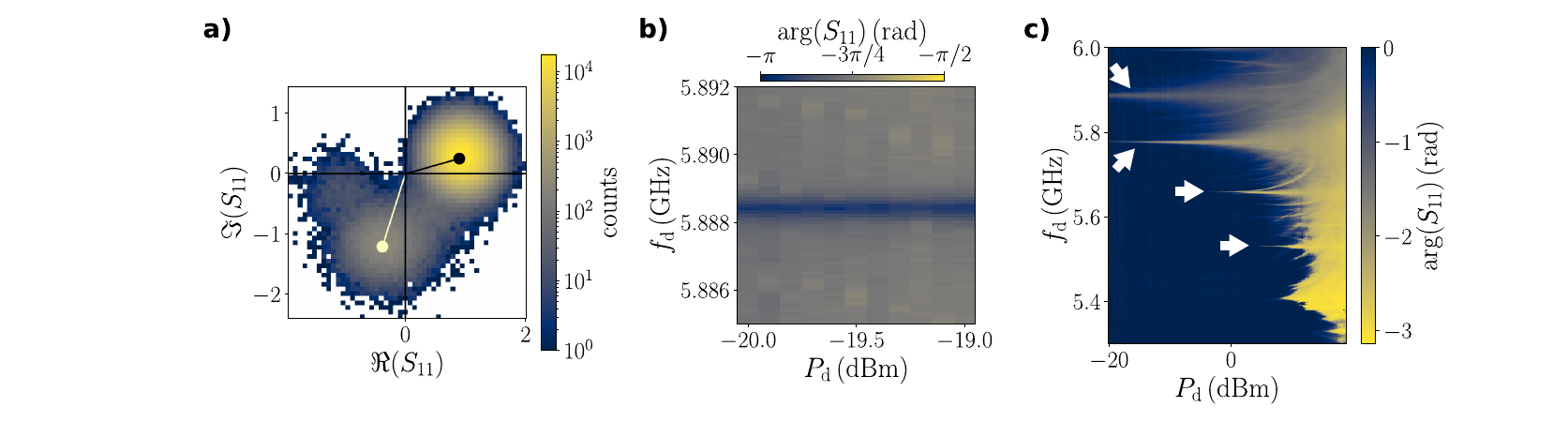}
  \caption{\textbf{Spectroscopic measurements of the KIT sample (Cooldown 2).} \textbf{a)}~Reflection coefficient $S_{11}$ of the readout resonator measured at the constant readout frequency $f_{\mathrm{s}} = 7.4607\,\mathrm{GHz}$ and shown in the complex plane. The two pointer states visible are associated to the ground and first excited state of the transmon. From their relative phase, we extract the dispersive shift $\chi_{01} = - 2 \pi \times 2.6\pm0.1\,\mathrm{MHz}$. \textbf{b)}~Two-tone spectroscopy around the fundamental transition $f_{01} = 5.8885\,\mathrm{GHz}$. In comparison to the first cooldown, the frequency of the transmon changed by around $160\,\mathrm{MHz}$. \textbf{c)}~Transmon spectrum extracted from two-tone spectroscopy by changing the drive frequency $f_{\mathrm{d}}$ and drive power $P_{\mathrm{d}}$. The features associated to transitions in the transmon are indicated by white arrows.}
  \label{suppfig:KIT_CD2}
\end{figure*}

In comparison to cooldown 1, a Dimer-Josephson-Junction-Array-Amplifier (DJJAA) \cite{Winkel_2020_DJJAA} was added to the readout line to enable fast qubit readout and the measurement of quantum jump traces by monitoring the reflection coefficient in the vicinity of the readout resonator. At finite temperature, a histogram of such a trace reveals the pointer states associated to different transmon states, from which we can extract the dispersive shift using their phase difference. For the calculation of the dispersive shift between the ground state and the j-th transmon state, we use 
\begin{equation}
    \chi_{0j} = - \tan\left(\frac{\varphi_0 - \varphi_j}{2}\right) \frac{1 + \left(\frac{2 (\omega_{\mathrm{r}} - \omega_{\mathrm{s}})}{\kappa +\gamma} \right)^2}{2 + \frac{4 (\omega_{\mathrm{r}} - \omega_{\mathrm{s}})}{\kappa + \gamma} \tan\left(\frac{\varphi_0 - \varphi_j}{2}\right)} (\kappa + \gamma)\,,
    \label{EQ:chi_phis}
\end{equation}
where $\varphi_0$ and $\varphi_j$ are the phase of the pointer state associated with the ground and j-th excited state, respectively, $\kappa$ and $\gamma$ are the external and internal decay rates of the readout resonator, $\omega_{\mathrm{r}}$ is the readout resonator frequency, and $\omega_{\mathrm{s}}$ is the frequency of the readout signal. From the pointer states shown in Fig.~\ref{suppfig:KIT_CD2}a, which are associated to the ground and first excited state, we extract a dispersive shift $\chi_{01} = - 2 \pi \times 2.6\pm0.1\,\mathrm{MHz}$. The internal and external decay rates of the readout resonator, $\gamma = 2 \pi \times 190\,\mathrm{kHz}$ and $\kappa = 2\pi \times 3.90\,\mathrm{MHz}$, respectively, are extracted from circle fits to the complex reflection coefficient similar to Fig.~\ref{suppfig:KIT_CD1}a and Fig.~\ref{suppfig:KIT_CD3_1}a. The frequencies of transitions into higher excited states of the transmon spectrum are extracted from two-tone spectroscopy, exemplified for the fundamental transition in Fig.~\ref{suppfig:KIT_CD2}b, and on a larger scale in Fig.~\ref{suppfig:KIT_CD2}c.

\subsubsection*{Cooldown 3}
\begin{figure*}
  \centering
  \includegraphics[width=\textwidth]{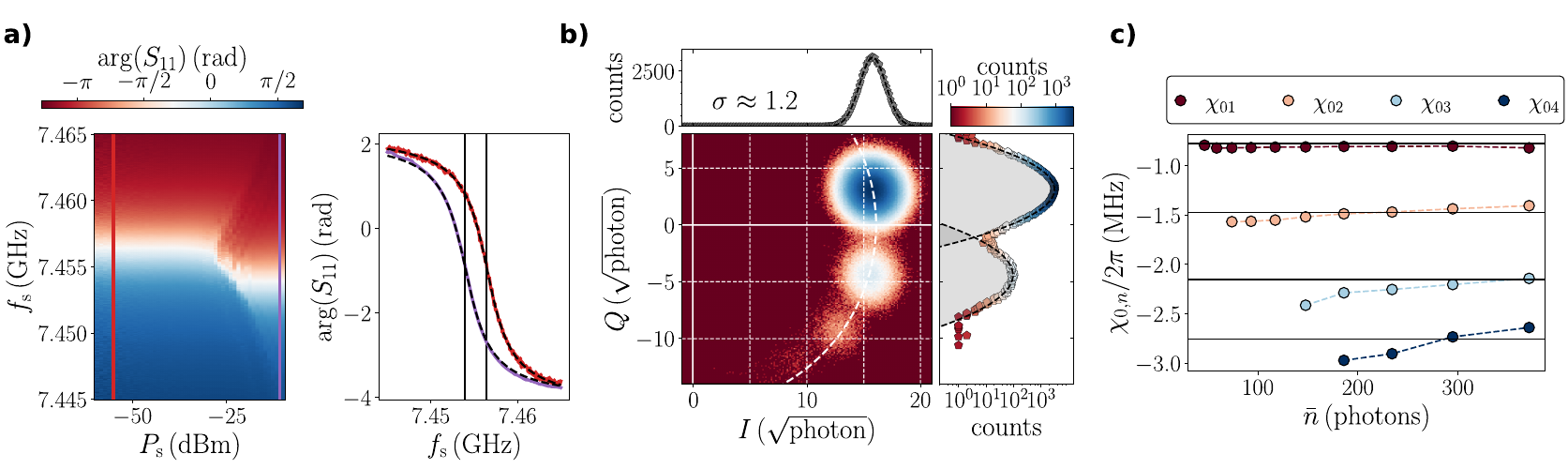}
  \caption{\textbf{Spectroscopic measurements of the KIT sample (Cooldown 3).} \textbf{a)}~Power dependence of the reflection coefficient in the vicinity of the readout resonator, similar to Fig.~\ref{suppfig:KIT_CD1}a. At high signal power, the readout resonator effectively decouples from the transmon as it is highly excited, resulting in a frequency difference between low power ($f_{\mathrm{r}} = 7.4564\,\mathrm{GHz}$) and high power ($f_{\mathrm{r}} = 7.4539\,\mathrm{GHz}$), from which we extract the Lamb shift $\tilde{\chi}_{12} / 2 = - 2 \pi \times 2.51 \, \mathrm{MHz}$. The shift is significantly smaller compared to CD1, since the transmon-resonator detuning increased significantly between the two cooldowns. From circle fits to the complex reflection coefficient (right panel), we extract the internal and external coupling rates $\gamma = 2 \pi \times 63\,\mathrm{kHz}$ and $\kappa = 2 \pi \times 4.13\,\mathrm{MHz}$, respectively. \textbf{b)}~Histogram of the reflection coefficient in log scale measured at the readout frequency $f_\mathrm{s} = 7.4560\,\mathrm{GHz}$. Due to the finite temperature of the transmon, multiple pointer states associated to different transmon states are visible in the complex plane spanned by the in-phase and quadrature components $I$ and $Q$, respectively. As expected from a Boltzmann distribution, the occupation probability significantly decreases with the level index. The top panel shows a horizontal slice along the in-phase component $I$ through the maximum of the ground state in linear scale, while the right-hand panel shows a slice along the quadrature component $Q$ in log scale. The signal strength is calibrated from the measurement induced dephasing of the transmon, and expressed in units of measurement photons. From the standard deviation of the pointer states, we can calculate the measurement efficiency $\eta = 1 / (2 \sigma^2) = 0.35$ \cite{Winkel_2020_DJJAA}. \textbf{c)}~Dispersive shift of the multi-photon transitions $\chi_{0j}$, extracted from the angles between the pointer states shown in panel b) using Eq.~\eqref{EQ:chi_phis}, measured as a function of the mean number of photons $\bar{n}$ in the readout resonator. For comparison reasons, the solid black lines indicate the prediction of the simple transmon model under the Schrieffer-Wolff transformation \cite{koch2007transmon,Braumueller2015_multiphoton} using $g_{01} = 2 \pi \times 89\,\mathrm{MHz}$, $E_{\mathrm{J}} / h = 13.82\,\mathrm{GHz}$ and $E_{\mathrm{C}}/h = 223\,\mathrm{MHz}$.}
  \label{suppfig:KIT_CD3_1}
\end{figure*}

At the end of cooldown 2, the sample was accidentally annealed during the warm-up of the measurement setup, which brought the fridge to approx.~$100^\circ \mathrm{C}$. While the readout resonator changed only insignificantly in frequency ($f_{\mathrm{r}} = 7.4564\,\mathrm{GHz}$), the fundamental transition of the transmon sample shifted down in frequency by more than $1\,\mathrm{GHz}$. As a consequence of the increased detuning between transmon and readout, the dispersive shift decreased noticeably, resulting in a smaller Lamb shift $\Delta \omega_{\mathrm{r}} = - 2 \pi \times 2.51\,\mathrm{MHz}$ (see Fig.~\ref{suppfig:KIT_CD3_1}a). The extracted shift in the resonator frequency corresponds to a coupling rate $g_{12} = 2 \pi \times 122\,\mathrm{MHz}$, when using the simple transmon Hamiltonian and the Schrieffer-Wolff transformation to map the transmon onto an effective two-level system. The decrease in the coupling rate compared to CD1 is not surprising, since the transition matrix elements from which the coupling rates are calculated depend on the ratio $E_{\mathrm{J}} / E_{\mathrm{C}}$ \cite{koch2007transmon}, and even change for the simple transmon Hamiltonian. The decrease in the dispersive shifts enabled the discrimination of multiple pointer states in the histogram of quantum jump traces measured at the same readout frequency (see Fig.~\ref{suppfig:KIT_CD3_1}b), from which we extracted the multi-photon dispersive shifts $\chi_{0j}$ from the relative angle of the pointer states using Eq.~\eqref{EQ:chi_phis} (see Fig.~\ref{suppfig:KIT_CD3_1}c). The internal and external decay rates of the readout resonator are $\gamma = 2 \pi \times 63\,\mathrm{kHz}$ and $\kappa = 2 \pi \times 4.13\,\mathrm{MHz}$, respectively. We observe a weak power dependence of the extracted dispersive shifts, which is increasingly more pronounced for transitions into higher levels. We compare the measurement outcome to the prediction of the simple transmon Hamiltonian under the Schrieffer-Wolff transformation, which predicts these dispersive shifts from the partial dispersive shifts $\chi_{0j}= \tilde{\chi}_{01} + \tilde{\chi}_{j-1,j} - \tilde{\chi}_{j,j+1}$~\cite{koch2007transmon,Braumueller2015_multiphoton}. In our calculation, we use $g_{01} = 2 \pi \times 89\,\mathrm{MHz}$ for the fundamental transition, $E_{\mathrm{J}} / h = 13.82\,\mathrm{GHz}$ and $E_{\mathrm{C}}/h = 223\,\mathrm{MHz}$ (see Tab.\,\ref{tab:modelparameters}). The numerical results are shown in Fig.~\ref{suppfig:KIT_CD3_1}c as horizontal solid black lines. While the agreement is good for the two lowest dispersive shifts, we observe increasing deviations for the higher levels $j \geq 2$.   

\begin{figure*}
  \centering
  \includegraphics[width=\textwidth]{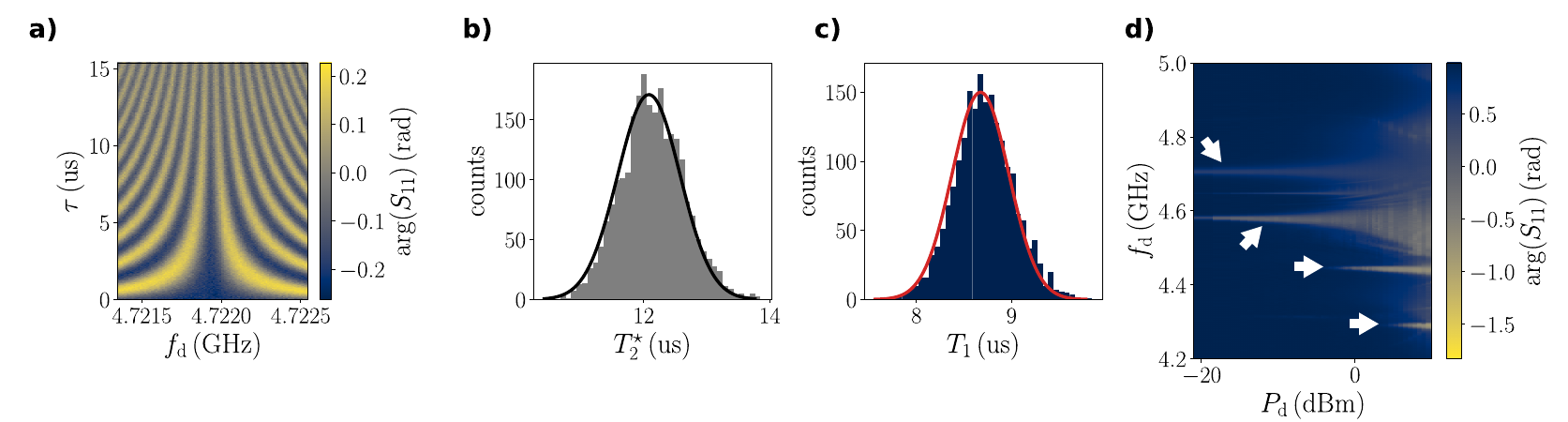}
  \caption{\textbf{Coherence and spectrum of the KIT sample (Cooldown 3).} \textbf{a)}~Ramsey-fringes experiment with varying relative detuning between the microwave pulse to prepare the transmon in a superposition state and the fundamental transition $0 \rightarrow 1$. Since a finite detuning induces a deterministic time evolution during the idling time $\tau$ between the $\pi/2$-pulses in the Ramsey sequence, we can determine the fundamental transition frequency $f_{01} = 4.72193\,\mathrm{GHz}$ with high accuracy. \textbf{b)}~By repeating the Ramsey fringes measurement, we extract the coherence time $T_2^\star = 12\pm 0.5\,\si{\micro\second}$ from the statistics of the exponentially decaying envelope. \textbf{c)}~The energy relaxation time of the transmon $T_1 = 8.7\pm0.3\,\si{\micro\second}$ is measured in a separate experiment. \textbf{d)}~Transmon spectrum measured in a two-tone experiment by applying a drive to the transmon which induces (multi-photon) transitions into higher excited states, resulting in a shift of the resonator frequency. The white arrows indicate the transitions we can unambiguously identify from their pointer state distribution in the complex plane (cf.~Fig.~\ref{suppfig:KIT_CD3_1}b).}
  \label{suppfig:KIT_CD3_2}
\end{figure*}

The frequencies of transitions into higher excited states of the transmon spectrum are obtained from a combination of time-domain measurements and two-tone spectroscopy. The fundamental transition is extracted from a Ramsey-fringes measurement, which is particularly sensitive to the detuning with respect to a reference drive (see Fig.~\ref{suppfig:KIT_CD3_2}a). Moreover, we measure an average Ramsey coherence time of ${T_2^*=\SI{12\pm0.5}{\micro\second}}$ (see Fig.~\ref{suppfig:KIT_CD3_2}b) and an energy relaxation time of $T_1=\SI{8.7\pm0.3}{\micro\second}$ (see Fig.~\ref{suppfig:KIT_CD3_2}c). The other transitions are obtained from two-tone spectroscopy (see Fig.~\ref{suppfig:KIT_CD3_2}d).


\subsection{ENS}
\label{sec:samplesENS}
The ENS 3D transmon sample is the same as documented in refs.~\cite{Ficheux2018LescanneQubit, Lescanne2019EscapeUnconfinedStatesFloquet}. The transmon contains a single JJ shunted by an in-plane plate capacitor with rectangular pads, which are capacitively coupled to a copper cavity. The cavity is read out in transmission. 
It was fabricated in a single double-shadow evaporation step on a sapphire substrate. A MMA/PMMA resist stack was used. Before the metal deposition in a Plassys evaporation system, the sample was cleaned in-situ using an Ar/O$_2$ descum process. The two aluminum layers were grown with a target thickness of \SI{35}{\nano\meter} and \SI{100}{\nano\meter} at \SI{\pm35}{\degree} angles. Between the two evaporations, the oxide barrier is grown under static oxidation at a pressure of \SI{20}{\milli\bar} with a 4:1 Ar:O$_2$ mixture for \SI{7}{\minute}. The JJ size is \SI{260}{\nano\meter} x \SI{200}{\nano\meter}. 
The measured qubit transition and readout resonator frequencies are listed in Tab.~\ref{supptab:measured_frequencies}. The qubit has an average energy relaxation time of $T_1=\SI{15}{\micro\second}$ and Ramsey coherence time of ${T_2^*=\SI{11}{\micro\second}}$.

\clearpage

\subsection{K\"oln}
\label{sec:samplesKoeln}

\begin{figure*}
    \centering
    \includegraphics[width=\textwidth]{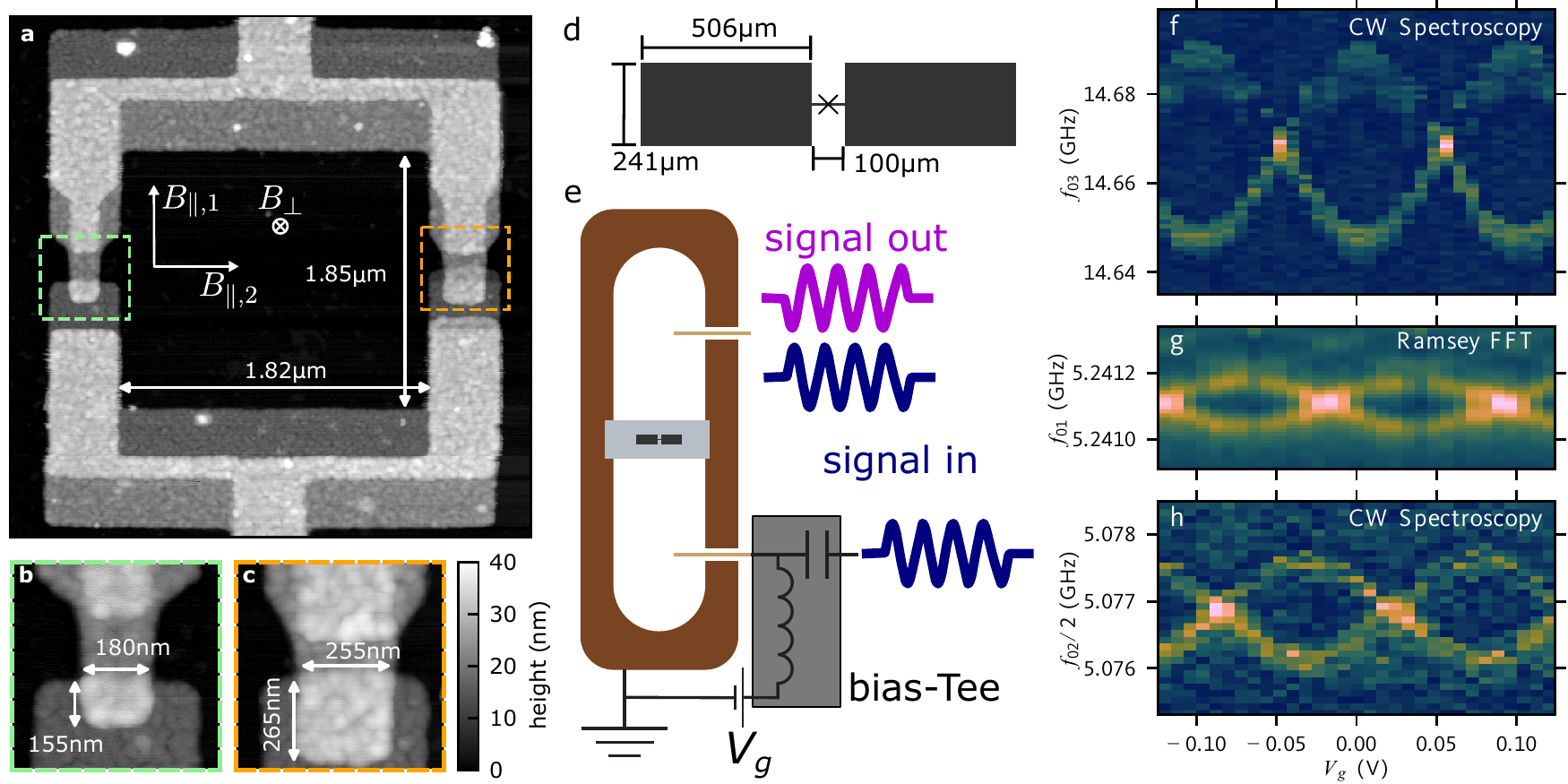}
    \caption{\textbf{Overview of the K\"oln device, setup and charge dispersion measurements.}
    \textbf{a} Atomic force microscope (AFM) image of the SQUID area with zoom-ins on the JJs (\textbf{b} and \textbf{c}).
    The magnetic field coordinate system is shown relative to the JJs and measurements of the relevant dimensions are given.
    \textbf{d} Transmon electrode dimensions.
    \textbf{e} Schematic of the transmon in a 3D cavity with two symmetrically coupled ports. A voltage $V_\mathrm{g}$ is applied to one of the cavity pins relative to ground in order to change the charge offset $n_\mathrm{g}$ of the floating transmon.
    \textbf{f} Two-tone spectroscopy of the $f_{03}$ transition as a function of $V_\mathrm{g}$ showing the characteristic two parity branches and sinusoidal dependence.
    \textbf{g} Fourier transform of Ramsey datasets (excited-state population as a function of wait time) measured for the first transition $f_{01}$ and at different $V_\mathrm{g}$.
    \textbf{h} Two-tone spectroscopy of the $f_{02}/2$ transition.
    The datasets in \textbf{f}-\textbf{h} are all measured at $B_{\mathrm{\parallel}, 1}=\SI{0.15}{\tesla}$.
    The charge offset is subject to slow drift in time explaining the different offset in $V_\mathrm{g}$ for the three datasets.}
  \label{suppfig:cologne_sample_figure}
\end{figure*}

The K\"oln setup and sample geometry is shown in Fig.~\ref{suppfig:cologne_sample_figure}a-e. The sample is a SQUID transmon with rectangular capacitor pads, which are capacitively coupled to a copper cavity. It was fabricated with the same capacitor geometry and in the same batch as the device documented in~\cite{Krause2022KoelnQubit}. The cavity is measured in reflection, but two-tone signals as well as time-domain pulses at the qubit frequency are applied to the other port. The energy relaxation time of the first excited state is on the order of ${T_1=\SI{10}{\micro\second}}$, the Ramsey coherence time on the order of ${T_2^*=\SI{3}{\micro\second}}$ and $T_2^{\mathrm{echo}}$ is usually similar to $T_1$.

\subsubsection*{Fabrication}

The sample is fabricated with a single electron-beam lithography step on a sapphire substrate.
The Dolan bridge~\cite{Dolan1977JunctionTechnique} mask is made using an MMA-PMMA resist stack. 
After developement, a weak oxygen plasma is applied to remove resist residues. 
The JJs are fabricated using double-shadow evaporation (angles are $\pm\SI{20}{\degree}$) in a Plassys MEB 550S system without substrate cooling or heating. 
The target film thicknesses of the first and second aluminum layer are \SI{10}{\nano\meter} and \SI{18}{\nano\meter}, respectively.
The aluminum oxide barriers are grown under static oxidation at an oxygen pressure of \SI{1}{\milli\bar} for \SI{6}{\minute}. 
To measure the film thicknesses and JJ geometry, we took an atomic force microscope (AFM) image of the SQUID region and close-up pictures of the JJs (Fig.~\ref{suppfig:cologne_sample_figure}c-e). 
The AFM measurements for the first and second layer thickness yield \SI{15}{\nano\meter} and \SI{21}{\nano\meter} respectively (including the oxide layer), while the combined thickness gives \SI{32}{\nano\meter} (this should include two oxide layers). 
The Dolan-style  JJs in the SQUID have approximate areas $(\SI{160}{\nano\meter})^2$ and $(\SI{260}{\nano\meter})^2$ and the JJ-widths are compatible with the Fraunhofer dependence of $E_\mathrm{J}$ that we measure. 
The granularity of the thin-film aluminum is clearly visible in the AFM picture and the JJ area comprises many grains.

\subsubsection*{Tuning the $E_\mathrm{J}$ with an in-plane magnetic field}

A three-axis vector magnet is used to measure the field dependence of the transmon (cf.~Ref.~\cite{Krause2022KoelnQubit}). 
It changes the transmon $E_\mathrm{J}$ (and harmonics) due to a combination of a geometric Fraunhofer contribution and the suppression of the superconducting gap. 
The SQUID can be tuned with a small out-of-plane field $B_{\mathrm{\perp}}$.
Based on the SQUID oscillations, we align the magnetic field in the in-plane direction using the vector magnet.

The sample is fabricated with comparatively thin aluminum films and narrow leads to the JJ, in order to make it more magnetic field resilient. Moreover, we chose a relatively small SQUID area (cf.~Fig.~\ref{suppfig:cologne_sample_figure}a) to be less sensitive to noise from the magnet. While we estimate a critical field above \SI{0.86}{T}, detailed measurements for this sample were only possible up to \SI{0.4}{T}, because for higher fields the SQUID starts to be unstable with flux jumps on a timescale of minutes. 

Following the approach in Ref.~\cite{Krause2022KoelnQubit}, we measure the field dependence of $E_\mathrm{J}$.
The $E_\mathrm{J}$s of the individual junctions can be extracted from the SQUID dependence with out-of-plane field which is measured at different in-plane fields.
We estimate the Fraunhofer effect to dominate the in-plane field dependence of $E_\mathrm{J}$ at low fields, thus the geometric asymmetry of the two JJs comprising the SQUID matters (cf.~Fig.~\ref{suppfig:cologne_sample_figure}b-c).
While $E_\mathrm{J}$ of the larger JJ changes from $\approx\SI{19.4}{\giga\hertz}$ at zero field to $\approx\SI{12.3}{\giga\hertz}$ at \SI{0.4}{\tesla}, the small JJ changes from \SI{6.0}{\giga\hertz} to \SI{4.9}{\giga\hertz}.
At \SI{0.4}{\tesla}, the expected change in the superconducting gap is on the order of \SI{10}{\percent}.
The data presented here is predominantly measured at the bottom sweetspot of the asymmetric SQUID, because the top sweetspot is close to the cavity at low magnetic field. 

The fact that the K\"oln device is a SQUID transmon affects the Josephson harmonics.
At the bottom sweetspot of a SQUID, even harmonics would be enhanced, while odd harmonics would be suppressed. 
However, likely due to the fact that we only use data close to the bottom sweetspot at different in-plane field, our simplified model (cf.~Section~\ref{sec:iepkoeln}) that considers only an effective $E_\mathrm{J}$ and fixed ratios for the harmonics describes the data well.

\subsubsection*{Measuring the charge dispersion}

Note that the charge dispersion grows for the higher levels and, if it cannot be explicitly resolved, it adds to the linewidth and can become a source of systematic errors in measurements.
If the offset charge cannot be explicitly controlled, it will slowly drift, particularly between measurements of different transitions~ \cite{Schreier2008SuppressingChargeNoise}, which could for example lead to systematic errors in estimating the anharmonicity.
For the K\"oln sample, the offset voltage can be explicitly controlled and for many transitions both $f_{ij}$ and $\delta f_{ij}$ are measured (see Section~\ref{sec:samplesKoeln}).
Then the frequency $f_{ij}$ can be extracted as the mean frequency of the two parity branches reducing systematic errors and $\delta f_{ij}$ is a separately measured quantity that can be used to better fit the Hamiltonian.

The gate-voltage $V_\mathrm{g}$ is applied to one of the cavity pins through a bias-tee to control the offset charge $n_\mathrm{g}$, such that spectroscopy as a function of frequency and $n_\mathrm{g}$ can be measured~\cite{Schreier2008SuppressingChargeNoise, Sun2012QPdynamicsBandgapEngineering}. This voltage is applied relative to the ground of the dilution refrigerator, which is connected to the 3D cavity. The voltage bias works because the pin is closer to one of the two floating transmon islands. The transitions $f_{01}$, $f_{02}/2$ and $f_{03}$ were measured in two-tone continuous-wave spectroscopy as a function of $V_\mathrm{g}$ to extract $f_{0j}$ and the corresponding charge dispersion $\delta f_{0j}$ (example data shown in Fig.~\ref{suppfig:cologne_sample_figure}f,h). Both parity branches are visible in our data, as the spectroscopy measurements are slow compared to the parity switching timescale, which we measured to be on the order of \SI{1}{\milli\second}. The $f_{0j}$ and $\delta f_{0j}$ were extracted from the 2-D plots by peak finding and fitting the two branches as combined $\sin$-functions to the data. 

In addition to the spectroscopy data, Ramsey measurements on the 0-1, 1-2 and 0-2 transitions were performed as a function of $V_\mathrm{g}$, showing the characteristic beating for the two parities. Fits to the Ramsey fringe data can be used to extract $f_{ij}$ and $\delta f_{ij}$. Alternatively, a Fourier transform of the Ramsey data for $f_{01}$ as a function of $V_\mathrm{g}$ closely resembles the spectroscopy data and is plotted in Fig.~\ref{suppfig:cologne_sample_figure}g. The frequency resolution of the Ramsey measurements should be limited only by $T_2^*$ and there is no power broadening, Stark shifts and Photon-number broadening or splitting to consider. We confirm that continuous-wave spectroscopy and time-domain datasets are consistent with each other. For the magnetic field setting $B=\SI{0.2}{\tesla}$ used in Fig.~\ref{fig:3} in the main text, the measured qubit transition and readout resonator frequencies are listed in Tab.~\ref{supptab:measured_frequencies}. The full spectroscopy and charge dispersion data at different in-plane magnetic field is documented in the repository~\cite{repository} accompanying this manuscript.

\subsection{IBM}
The IBM data was measured on the Hanoi, Falcon r5.11 processor for 20 out of the 27 qubits. The transition frequencies of the IBM transmons were obtained by multi-mode spectroscopy (at a single probe frequency) enabled by Qiskit Pulse~\cite{McKay2018OpenPulse,Alexander2020QiskitPulse} to measure $f_{0j}/j$ for $j=1,2,3,4$. Since the $j=4$ transition frequency was often near the bandwidth limit imposed by Qiskit ($\SI{\pm500}{\mega\hertz}$ from the $j=1$ transition), additional sideband modulation was applied at the pulse level to probe those frequencies. The measured qubit transition and readout resonator frequencies for qubits 0 and 13 are listed in Tab.~\ref{supptab:measured_frequencies}. Spectroscopy data for all qubits is documented in the repository~\cite{repository} accompanying this manuscript.

\begin{table}
\begingroup
\setlength{\tabcolsep}{3pt}
\renewcommand{\arraystretch}{1.7}
\begin{tabular}{ll|cccccc|ccccccc}
\hline\hline
\multicolumn{2}{c|}{\multirow{2}{*}{Sample}}& \multicolumn{6}{c}{$f_{0j}/j$ (GHz)}              & \multicolumn{7}{c}{$f_{\mathrm{res},j}$ (GHz)}                            \\
\multicolumn{2}{c|}{}                       & 1        & 2      & 3      & 4      & 5      & 6      & 0       & 1         & 2       & 3       & 4       & 5       & 6       \\\hline
\multirow{3}{*}{KIT} & CD1                  & 6.0391   & 5.934  & 5.819  & 5.6945 & 5.5588 & --     & 7.4613  & 7.4587    & --      & --      & --      & --      & --      \\
                     & CD2                  & 5.8884   & 5.7777 & 5.6596 & 5.5305 & -- & --     & 7.4615  & 7.4589    & --  & --  & --      & --      & --      \\
                     & CD3                  & 4.7219   & 4.5966 & 4.4590 & 4.3066 & --     & --     & 7.4564  & 7.45561   & 7.45495 & 7.45415 & --      & --      & --      \\\hline
\multicolumn{2}{l|}{ENS}                    & 5.3548   & 5.2678 & 5.1763 & 5.0792 & 4.9758 & 4.8648 & 7.76131 & 7.75608   & 7.75135 & 7.747   & 7.74276 & 7.73902 & 7.73385 \\\hline
\multirow{2}{*}{K\"oln}& \SI{0}{\tesla}       & 7.1095    & 6.977  & --  & --     & --     & --     & 7.5658      & 7.5838        & --      & --      & --      & --      & --      \\
                     & \SI{0.2}{\tesla}     & 5.079    & 4.912  & 4.722  & --     & --     & --     & --      & --        & --      & --      & --      & --      & --      \\\hline
\multirow{2}{*}{IBM} & Q0                   & 5.0354   & 4.8598 & 4.6698 & 4.535  & --     & --     & 7.167   & 7.1636   & --      & --      & --      & --      & --      \\
                     & Q13                  & 4.9632   & 4.791  & 4.599  & 4.4251 & --     & --     & 7.222   & 7.219   & --      & --      & --      & --      & --      \\\hline\hline
\end{tabular}
\endgroup
\caption{\textbf{Spectroscopy data for the measured samples.} The qubit frequencies are listed as multi-photon transitions $f_{0j}/j$ between ground state and state $j$. The resonator frequencies $f_{\mathrm{res},j}$ are given for the transmon in state $j$. For K\"oln and IBM, we only show two selected field settings and qubits, respectively. The complete set of frequencies is available in the repository~\cite{repository} accompanying this manuscript.}
\label{supptab:measured_frequencies}
\end{table}

\clearpage

\section{Molecular dynamics simulation of junction growth}
\label{sec:moleculardynamics}

All molecular dynamics (MD) calculations are based on ReaxFF force fields \cite{ReaxFF} developed by Hong and van Duin \cite{duin}, as implemented in {\sc Gulp}  \cite{gulp}. ReaxFF allows for both the description of oxygen molecules dissociation on the surface as well as the formation of new bonds (i.e.~Al-O).
The visual representation of the structures is made with xcrysden \cite{xcrysden} and gdis \cite{gdis}.

\subsection{System construction}

We consider two distinct cases for the geometric structure of the surfaces, namely Al(100) and Al(111). Both represent opposite extremes that we average over because the orientation of the Al crystals in the sample is not known. The spacing between layers in terms of the lattice constant $a$ in Al(100) (Al(111)) is $a$ ($a/\sqrt{3}$).
A similar approach using Al(100) and Al(111) to investigate junction properties was recently proposed in~\cite{Cyster2021SimulatingFabricationAlOxJunctions}.

\begin{figure}[htp]
	\begin{center}
		\includegraphics[height=3.5 cm, angle=0]{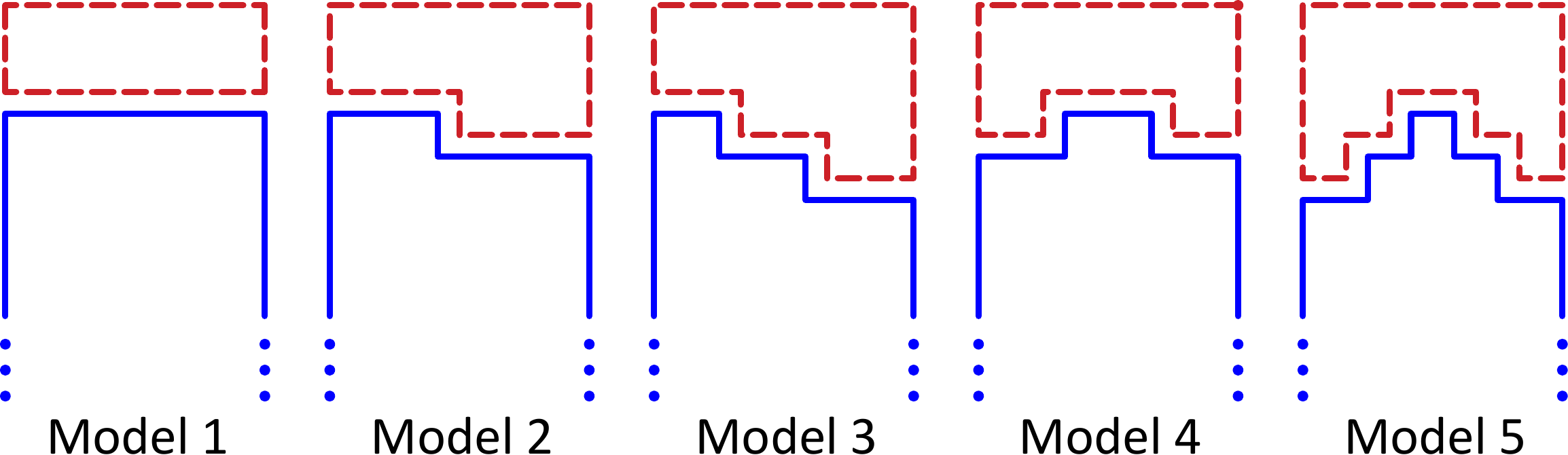}
		\caption{\textbf{Schematic representation of the five surface models considered.}  We illustrate the starting conditions for the MD. The aluminum volume is schematized in blue and the oxygen volume in red. Model 1 describes an ideal smooth surface while Models 2 to 5 simulate closer to realistic geometries of rough aluminum surfaces, with step-like and respectively island-like defects.}    
	\label{comp_models}
	\end{center}	
\end{figure}

\begin{figure}[htp]
	\begin{center}
		\includegraphics[width=\textwidth]{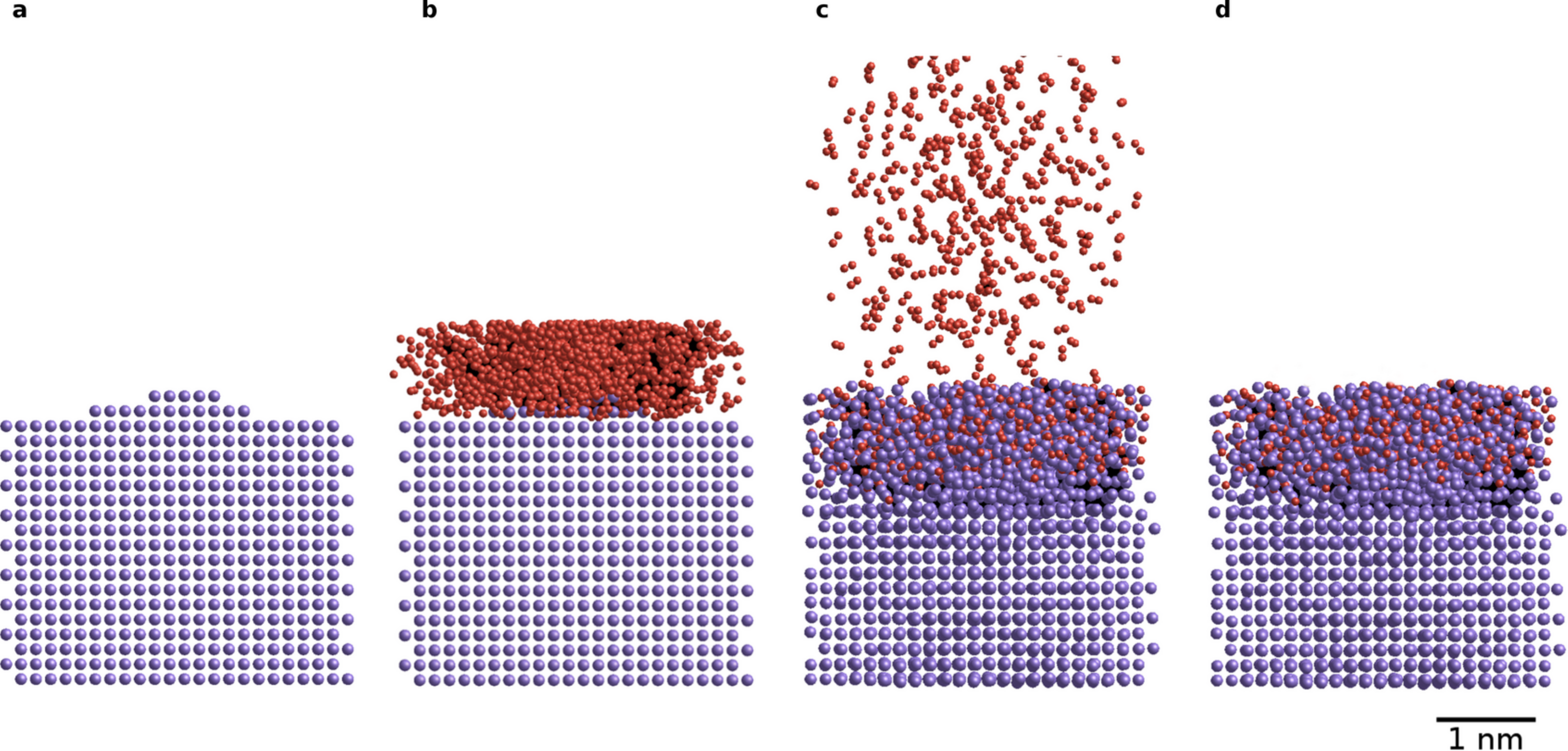}
		\caption{\textbf{The stages of creating the molecular dynamics models (aluminum in violet and oxygen in red).} The image shows an Al(100) model. \textbf{a} The initial construction of a Model 5 (double island) from Al atoms in a face-centered cubic crystal structure. \textbf{b} Oxygen atoms are randomly added to the model; the minimum distance between Al and O atoms is \SI{1}{\text{\AA}} and the maximum is \SI{10}{\text{\AA}}. \textbf{c} The system is propagated for 1500 steps of Verlet molecular dynamics at \SI{300}{\kelvin}, with a time step length of \SI{2}{\femto\second}. \textbf{c} The remaining free oxygen is removed from the system, resulting in aluminum structures wrapped in a layer of AlO$_\text{x}$ that can be further used in surface-like or junction-like configurations for the analysis of geometric parameters.}
	\label{suppfig:md_model_stages}
	\end{center}	
\end{figure}

Al(111) and Al(100) are generated with a bulk parameter of $a=\SI{4.027}{\text{\AA}}$ in both ideal (flat) configuration as well as with several types of defects (steps or islands) as schematically represented in Fig.~\ref{comp_models}. They were allowed to interact with gaseous oxygen atoms placed on top. The initial ratio between oxygen and aluminum is approximately 1:2, i.e.~the initial oxygen represents a third of the total number of atoms in the system (12x12x20=2880 Al atoms and 1500 O atoms). For each model the system was propagated at \SI{300}{\kelvin} using a Verlet-type algorithm for \SI{3}{\pico\second}, with a time step of \SI{2}{\femto\second}, as illustrated in Fig.~\ref{suppfig:md_model_stages} for model~5 (cf.~Fig.~\ref{comp_models}). During propagation, the aluminum layer incorporates oxygen atoms (Fig.~\ref{suppfig:md_model_stages}c). Since we are only interested in the resulting solid structure, the oxygen atoms that remain unbound are removed after the Verlet propagation, as shown in Fig.~\ref{suppfig:md_model_stages}d.

The single layer configuration assumes periodic boundary conditions of the unit cell in the x and y directions; in this case z=infinite (2D type simulation), which means that there is no interaction between the last Al atoms from bulk and the first AlO$_\text{x}$ layer. The multi-layer configuration used to construct Josephson junctions assumes periodic boundary conditions of the unit cell in all three directions: x, y, z. In this case $\text{z}=\SI{40}{\text{\AA}}$, which implies dynamic interaction between the last Al layer and the AlO$_\text{x}$ layer in the neighboring unit cell. All models were further propagated for \SI{4}{\pico\second} in Verlet molecular dynamics at \SI{300}{\kelvin} with a time step of \SI{2}{\femto\second}.

\subsection{Molecular dynamics results}

After performing the MD relaxation, we observe differences between Al(100) and Al(111). The data is tabulated in Table~\ref{tab_Al_O_ratio}. On average, Al(100) accepts up to \SI{2}{\percent} more oxygen, where the percentage refers to the total number of atoms. We also see that surfaces with defects generally accept more oxygen.

\begin{table}[htp]
	\begin{tabular}{c|cc}
		\hline
		Model           & Al(100) & Al(111)    \\
		\hline
		Model 1         & 16.0  &	13.9      \\
		Model 2         & 14.3  &	15.0      \\
		Model 3         & 15.4  &	13.8      \\
		Model 4         & 15.7  &	14.7      \\
		Model 5         & 20.0  &	15.4      \\
		\hline
		Average         & 16.3  &	14.6      \\
	\end{tabular}
	\caption{\textbf{Fraction of oxygen atoms incorporated in the oxide layer, with respect to the total number of atoms in the model}. Values are expressed in \%.}
	\label{tab_Al_O_ratio}
\end{table}

\begin{figure}[htp]
	\begin{center}
		\includegraphics[height=10.0 cm, angle=0]{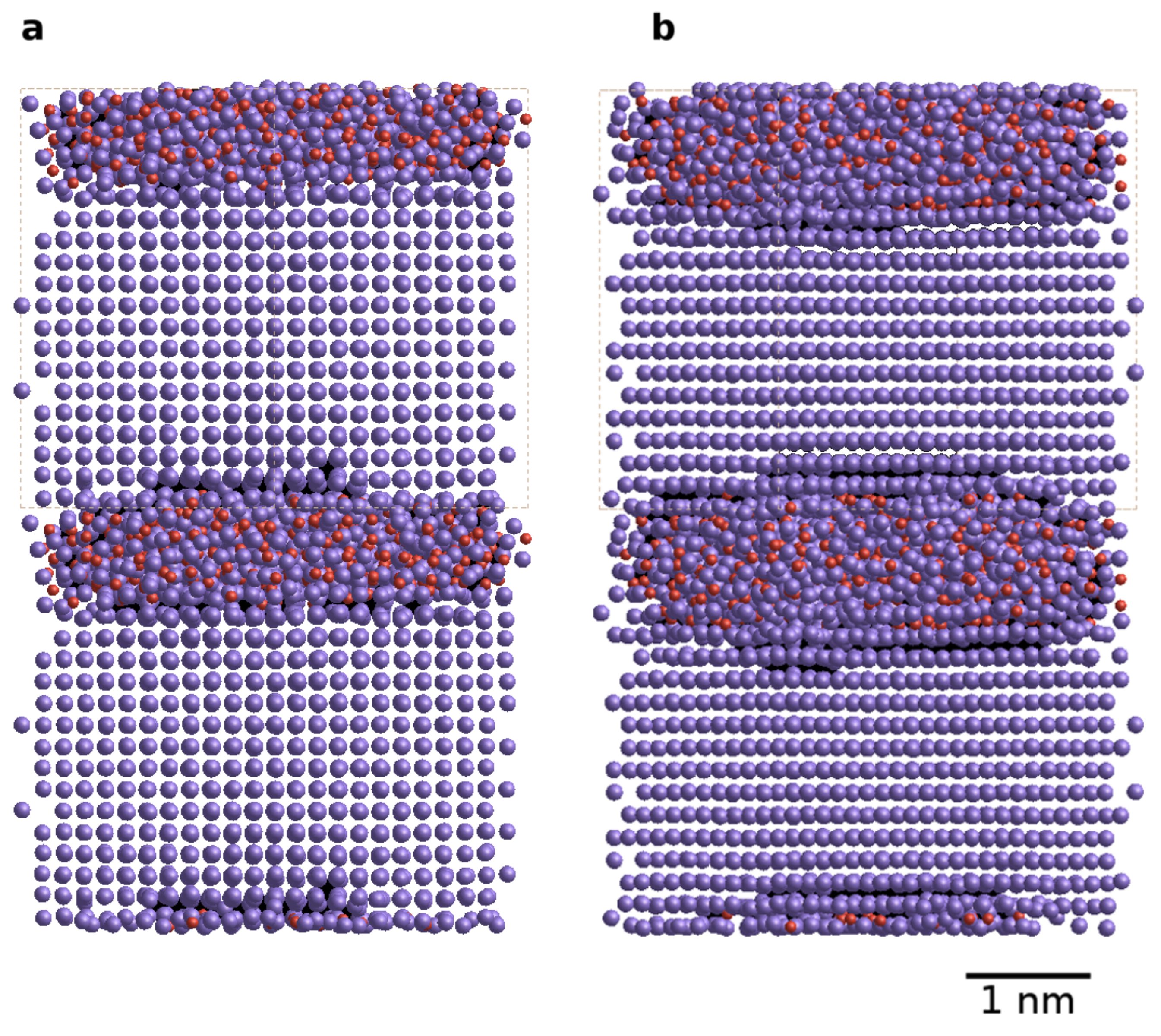}
		\caption{\textbf{Molecular model (aluminum in violet and oxygen in red) of the junction barrier thickness for different crystalline orientations.} Model 1---ideal smooth surface---in multi-layer (junction) for (a) Al(100) and (b) Al(111). The layer formed on the Al(111) surface is up to \SI{2}{\text{\AA}} thicker compared to the one formed on Al(100).}
	\label{fig_JJoxide_model_comparison}
	\end{center}	
\end{figure}

The thickness of the AlO$_\text{x}$ layer was calculated for open surfaces and junctions in the case of all five surfaces models for both Al(100) and Al(111). The barrier thickness is defined as the difference between the coordinates of the first and the last atom that forms it.
On average, the AlO$_\text{x}$ layer formed on Al(111) is \SI{2}{\text{\AA}} thicker than the one formed on Al(100), as documented in table~\ref{tab_layer_thickness} and illustrated in Fig.~\ref{fig_JJoxide_model_comparison}. This is surprising considering that Al(111) accepts generally fewer O atoms than the Al in symmetry (100). If we compare the Al-O bond lengths, we obtain only a slightly larger length for Al(111), i.e.~$(1.96 \pm 0.16)\,\text{\AA}$ compared to $(1.95 \pm 0.17)\,\text{\AA}$ for Al(100). Therefore the difference in the length of the Al-O bonds formed on the two types of bulk does not explain the difference in the oxide thickness of up to \SI{2}{\text{\AA}}.

\begin{table}[htp]
	\begin{tabular}{c|cc|cc}
		\hline
		Model  &Al(100) junction  &Al(100) surface  & Al(111) junction & Al(111) surface      \\
		\hline
		Model 1         & 11.9 &	10.2 & 15.6 & 10.6\\
		Model 2         & 13.3 &	8.4  & 16.4 & 10.1\\
		Model 3         & 13.5 &	9.8  & 15.8 & 9.8\\
		Model 4         & 13.8 &	10.1 & 16.9 & 9.9\\
		Model 5         & 14.8 &	10.8 & 17.1 & 11.9\\
		\hline
		Average         & 14.5 &	9.9  & 16.4 & 10.5\\
	\end{tabular}
	\caption{\textbf{AlO$_\text{x}$ layer thickness for all models.} It is calculated as the difference between the position on the z axis of the first and last atom that forms the oxide layer. All values are expressed in \AA.} 
	\label{tab_layer_thickness}
\end{table}

However, the analysis of the structure and composition of the systems shows that different chemical bonds are formed between aluminum and oxygen in the AlO$_\text{x}$ layer, depending on the surface orientation. In Al(111), even though it incorporates less oxygen compared to Al(100), a higher percentage of oxide and sub-oxide is formed, resulting in a thicker barrier. This fact, correlated with the bond lengths and the penetration depth of oxygen, indicates that the thickness of the AlO$_\text{x}$ layer depends on the type and quality of the surface. The rougher the surface, the more oxygen it absorbs and consequently the larger the barrier thickness. Since the typical Josephson junction barrier is obtained by oxidizing a poly-crystalline film with grains much smaller than the junction and with random orientations, we expect the barrier to be inhomogeneous: conduction channels have different transparencies, corresponding to the crystalline orientation and oxide thickness at their respective positions.  

\subsection{Additional STEM images of JJ barriers}\label{sec:TEM}

\begin{figure}[htp]
  \centering
  \includegraphics[width=\textwidth]{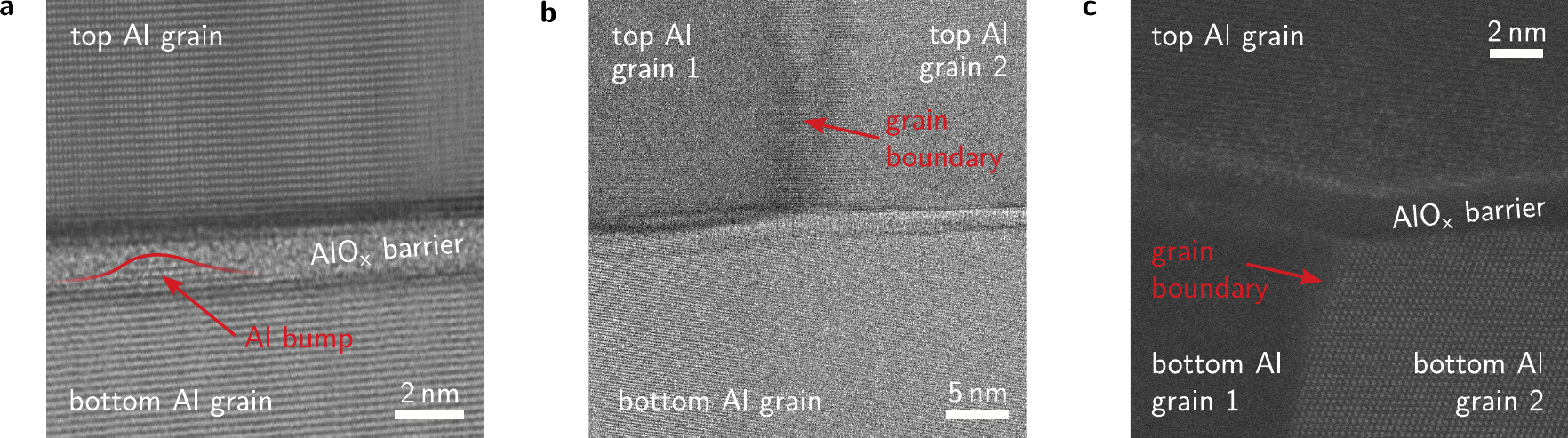}
  \caption{\textbf{Additional STEM images of Al-AlO$_x$-Al junctions.}
  \textbf{a}~Bright field (BF) STEM image of a barrier grown under static oxidation at \SI{1}{\milli\bar} for \SI{30}{\minute}. As the zone axes of both electrodes are pointing in different directions, the image is acquired with both axes misaligned, confirming the crystallinity of both the upper and lower electrode simultaneously. The rotational misalignment of the crystals explains the linear patterns compared to the dotted pattern in Fig.~\ref{fig:2}c in the main text. In this image, the Al grain of the bottom electrode is not homogeneously oxidized. This leads to crystalline Al reaching into the barrier region (cf.~linear pattern indicated by the red arrow and line) -- thus reducing the barrier thickness locally.
  \textbf{b}~A zoomed-out BF-STEM image of a different part of the barrier in panel a showing a grain boundary in the top Al electrode, indicated by the red arrow.
  \textbf{c}~ High-angle annular dark field (HAADF) STEM image of a different region of the barrier shown in Fig.~\ref{fig:2}c in the main text. This image shows a grain boundary of the bottom Al electrode, indicated by the red arrow.
  }
  \label{suppfig:tem}
\end{figure}

The Al-AlO$_x$-Al barriers resulting from MD models are in qualitative agreement with images obtained from STEM on JJ barriers fabricated by e-beam deposition of aluminum and thermal oxidation, as illustrated in Fig.~\ref{suppfig:tem}. 

\end{document}